\documentclass[aps,amsmath,amssymb,superscriptaddress,twocolumn,nofootinbib,showpacs,preprintnumbers,notitlepage,prd]{revtex4-2}
\allowdisplaybreaks

\usepackage{amsmath}
\usepackage{amsfonts}
\usepackage{amssymb}
\usepackage{physics}
\usepackage{slashed}
\usepackage{simplewick}
\usepackage{mathtools}
\usepackage{braket}
\usepackage{bm} 
\usepackage{dsfont}

\usepackage{siunitx}
\usepackage{orcidlink}
\usepackage{placeins}
\usepackage{graphicx}
\usepackage{float}
\usepackage[caption=false]{subfig}

\usepackage{listings}

\usepackage{hyperref}
\hypersetup{ 
    pdfnewwindow=true,      
    colorlinks=true,       
    allcolors=[RGB]{31 119 180}
} 
\usepackage{tikz}
\usetikzlibrary{calc}
\usetikzlibrary{patterns}
\usetikzlibrary{patterns}
\usetikzlibrary{positioning,decorations.pathmorphing,decorations.markings,snakes}

\usepackage{xcolor}
\definecolor{jpac-blue}{rgb}{0.12,0.47,0.71}
\definecolor{jpac-orange}{rgb}{1,0.5,0.05}
\definecolor{jpac-green}{rgb}{0,0.62,0.45}
\definecolor{jpac-red}{rgb}{0.84,0.15,0.15}
\definecolor{jpac-purple}{rgb}{0.58,0.40,0.71}
\definecolor{jpac-brown}{rgb}{0.54,0.34,0.29}
\definecolor{jpac-pink}{rgb}{0.89,0.47,0.76}
\definecolor{jpac-grey}{rgb}{0.5,0.5,0.5}
\definecolor{jpac-gold}{rgb}{0.74,0.74,0.13}
\definecolor{jpac-aqua}{rgb}{0.09,0.75,0.81}
\definecolor{jpac-white}{rgb}{1,1,1}

\usepackage[capitalise]{cleveref}

\usepackage{mathrsfs}
\usepackage{pgfplots,pgfplotstable}
\usetikzlibrary{pgfplots.groupplots}
\usetikzlibrary{decorations.markings}
\usepgfplotslibrary{fillbetween}
\pgfplotsset{compat=1.14}
\usepackage{tikz}
\usepackage{tikz-3dplot}
\usetikzlibrary{shapes.geometric, arrows.meta, positioning}
\usetikzlibrary{calc}
\tikzset{->-/.style={decoration={
  markings,
  mark=at position .5 with {\arrow{>}}},postaction={decorate}}}
\tikzset{>=stealth}
\makeatletter
\tikzoption{canvas is plane}[]{\@setOxy#1}
\def\@setOxy O(#1,#2,#3)x(#4,#5,#6)y(#7,#8,#9)%
  {\def\tikz@plane@origin{\pgfpointxyz{#1}{#2}{#3}}%
   \def\tikz@plane@x{\pgfpointxyz{#4}{#5}{#6}}%
   \def\tikz@plane@y{\pgfpointxyz{#7}{#8}{#9}}%
   \tikz@canvas@is@plane
  }
\makeatother 

\newcommand\bsub{\begin{subequations}}
\newcommand\esub{\end{subequations}}

\usepackage{multirow}
\usepackage{dcolumn}
\usepackage{tabularx}

\usepackage{array}   
\newcolumntype{L}{>{$}l<{$}} 
\newcolumntype{R}{>{$}r<{$}}
\newcolumntype{C}{>{$}c<{$}}
\usepackage{xspace}
\usepackage[normalem]{ulem}
\usepackage[format=plain,justification=raggedright,singlelinecheck=false]{caption}
\usepackage[format=plain,justification=raggedright,singlelinecheck=false]{subcaption}
\usepackage{nameref}

\newcommand{\mevnospace}{\ensuremath{{\mathrm{\,Me\kern -0.1em V}}}}
\newcommand{\gevnospace}{\ensuremath{{\mathrm{\,Ge\kern -0.1em V}}}}
\newcommand{\tevnospace}{\ensuremath{{\mathrm{\,Te\kern -0.1em V}}}}

\newcommand{\gev}{\gevnospace\xspace}

\newcommand{\eg}{{\it e.g.}\xspace}
\newcommand{\Eg}{{\it E.g.}\xspace}
\newcommand{\cf}{{\it cf.}\xspace}

\newcommand{\ie}{{\it i.e.}\xspace}

\newcommand{\comment}[1]{}

\usepackage{diagbox}
\usepackage{todonotes}
\usepackage{enumitem}

\newcommand{\DelBar}{\Delta(1232)}

\newcommand{\helgamma}{\lambda_\gamma}
\newcommand{\helDel}{\lambda_\Delta}
\newcommand{\helproton}{\lambda_N}
\newcommand{\helgammat}{\mu_\gamma}
\newcommand{\helDelt}{\mu_\Delta}
\newcommand{\helprotont}{\mu_{\bar{N}}}
\DeclareMathOperator{\im}{Im}
\DeclareMathOperator{\re}{Re}
\newcommand{\oneh}{\frac{1}{2}}
\newcommand{\threeh}{\frac{3}{2}}

\newcommand{\AGH}{AGH University of Krakow, Faculty of Physics and Applied Computer Science, PL-30-059 Krak\'ow, Poland}
\newcommand{\catania}{INFN Sezione di Catania, I-95123 Catania, Italy}
\newcommand{\ceem}{Center for  Exploration  of  Energy  and  Matter, Indiana  University, Bloomington,  IN  47403,  USA}
\newcommand{\indiana}{Department of Physics, Indiana  University, Bloomington,  IN  47405,  USA}
\newcommand{\jlab}{Theory Center, Thomas  Jefferson  National  Accelerator  Facility, Newport  News,  VA  23606,  USA}
\newcommand{\messina}{Dipartimento di Scienze Matematiche e Informatiche, Scienze Fisiche e Scienze della Terra, Universit\`a degli Studi di Messina, I-98122 Messina, Italy}

\newcommand{\ub}{Departament de F\'isica Qu\`antica i Astrof\'isica and Institut de Ci\`encies del Cosmos, Universitat de Barcelona, E-08028 Barcelona, Spain}
\newcommand{\uned}{Departamento de F\'isica Interdisciplinar, Universidad Nacional de Educaci\'on a Distancia (UNED), E-28040 Madrid, Spain}

\newcommand{\odu}{Department of Physics, Old Dominion University, Norfolk, VA 23529, USA}
\newcommand{\MIT}{Center for Theoretical Physics - A Leinweber Institute, Massachusetts Institute of Technology, Cambridge, MA 02139, USA}
\newcommand{\icn}{Instituto de Ciencias Nucleares,
Universidad Nacional Aut\'onoma de M\'exico, Ciudad de M\'exico 04510, Mexico}

\begin{document}
\preprint{JLAB-THY-26-4685}
\preprint{MIT-CTP/6028}

\title{Mechanisms of high energy polarized photoproduction of \texorpdfstring{$\pi^{-}\Delta^{++}$}{}}%

\author{Vanamali~\surname{Shastry}\orcidlink{0000-0003-1296-8468}}
\email{vanamalishastry@gmail.com}
\affiliation{\ceem}
\affiliation{\indiana}
\author{{\L}ukasz Bibrzycki\orcidlink{0000-0002-6117-4894}}
\affiliation{\AGH}
\author{Vincent~Mathieu\orcidlink{0000-0003-4955-3311}}
\affiliation{\ub}
\author{Gl\`oria~\surname{Monta\~na}\orcidlink{0000-0001-8093-6682}}
\affiliation{\ub}
\author{Alessandro~\surname{Pilloni}\orcidlink{0000-0003-4257-0928}}
\affiliation{\messina}
\affiliation{\catania}
\author{C\'esar~\surname{Fern\'andez-Ram\'irez}\orcidlink{0000-0001-8979-5660}}
\affiliation{\uned}
\author{Robert~J.~\surname{Perry}\orcidlink{0000-0002-2954-5050}}
\affiliation{\MIT}
\author{Arkaitz~\surname{Rodas}\orcidlink{0000-0003-2702-5286}}
\affiliation{\jlab}
\affiliation{\odu}
\author{Adam~P.~\surname{Szczepaniak}\orcidlink{0000-0002-4156-5492}}
\affiliation{\ceem}
\affiliation{\indiana}
\affiliation{\jlab}
\author{Daniel~\surname{Winney}\orcidlink{0000-0002-8076-243X}}
\affiliation{\icn}
\collaboration{Joint Physics Analysis Center}

%
%
\begin{abstract}
    We present an amplitude analysis of high-energy polarized photoproduction of $\pi^-\Delta^{++}$ within a Regge exchange framework. A Regge amplitude model incorporating $\pi$, $\rho$, $b_1$, and $a_2$ trajectory exchanges is fit simultaneously to spin density matrix elements measured by the GlueX experiment at photon energies of $E_\gamma = 8.2$--$8.8$ GeV and differential cross section data from SLAC. By including SDME data, the fit constrains not only the magnitudes but also the relative phases of the helicity amplitudes. The results confirm the dominance of pion exchange at small momentum transfer, while natural parity exchanges become significant at larger $t$. We analytically continue the $s$-channel amplitude to the $t$-channel, taking care of the kinematical singularities, and isolate the dynamical residues at the meson poles. The extracted $\pi N\Delta$ coupling constant is found to be consistent with the value obtained from the decay width of the $\Delta(1232)$. For the $\rho N\Delta$, $b_1 N\Delta$, and $a_2 N\Delta$ vertices, first extractions of the relevant coupling constants are provided.
\end{abstract}

\maketitle

\section{Introduction}

A central goal of hadron spectroscopy is to understand how the observed spectrum arises from interactions among the fundamental constituents of quantum chromodynamics (QCD). In recent years, the focus has shifted to searching for {\it exotic states} -- hadrons that do not fit into the conventional quark model classification -- and to understanding their properties. While most evidence for unconventional states has been found in the heavy quark sector~\cite{Brambilla:2019esw,Mezzadri:2022loq,Chen:2022asf,Husken:2024rdk}, evidence from both theory and experiment now exists for the presence of light exotics. A prominent role is played by the $\pi_1(1600)$, a candidate for a \textit{hybrid} meson -- a state with a valence gluon~\cite{Meyer:2015eta}.\par
To date, all experimental evidence for the $\pi_1(1600)$ has come from hadroproduction experiments~\cite{ParticleDataGroup:2024cfk}. It is therefore of interest to study this state using other production mechanisms. The GlueX experiment at Jefferson Lab aims to study hybrid mesons like the $\pi_1(1600)$ using diffractive photoproduction of multi-meson final states~\cite{GlueX:2012idx}. It was recently shown that in GlueX, the hybrid was more likely to be produced in a charge exchange process with a $\DelBar$ in the final state than in a neutral exchange process with a recoil proton~\cite{GlueX:2024erj}. Thus, it becomes essential to understand the mechanism behind the charge exchange photoproduction process, starting with the simplest $\pi\Delta$ final state. \par

At GlueX energies, meson photoproduction is naturally described by Regge theory. In this approach, many key features of reaction amplitudes can be understood efficiently in terms of exchange of ``Reggeons'' in the crossed channel, \ie, quasiparticles characterized by their Regge trajectories and Regge couplings~\cite{Winney:2025tla}. As with low-energy effective Lagrangian approaches, Regge theory does not predict interaction strengths, which therefore must be determined from analysis of experimental data or estimated from other theoretical inputs. However, if Regge poles dominate, the factorization of vertices in high-energy processes allows us to use couplings derived from one process to determine the corresponding vertices in other processes.

Previous attempts to determine Regge couplings relied on Lagrangian models that successfully explained the general features of the cross section and beam spin asymmetry (BSA)~\cite{Barbour:1974nr,Clark:1974xj,Gotsman:1969rsi,Yu:2016jfi,JointPhysicsAnalysisCenter:2017del}. However, these observables do not constrain the relative phases between the helicity amplitudes. This ambiguity can be resolved by examining the angular distributions of the decay products of the produced resonances.

This information is encoded in the spin density matrix elements (SDMEs) of the reaction, which have recently been measured with high precision by GlueX for the photoproduction of $\pi\Delta$ at photon energy $E_\gamma = 8.2$--$8.8\gev$~\cite{GlueX:2024dbr}. We use this data, along with the SLAC cross-section data~\cite{Boyarski:1968dw}, to determine the aforementioned couplings. 
This constrains both the magnitudes and the relative phases of the helicity amplitudes.
We then analytically continue the amplitude into the crossed channel, where the exchanges can be related to physical states, and extract the corresponding residues.
To do so, one needs to construct partial-wave amplitudes of definite parity by removing the kinematical singularities that would hinder the analytical continuation. This is a nontrivial task that we describe in this paper.

The structure of the paper is as follows. In Sec.~\ref{sec:KPO}, we introduce the reaction kinematics. In Sec.~\ref{sec:formalism}, we develop the formalism required to continue to the crossed channel. In Sec.~\ref{sec:model}, we describe the Regge model used to analyze the experimental data. The results of the work are presented, and their implications and consequences are discussed in Sec.~\ref{sec:RnD}. Our conclusions are provided in Sec.~\ref{sec:Conclusions}.

\section{Kinematics at large $s$\label{sec:KPO}}
We define the reference frames corresponding to the $s$-channel and $t$-channel center-of-mass 
and the associated kinematical variables. 
\begin{figure}
    \centering
    \begin{tikzpicture}
        \draw[pattern=north east lines,thick] (0,0) circle (0.75cm);
        \draw[thick,decorate,decoration={snake}] (-2.12,1.41) -- (-0.53,0.53);
        \draw[thick] (-2.12,-1.41) -- (-0.53,-0.53);
        \draw[thick,dashed] (2.12,1.41) -- (0.53,0.53);
        \draw[thick,double] (2.12,-1.41) -- (0.53,-0.53);
        \node at (-2.5,1.5) {$\gamma$};
        \node at (-2.5,-1.5) {$p$};
        \node at (2.5,1.5) {$\pi^-$};
        \node at (2.5,-1.5) {$\Delta^{++}$};
        \draw[thick,->] (-3,0) -- (-2,0);
        \draw[thick,->] (0,2.5) -- (0,1.5);
        \node at (-2.5,0.25) {$s$};
        \node at (0.25,2) {$t$};
        \draw[thick,->] (-1.5,1.5) -- (-1,1.25);
        \draw[thick,->] (-1.5,-1.5) -- (-1,-1.25);
        \draw[thick,->] (1,1.25) -- (1.5,1.5);
        \draw[thick,->] (1,-1.25) -- (1.5,-1.5);
        \node at (-1.25,1.65) {$p_\gamma$};
        \node at (-1.25,-1.65) {$p_N$};
        \node at (1.25,1.65) {$p_\pi$};
        \node at (1.25,-1.65) {$p_\Delta$};
    \end{tikzpicture}
    \caption{    Schematic diagram for the process $\gamma p \to \pi^-\Delta^{++}$. In the physical region for this process, $s\geq(m_\Delta + m_\pi)^2$, and negative $t$. This process is related to the process $\gamma \pi^+ \to \bar{p} \Delta^{++}$ via crossing symmetry, when continued to $t \geq(m_\Delta + m_p)^2$ at negative $s$.  }
    \label{fig:schematics}
\end{figure}
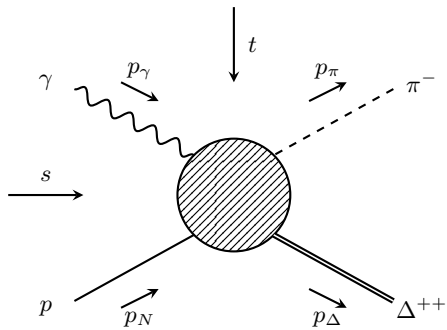

The kinematics of the photoproduction process are shown in Fig.~\ref{fig:schematics}. 
The four-momenta of the external states are labeled as $p_\gamma$, $p_\pi$, $p_N$\, $p_\Delta$ where the subscript denotes the particle. The Mandelstam variables are defined as,
\bsub
\begin{align}
    s &= (p_\gamma + p_N)^2\, = (p_\Delta + p_\pi)^2~, \\
    t &= (p_\gamma - p_\pi)^2\, = (p_\Delta - p_N)^2 \text{ .}
\end{align}\label{eq:Mands}
\esub
The scattering angle in the $s$-channel CM frame ($\theta_s$) is related to the Mandelstam variables as,
\begin{align}
    z_s \equiv\cos\theta_s &= \frac{1}{2\,p^s\,q^s}(t - m_\pi^2 + 2\,q^s\,E_\pi^s)
\end{align}
where $q^s$ and $p^s$ are the magnitudes of the $3$-momenta of the initial and final states respectively and are given by,
\begin{align}
    q^s &= \frac{\lambda^{1/2}(s,0,m_N^2)}{2\sqrt{s}}\,, & p^s &= \frac{\lambda^{1/2}(s,m_\pi^2,m_\Delta^2)}{2\sqrt{s}}\label{eq:moms}
\end{align}
and $E_\pi^s = \sqrt{(q^s)^2 + m_\pi^2}$ is the energy of the pion. In the above expressions $\lambda(a,b,c)$ is the K\"allen function.
 We also define the so-called half-angle factor as,
\begin{align}
    \xi^{(s)}_{\helgamma\helproton\helDel} &= \sqrt{s\frac{1 - z_s}{2}}^{|\helgamma - \helproton + \helDel|}~\sqrt{\frac{1 + z_s}{2}}^{|\helgamma + \helproton - \helDel|} \text{ .}
\end{align} 
We insert an extra factor of $s$ in the definition to simplify the form in the large-$s$ limit.
In the $t$-channel, the reaction is $\gamma  \pi^+ \to \bar{p}  \Delta^{++}$,
and $s$ and $t$ play the role of momentum-transfer and center of mass energy variables respectively.
The scattering angle can be written as,
\begin{align}
    z_t \equiv\cos\theta_t &= \frac{1}{2\,p^t\,q^t} (s - m_N^2 + 2\,q^t\,E_{\bar{N}}^t)
\end{align}
where, $q^t$ and $p^t$ are the magnitudes of the $3$-momenta of the initial and final states respectively and are given by,
\begin{align}
    q^t &= \frac{\lambda^{1/2}(t,0,m_\pi^2)}{2\sqrt{t}}\,, & p^t &= \frac{\lambda^{1/2}(t,m_N^2,m_\Delta^2)}{2\sqrt{t}}\label{eq:momt}
\end{align}
and $E_{\bar{N}}^t = \sqrt{(q^t)^2 + m_N^2}$ is the energy of the antiproton. The $t$-channel $\Delta\bar{p}$ threshold and pseudothreshold are,
\begin{align}
    t_{th} &= (m_\Delta + m_N)^2 \,, & t_{pth} &= (m_\Delta - m_N)^2
\end{align}
respectively, so that $p^t = \sqrt{(t-t_{th})(t-t_{pth})}\Big/2\sqrt{t}$. The $\pi\gamma$ threshold and pseudothreshold are both equal to $m_\pi^2$. The half-angle factor is defined as,
\begin{align}
    \xi^{(t)}_{\helgammat\helprotont\helDelt} &= \sqrt{\frac{1 + z_t}{2}}^{|\helgammat - \helprotont + \helDelt|}~\sqrt{\frac{1 - z_t}{2}}^{|\helgammat + \helprotont - \helDelt|} \text{ .}
\end{align}\par
Since all particles except the pion have spin, the reaction amplitude can be written as:
    \begin{equation}
    T^{(s)}_{\helgamma,\helproton,\helDel}(s,t) \equiv \mel{\helDel}{T}{\helgamma, \, \helproton}~,
    \end{equation}
which not only depends on invariant variables but also on the helicity projection of each state. In this case, $\lambda_i$ are the defined as the helicities in the $s$-channel frame which we denote by the superscript. We can define the analogous amplitude for the $t$-channel reaction:
    \begin{equation}
    T^{(t)}_{\helgammat,\helprotont,\helDelt}(t,s) \equiv \mel{\helprotont \,\helDelt}{T}{\helgammat}~,
    \end{equation}
in terms of the $t$-channel helicities, $\mu_i$.
Since helicities are not Lorentz invariant, the amplitudes mix under crossing to the $t$-channel~\cite{Trueman:1964zzb}. The transformation of the amplitude from the $s$-channel frame to the $t$-channel frame is given by the crossing relations as~\cite{Trueman:1964zzb,Martin:1970hmp,Collins:1971ff},
\begin{widetext}
\begin{align}
    T^{(t)}_{\helgammat\helprotont\helDelt} (t,s) &= \sum_{\helgamma,\helproton,\helDel}\delta_{\helgamma,\helgammat}\, d^\oneh_{\helprotont,\helproton}(\omega_N)\, d^\threeh_{\helDelt,\helDel}(\omega_\Delta)\, T^{(s)}_{\helgamma\helproton\helDel}(s,t) \label{eq:xingrel}
\end{align}
\end{widetext}
where $T^{(t)}_{\helgammat\helprotont\helDelt} (t,s)$ and $T^{(s)}_{\helgamma\helproton\helDel}(s,t)$ are the helicity amplitudes in the $t$-channel and $s$-channel frame respectively, $d^\oneh_{\helprotont,\helproton}(\omega_N)$, and $d^\threeh_{\helDelt,\helDel}(\omega_\Delta)$ are the Wigner-$d$ matrices, and the crossing angles are defined as,
\begin{align}
    \cos\omega_N &= \frac{-4\sqrt{s\,t}\,E_N^s\,E_{\bar{N}}^t - 2m_N^2 S}{4\sqrt{s\,t}\,q^s\,p^t} \\
    \cos\omega_\Delta &= \frac{4\sqrt{s\,t}\,E_\Delta^s\,E_\Delta^t - 2m_\Delta^2 S}{4\sqrt{s\,t}\,p^t\,p^s}
\end{align}
where, $S=m_\Delta^2 - m_N^2 - m_\pi^2$, $E_i^j$ are the energy of the state $i$ in the frame $j$. These can be obtained from Eq.~\eqref{eq:moms} and Eq.~\eqref{eq:momt} using the energy-momentum relations. 

Since we are interested in studying the process in the Regge region, \ie at large $s$ and small $t$, we expand the expressions for the kinematical quantities as,
\begin{subequations}
\begin{align}
    q^s &\simeq p^s \simeq \frac{\sqrt{s}}{2} + \mathcal{O}\left(\frac{1}{\sqrt{s}}\right)\text{ ,}\\
    z_s &\simeq 1 + \frac{2t}{s} + \mathcal{O}\left(\frac{1}{s^{2}}\right)\text{ ,}\\
    \xi^{(s)}_{\helgamma\helproton\helDel} &\simeq \sqrt{-t}^{|\helgamma - \helproton + \helDel|}+ \mathcal{O}\left(\frac{1}{s}\right)\text{ ,}\label{eq:halfangleslarges}\\
    z_t &\simeq \frac{s}{2p^t q^t} + \mathcal{O}\left(s^{0}\right)\text{ ,}\\
    \cos\omega_N &\simeq  -\frac{E_{\bar{N}}^t}{p^t}+ \mathcal{O}\left(\frac{1}{s}\right)\text{ ,}\\
    \cos\omega_\Delta &\simeq  \frac{E_\Delta^t}{p^t}+ \mathcal{O}\left(\frac{1}{s}\right)\text{ .}
\end{align}
\end{subequations}

\section{Formalism\label{sec:formalism}}

As mentioned, we perform the fit in the $s$-channel, where the physical reaction occurs, and aim to analytically continue it to the $t$-channel, where we can naturally relate the exchanges to physical states and extract information on how they couple to the $\Delta \bar{p}$ final state.
Helicity amplitudes contain singularities dictated by the angular-momentum structure. The Wigner matrices that relate the amplitude between the $s$-channel and the $t$-channel also contain singularities that must be understood. All these singularities have kinematical origins and would lead to unphysical results if the amplitude were naively evaluated at the resonance poles. To do it correctly,
we need to define the corresponding parity conserving helicity amplitudes (PCHAs), whose kinematical singularity structure has been characterized~\cite{Hara:1964zza,Wang:1966zza,Cohen-Tannoudji:1968lnm,Jackson:1968rfn,Collins:1971ff}.
Removing these singularities also reveals the residual behavior of the amplitude, which contains the dynamical information. In the following, we will first show how to extract the residues from the $t$-channel amplitude, and then show that crossing the $s$-channel amplitude to the $t$-channel does not introduce spurious singularities, making the extraction of residues well posed.

\subsection{Amplitude in the $t$-channel frame and extraction of residues\label{sec:pwagen}}
The definite-parity partial-wave helicity amplitudes in the $t$-channel, also called parity-conserving helicity amplitudes (PCHAs), are defined as,\footnote{The precise expressions for the PCHAs and $\hat{d}^{J\pm}$ depend on the convention chosen for the Wigner-$d$. We adopt the same convention as  Ref.~\cite{Martin:1970hmp,JPAC:2018dfc}.}
\begin{align}
    &\hat{T}^{\eta}_{\helgammat\helprotont\helDelt} (t,s) \notag\\
    &= \hat{T}^{(t)}_{\helgammat\helprotont\helDelt} - \eta (-1)^{\helprotont - \helDelt + M}~\hat{T}^{(t)}_{-\helgammat\helprotont\helDelt}\text{ ,}\label{eq:ksfpchadef}
\end{align}
where $M=\text{max}(|\helgammat|,|\helprotont-\helDelt|)$, $\eta$ is naturality  and 
\begin{equation}
   \hat{T}^{(t)}_{\helgammat\helprotont\helDelt}(t,s) = T^{(t)}_{\helgammat\helprotont\helDelt}(t,s)\Big/\xi^{(t)}_{\helgammat\helprotont\helDelt}(z_t)\text{ .}\label{eq:ksftchannel}
\end{equation}
removes the kinematic singularities from $T^{(s)}$ associated with the half-angle factor $\xi^{(t)}$.
In the large-$s$ limit, the PCHAs have the structure (see Appendix~\ref{app:pcha} for details),
\begin{align}
    \hat{T}^{\eta}_{\helgammat\helprotont\helDelt} (t,s) &= K^\eta_{\helgammat\helprotont\helDelt} (t) \sum_{J\geq M} \frac{(2J+1)}{4\pi}s^{J-M} \hat{a}^{J\eta}_{\helgammat\helprotont\helDelt}(t)\nonumber\\
    &\qquad\qquad + \mathcal{O}\left( s^{J-M-1}\right)\text{ .} \label{eq:Lagksfpcha0}
\end{align}
where, $K^\eta_{\helgammat\helprotont\helDelt} (t)$ contain the kinematical singularities in $t$. The definite-parity partial-wave amplitudes, $\hat{a}^{J\eta}_{\helgammat\helprotont\helDelt}(t)$, are free of kinematical singularities and contain only dynamical singularities, including resonance poles. In the vicinity of a resonance pole, the residue given by 
\begin{align}
    \mathcal{R}^{J_R\eta_R}_{\helgammat\helprotont\helDelt} &= \lim_{t \to m_R^2} (t-m_R^2)\,\hat{a}^{J_R\eta_R}_{\helgammat\helprotont\helDelt}(t)\label{eq:residueLag}
\end{align}
is given by the product of couplings of the resonances or the external particles. Here $J_R$, $\eta_R = P_R(-1)^{J_R}$, $P_R$, and $m_R$ are the spin, naturality, parity, and mass of resonance of interest. This is valid for exchange of $J\geq M$. Because of the photon, $M \geq 1$. Thus, pion exchange must be treated separately. The procedure for analytically continuing the summation to $J=0$ is detailed in Ref.~\cite{JointPhysicsAnalysisCenter:2024kck}, and its implementation in this analysis is discussed in Appendix~\ref{app:pionEx}. \par

\subsection{Crossing from the $s$-channel}
\label{sec:cross}
In the problem of interest, the physical region of the process is defined in the $s$-channel frame. Hence, the formalism described above must be used in conjunction with the crossing relation given in Eq.~\eqref{eq:xingrel}.
The general form of the $s$-channel helicity amplitude is given by,

\begin{align}
    T^{(s)}_{\helgamma\helproton\helDel}(s,t) &= \xi^{(s)}_{\helgamma\helproton\helDel}(t)~\hat{T}^{(s)}_{\helgamma\helproton\helDel}(s,t)\label{eq:amp}
\end{align}
Since we are interested in extracting the residues of the poles of the $t$-channel, we can write the amplitude as
\begin{align}
    \hat{T}^{(s)}_{\helgamma\helproton\helDel}(s,t) &\sim 
    \mathcal{B}^{J_R\eta_R}_{\helgamma\helproton\helDel} (t)\,\frac{s^{J_R}}{t-m_{R}^2}
        \label{eq:Bsumpoles}
\end{align}
in the vicinity of the pole. Combining this behavior with Eqs.~\eqref{eq:xingrel},\eqref{eq:Lagksfpcha0}, and \eqref{eq:residueLag} we get the residues as,
\begin{widetext}
\begin{align}
     \mathcal{R}^{J_R\eta_R}_{\helgammat\helprotont\helDelt}
      &= \lim_{t \to m_R^2} \Bigg[\left(\frac{4 p^t}{\sqrt{t}}\right)^{\frac{1-\eta_R}{2}}\frac{2p^t\sqrt{t-t_{pth}}^{\eta_R}}{2^{M} \sqrt{t}^{M+N-2+\eta_R}}~\frac{4\pi}{2J_R + 1}\frac{1}{F^{J_R}_{\helgammat\helprotont\helDelt}}\notag\\
      &\qquad\times\sum_{\helgamma,\helproton,\helDel} \delta_{\helgammat,\helgamma} 
      \mathcal{C}^{\eta_R}_{\helprotont\helDelt;\helproton\helDel}(t)\frac{\sqrt{-t}^{|\helgamma - \helproton + \helDel|}}{i^{|\helgammat + \helprotont - \helDelt|}} \mathcal{B}^{J_R\eta_R}_{\helgamma\helproton\helDel}(t)\Bigg]\text{ .} \label{eq:residueFinal}
\end{align}
\end{widetext}
The details of the derivations and the descriptions of the various terms are given in Appendix~\ref{app:xingapp}. Despite the appearance of square roots in the expression for the residues, they all cancel for all helicities, and the coefficient of $\mathcal{B}^{J_R\eta_R}_{\helgamma\helproton\helDel}$ is free of kinematical singularities. Additional simple poles at $t=0$ may remain for some helicities, as discussed in a previous footnote. Being an isolated singularity, this does not spoil the analytic continuation. Similarly, we do not take into account additional kinematical constraints that make the different $t$-channel helicity amplitudes not independent at threshold and pseudothreshold, and that would appear here as kinematic zeroes that do not affect the analytic continuation.

\section{Regge model for $\gamma p \to \pi^- \Delta^{++}$\label{sec:model}}
\begin{figure}[b]
    \centering
    \begin{tikzpicture}
        \draw (0,0) -- (2,1);
        \draw[double,thick] (2,1) -- (4,0);
        \draw[decorate,decoration={snake}] (0,4) -- (2,3);
        \draw[dashed,thick] (2,3) -- (4,4);
        \draw[decorate,color=cyan,line width=3pt,decoration={snake,amplitude=2,segment length=25}] (2,1) -- (2,3);
        \filldraw[black!60,opacity=0.4] (2,1) circle (8pt);
        \filldraw[black!60,opacity=0.4] (2,3) circle (8pt);
        \filldraw[black] (2,1) circle (2pt);
        \filldraw[black] (2,3) circle (2pt);
        \node at (0.5,0.75) {$p$};
        \node at (0.5,3.25) {$\gamma$};
        \node at (3.5,0.75) {$\DelBar$};
        \node at (3.5,3.25) {$\pi$};
        \node at (3.5,2) {$R = \{\pi\,,\,\rho\,,\,b_1\,,\,a_2\}$};
        \node at (2,3.5) {$\beta^R_{\helgamma}(t)$};
        \node at (2,0.5) {$\beta^R_{\helproton\helDel}(t)$};
    \end{tikzpicture}
    \caption{The $t$-channel exchange diagram describing the photoproduction of the $\pi\Delta$.}\label{fig:feyndia}
\end{figure}
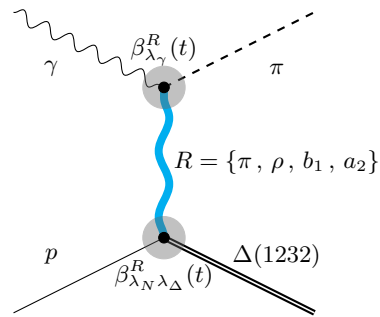

After the generalities on the amplitude, we now move to a specific dynamical model. 
At GlueX kinematics, photoproduction is naturally described using Regge theory, that imposes simultaneous analyticity of the amplitude in the energy and angular momentum plane. At large $s$ and small $t$, amplitudes are dominated by a finite number of crossed-channel Reggeon exchanges (\cf Fig.~\ref{fig:feyndia}). Assuming factorization of the the upper ($\gamma\pi$) and lower ($p\Delta$) vertices, the dynamical component of the amplitude reads as,

\begin{align}
    \hat{T}^{(s)}_{\helgamma\helproton\helDel} (s,t) &= \sqrt{-\frac{t}{s_0}}^{\,n}\sum_R \beta^R_{\helgamma}(t) \beta^R_{\helproton,\helDel}(t) \sqrt{s_0}^{J_R} \nonumber\\
    &\qquad \times \mathcal{P}_R(s,t) \mathcal{S}_R(t)\label{eq:ampB}
\end{align}
where $s_0$ is a constant which we take as $1\gev^2$, $J_R$ represents the spin of the parent state in the given trajectory, $\beta^R_{\helgamma}(t)$ represents the residue function of the upper vertex, $\beta^R_{\helproton\helDel}(t)$ represents the residue function of the lower vertex, $\mathcal{P}_R(s,t)$ the Regge propagator, the suppression factor $\mathcal{S}_R(t)$ includes an exponential factor for a given Reggeon exchange and a nonsense wrong signature zero canceling polynomial~\cite{Collins:1971ff,JointPhysicsAnalysisCenter:2017del}, and $n={|\helgamma| + |\helproton-\helDel|-|\helgamma-\helproton+\helDel|}$ is either $2$ or $0$. The summation is over Regge exchanges of all signatures and naturalities and includes the $\pi\,,\,b_1\,,\,\rho\,,\,a_2$ trajectories. 
While Regge poles can explain a majority of the features of the production process, unitarity in the crossed channels and rescattering effects introduce multi-Reggeon cuts~\cite{Collins:1971ff,Irving:1977ea}. These additional processes, collectively known as absorption corrections, are especially important in the case of pion exchange. From Eq.~\eqref{eq:amp} and Eq.~\eqref{eq:ampB}, where absorption corrections have not yet been implemented, we see that the amplitude vanishes in the limit $t\to0$ for every helicity combination, contradicting the forward peak observed in SLAC data on the $\pi\Delta$ photoproduction cross section~\cite{Boyarski:1968dw}. We implement these corrections in all the unnatural exchange amplitudes using the Williams' model or the poor man's absorption (PMA) scheme~\cite{Williams:1970rg}, which simply evaluates the factors of $\sqrt{-t}$ in \cref{eq:ampB} at $t=m_\pi^2$. Historically, absorption models attribute the forward peak to the suppression of lower partial waves via inelastic scattering. However, as we recently showed for pion photoproduction with nucleon recoil~\cite{JointPhysicsAnalysisCenter:2024kck}, the forward peak can instead be explained by the interference of electric and magnetic amplitudes, suggesting that true rescattering effects are indeed subleading. However, the PMA scheme happens to be basically equivalent to including the magnetic contribution of the Born diagrams, making it a practical prescription to implement in phenomenological calculations. 

The full amplitude, including PMA, takes the form,
\begin{widetext}
\begin{align}
    T^{(s)}_{\helgamma\helproton\helDel}(s,t) &= \sqrt{-\frac{t}{s_0}}^{|\helgamma|+|\helproton-\helDel|}\sum_{R \in \{\rho, a_2\}} \beta^R_{\helgamma}(t) \beta^R_{\helproton\helDel}(t)~\mathcal{P}_R(s,t) \mathcal{S}_R(t)\sqrt{s_0}^{J_R} \nonumber\\
    &+ \sqrt{-\frac{t}{s_0}}^{|\helgamma-\helproton+\helDel|}\sum_{R \in \{\pi, b_1 \}} \beta^R_{\helgamma}(t) \beta^R_{\helproton\helDel}(t)~\mathcal{P}_R(s,t) \mathcal{S}_R(t)\sqrt{s_0}^{J_R} \left(-\frac{m_\pi^2}{s_0}\right)^{\frac{n}{2}}\text{ .}\label{eq:sAmpwAbs}
\end{align}
\end{widetext}
The introduction of absorption brings in a peculiar feature to the amplitude {\it viz.,} the unnatural exchanges now contribute to positive reflectivity amplitudes as polynomial corrections (see Appendix~\ref{app:sdmeref} and Appendix~\ref{app:refAmp} for details). The consequences of this will be discussed when we discuss the observables. The explicit form of the Regge propagator is given by \cite{Nam:2011np,JointPhysicsAnalysisCenter:2017del,Mathieu:2020zpm},
\begin{align}
     \mathcal{P}_R(s,t) &= \pi \alpha'\left(\frac{s}{s_0}\right)^{\alpha(t)}\frac{\tau_R + \exp(-i\pi\alpha_R(t))}{2\sin(\pi\alpha_R(t))}
 \end{align}
where $\tau_R=\pm1$, the numerator cancels the poles of wrong signature arising in the given Regge trajectory, and $\alpha'=0.9\gev^{-2}$.  
Despite the use of Regge models to describe the spacelike $t\leq0$ region, the trajectory is itself is an analytic function and can thus be continued to arbitrary values of its argument. For positive $t$ it is related to the spectrum of resonances by $\alpha(m_J^2) = J$. In the vicinity of $t\sim m_J^2$, \cref{eq:Bsumpoles}, and therefore \cref{eq:amp}, recovers the expected angular structure of a spin-$J$ exchange in the high energy limit, \textit{cf.} \cref{eq:residueLag}. This will allow us to connect the Regge behavior of the lab reaction to specific resonance poles.
As is customary in the literature, we consider the Regge trajectories to be real and linear. In principle, one should consider trajectories which are complex above the threshold of the lowest-lying intermediate state (in this case, the $\pi\pi$ threshold) as a consequence of unitarity. In such a trajectory, physical particles occur at complex $m_J$ corresponding to resonance poles on unphysical Riemann sheets, e.g. as discussed in \cite{Stamen:2024gfz} and references therein. In the present application, the resonances of interest are low-lying and relatively narrow such that we expect the zero-width, linear approximation to be sufficient to constrain the moduli of the residues when continuing the amplitude to the resonance pole.\footnote{The phase of the residue is driven by the width of the resonance~\cite{Henyey:1971ah}. Anyway, it has been argued that it does not carry information about the resonance~\cite{Gribov:2009cfk,Ceci:2025gsm}.}
We further assume that the trajectories of Reggeons of like naturality are exchange degenerate. We thus take the Regge trajectories as $\alpha_U(t) = \alpha'(t - m_\pi^2)$ and $\alpha_N(t) = \alpha'(t - m_\rho^2) + 1$
with $m_\pi$ and $m_\rho$ the masses of pion and $\rho$-meson respectively.\footnote{We take $m_\pi = 0.1395\gev$ and $m_\rho = 0.775\gev$.}\par

The couplings of the upper vertex are extracted from model Lagrangians with the parameters fixed from the radiative decay widths and their explicit forms can be found in Ref.~\cite{JointPhysicsAnalysisCenter:2017del}
while those of the lower vertex are taken as zeroth- or first-degree polynomials, given in Table~\ref{tab:LVfactors}. Unlike Ref.~\cite{JointPhysicsAnalysisCenter:2017del}, the coefficients are extracted by fitting to the data. 
To improve the stability of fits, we choose the same residual polynomial but different form factors for $\pi$ and $b_1$ exchanges. The natural exchanges share the same form factor but have different polynomials. The suppression factors are given by,
\bsub\label{eq:S_R}\begin{align}
    \mathcal{S}_\pi &= e^{b_\pi t}(\alpha_U(t)+2)/2\\
    \mathcal{S}_{b_1} &= e^{b_{b_1} t} (\alpha_U(t) +1)\\
    \mathcal{S}_\rho &= e^{b_N t} (\alpha_N(t) +1)/2 \\
    \mathcal{S}_{a_2} &= e^{b_N t} \alpha_N(t) (\alpha_N(t) + 2)/3\text{ .}\label{eq:supa2}
\end{align}\esub
The suppression factors also contain polynomials that cancel the closest ``ghost poles'' that appear in the propagators corresponding to 
unphysical $J<0$~\cite{JointPhysicsAnalysisCenter:2017del}. These polynomials are normalized to 1 at $t=m_\rho^2$ for natural trajectories, and $t=m_\pi^2$ for unnatural trajectories.
\begin{table}[t]
    \begin{tabular}{|c|c|c|c|}
        \hline
         $\helDel$ & $\beta^{\pi}_{\oneh\helDel}(t) \left(= \beta^{b_1}_{\oneh\helDel}(t)\right)$ & $\beta^{\rho}_{\oneh\helDel}(t)$ & $\beta^{a_2}_{\oneh\helDel}(t)$\\\hline
         $-\threeh$ & $p^\pi_1$ & $p^\rho_1$ & $p^{a_2}_1$\\
         $-\oneh$ & $p^\pi_2 + p^\pi_5 \dfrac{t}{s_0}$ & $p^\rho_2 + p^\rho_5 \dfrac{t}{s_0}$ & $p^{a_2}_2 + p^{a_2}_5 \dfrac{t}{s_0}$\\
         $\oneh$ & $p^\pi_3$ & $p^\rho_3 + p^\rho_6 \dfrac{t}{s_0}$ & $p^{a_2}_3 + p^{a_2}_6 \dfrac{t}{s_0}$\\
         $\threeh$ & $p^\pi_4$ & $p^\rho_4$ & $p^{a_2}_4$\\\hline
    \end{tabular}
    \caption{Parameterization of the lower vertex factors for various exchanges. The factor of $s_0$ makes the coefficients dimensionless.}
    \label{tab:LVfactors}
\end{table}

\begin{figure*}[t]
    \centering
    \includegraphics[width=0.5\linewidth]{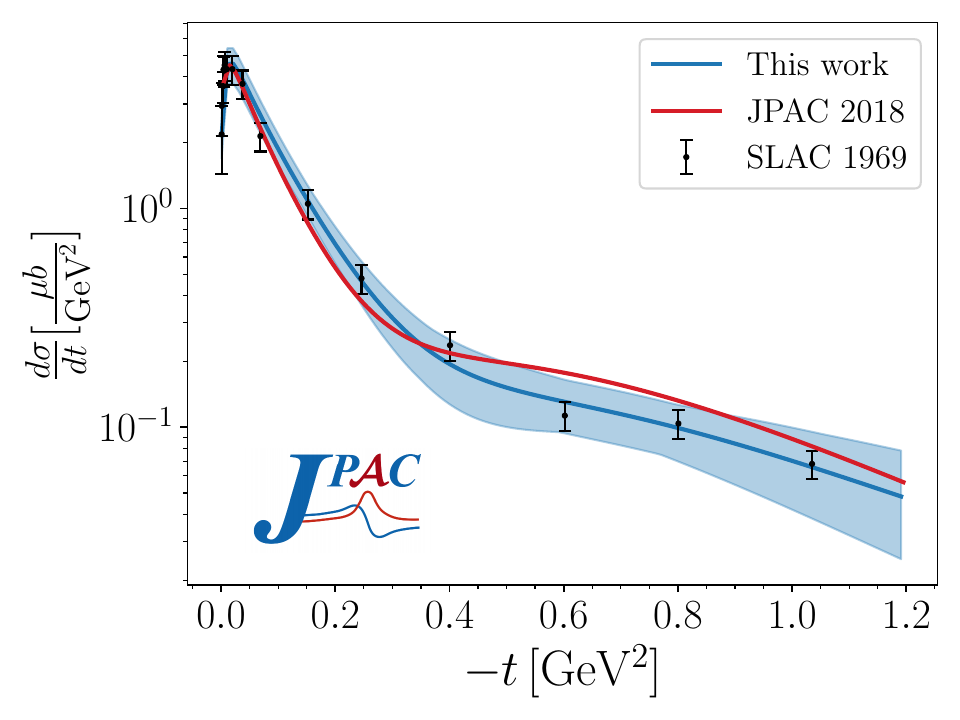}~\includegraphics[width=0.5\linewidth]{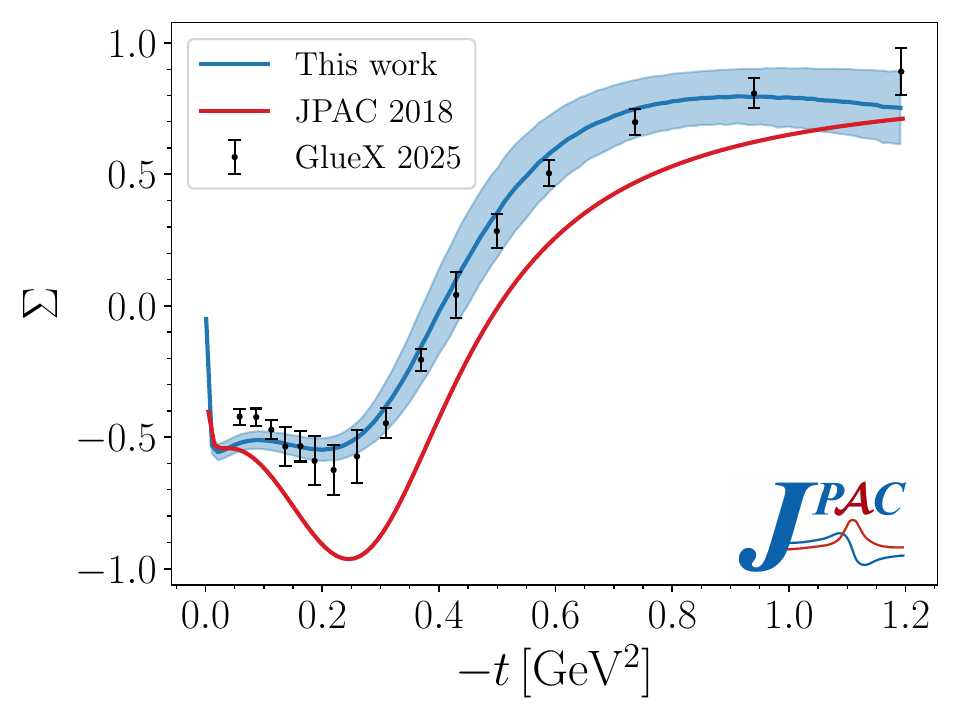}
    \caption{$\pi^-\Delta^{++}$ photoproduction cross section (left) and BSA (right). The cross section data are from Ref.~\cite{Boyarski:1968dw} and the BSA data are from Ref.~\cite{GlueX:2024dbr}. Since the BSA is not independent from the SDMEs, we show data for comparison but do not include in the fits.}
    \label{fig:xsec}
\end{figure*}
We now take a moment to deliberate on the parameterization of the residual polynomials that we have used in this work. Using the Lagrangian approach guarantees that the vertices have the correct singularity structure in both $s$ and $t$. However, the residual polynomials are specific to the spin of the exchanged states and hence are too restrictive to be used for a Regge exchange model. More specifically, the Lagrangian approach constrains the magnitudes of the exchanges to a good accuracy. This can be seen in their success in describing the differential cross section and BSA~\cite{Yu:2016jfi,JointPhysicsAnalysisCenter:2017del}. However, the angular distributions of the decay productions of the $\DelBar$ depend equally on the relative phases of the helicity amplitudes as well as the relative phases between the exchanges of different Reggeons. At a formal level, the residual polynomials represent the coupling of the poles in the crossed channel which needs to be extrapolated to the $t<0$ region (which is where the process happens). Regge theory does not constrain this extrapolation, other than prohibiting any kinematical singularities~\cite{Collins:1971ff,Irving:1977ea}. For these reasons and purposes, we model the residual behavior of the Regge couplings using polynomials and extract the values fo their coefficients from the data. \par

\section{Results and Discussions\label{sec:RnD}}
\begin{table*}[t]
    \centering
    \begin{tabular}{|c|c|c|c|c|c|c|c|}
          \hline
          $\pi$-exchange & Value & $b_1$-exchange & Value & $\rho$-exchange & Value & $a_2$-exchange & Value \\
          parameter & & parameter & & parameter & & parameter & \\\hline
          $b_\pi$ ($\gev^{-2}$) & $1.20 \pm 0.16$ & $b_{b_1}$ ($\gev^{-2}$) & $-0.80 \pm 0.19$ & $b_N$ ($\gev^{-2}$) & $-0.51 \pm 0.15$ & \multicolumn{2}{c|}{Same as $\rho$-exchange} \\\hline
          $p^\pi_1$  & $-21.01 \pm 2.36$ & \multicolumn{2}{c|}{\multirow{5}{*}{Same as $\pi$-exchange}} & $p^\rho_1$ & $24.7 \pm 4.6$ & $p^{a_2}_1$  & $3.82 \pm 1.57$ \\\cline{1-2}\cline{5-8}
          $p^\pi_2$ & $19.14 \pm 1.12$ & \multicolumn{2}{c|}{} & $p^\rho_2$ & $-11.8 \pm 2.7$ & $p^{a_2}_2$ & $ -21.7 \pm 2.5$  \\\cline{1-2}\cline{5-8}
          $p^\pi_3$ & $-7.22 \pm 0.21$ & \multicolumn{2}{c|}{} & $p^\rho_3$ & $-5.02 \pm 1.24$ & $p^{a_2}_3$ &  $-4.94 \pm 1.02$ \\\cline{1-2}\cline{5-8}
          $p^\pi_4$ & $-21.07 \pm 0.71$ & \multicolumn{2}{c|}{} & $p^\rho_4$  & $3.95 \pm 1.53$ & $p^{a_2}_4$ & $-20.6 \pm 1.7$ \\\cline{1-2}\cline{5-8}
          $p^\pi_5$ & $-22.58 \pm 7.12$ &\multicolumn{2}{c|}{} & $p^\rho_5$ & $-15.1 \pm 3.2$ & $p^{a_2}_5$ & $-7.15 \pm 1.93$ \\\hline
          \multicolumn{4}{c|}{} & $p^\rho_6$ & $-20.9 \pm 3.7$ & $p^{a_2}_6$ & $-5.61 \pm 1.73$ \\ \cline{5-8}
    \end{tabular}
    \caption{Values of parameters ($\chi^2/d.o.f = 153.6/140$). The uncertainties in the values of the parameters were extracted by bootstrapping over the uncertainties in the GlueX data~\cite{GlueX:2024dbr}. This parameter set represents one of the two degenerate solutions. The other solution can be obtained by flipping the signs of the parameters $p^\times_i$ keeping their relative signs fixed. This amounts to flipping the overall sign of the helicity amplitude and is an unremovable ambiguity.}
    \label{tab:parvals}
\end{table*}

\subsection{Fits and Observables}

To constrain the values of the parameters, we first calculate the SDMEs using the definition given in Eqs.~\eqref{eq:rho_all} and the $s$-channel amplitude in Eq.~\eqref{eq:amp}. We then rotate the resultant SDMEs to the $t$-channel rest frame (which is congruent to the Gottfried-Jackson (GJ) frame, modulo an unphysical Lorentz boost along the $+z$-axis) using the relation,
\begin{align}
    \rho^\text{GJ}_{\helDelt\helDelt^\prime} &= \sum_{\helDel,\helDel^\prime} d^\threeh_{\helDelt,\helDel}(\omega_\Delta) \rho^\text{H}_{\helDel\helDel^\prime} d^\threeh_{\helDelt^\prime,\helDel^\prime}(\omega_\Delta) \\ \nonumber
\end{align}
The parameters of the model are extracted by fitting the SDMEs to the GlueX data~\cite{GlueX:2024dbr} in the GJ-frame and the SLAC data on cross section~\cite{Boyarski:1968dw} ($160$ data points altogether) using the \texttt{ROOT} implementation of \texttt{MINUIT}~\cite{James:1975dr,Brun:1997pa}. The parameterization described in the previous subsection gives a good fit with 
$\chi^2/\text{dof}=153.6/140$. We estimate the uncertainties in the values of the parameters by bootstrapping~\cite{JPAC:2021rxu} the fit over the uncertainties in the GlueX and SLAC data assuming that the SDMEs and cross section are distributed normally about their mean values with the reported uncertainties as standard deviations. 
The bootstrapping was done using $\sim 11,000$ samples.
\begin{figure*}[t]
    \centering
    \includegraphics[width=0.3\linewidth]{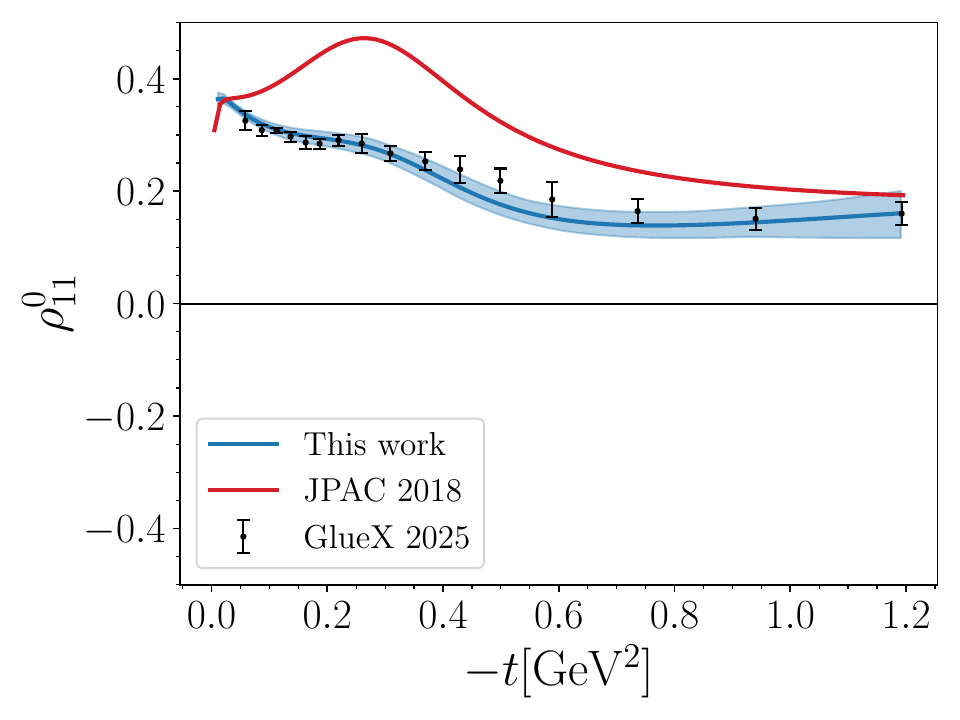}~\includegraphics[width=0.3\linewidth]{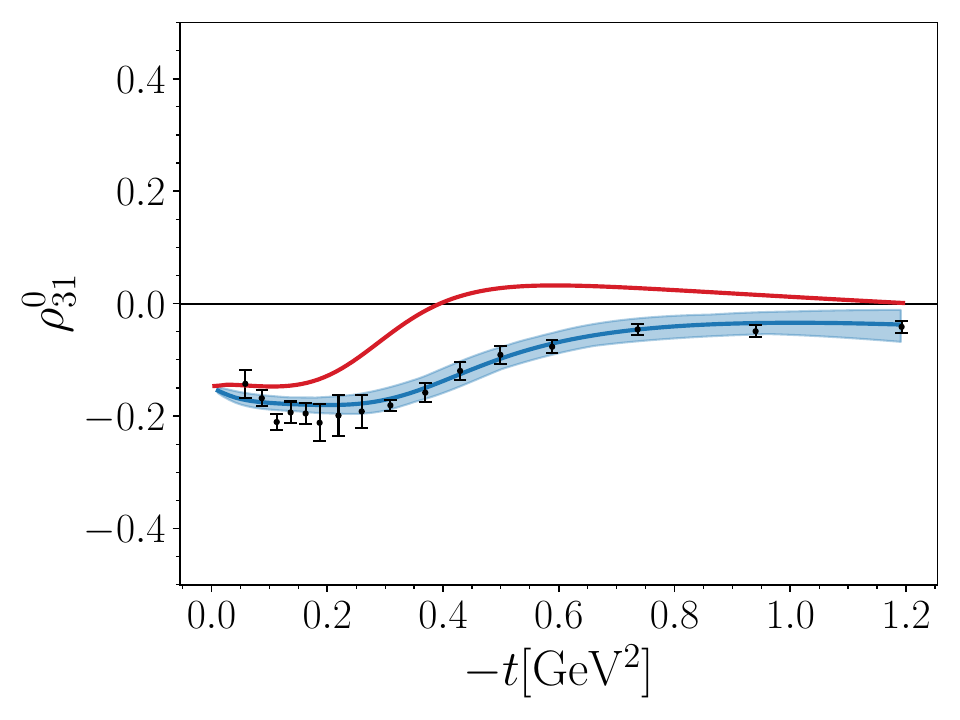}~\includegraphics[width=0.3\linewidth]{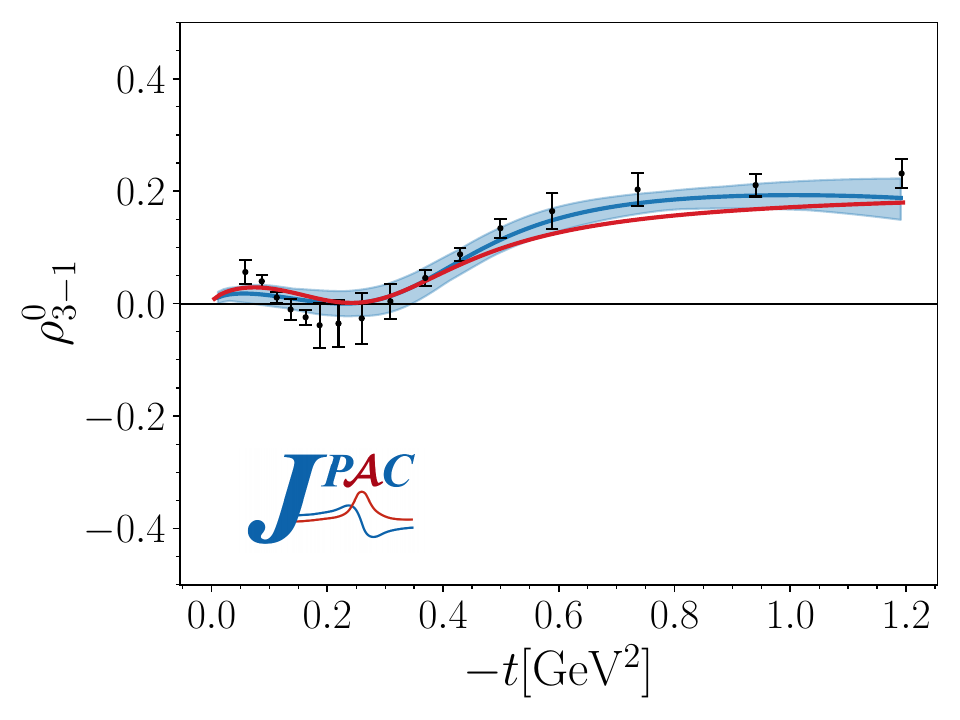}\\
    \includegraphics[width=0.3\linewidth]{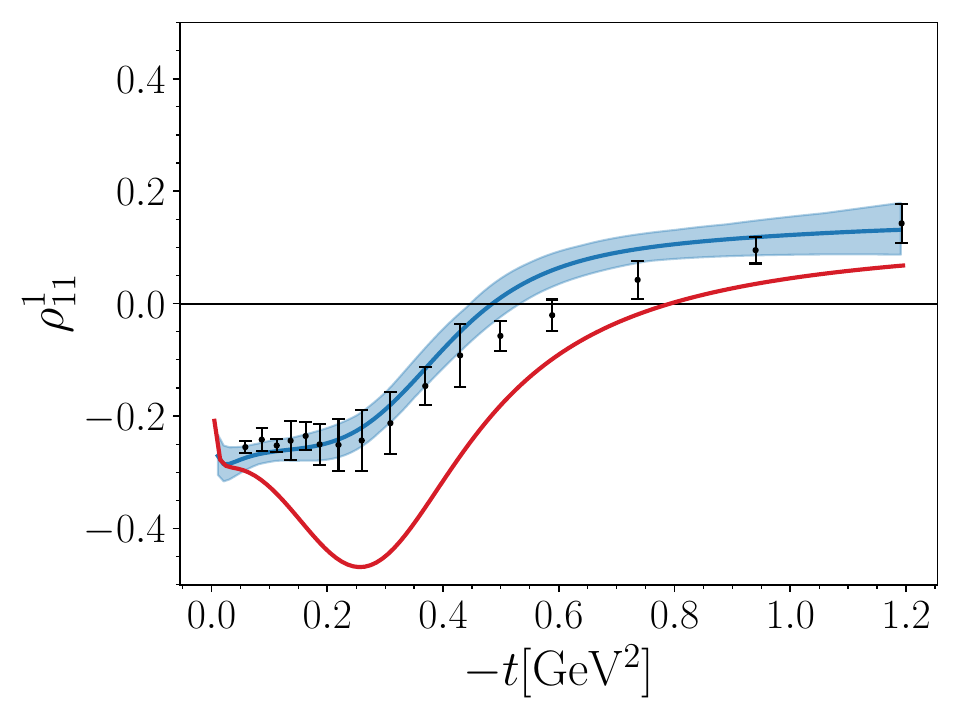}~\includegraphics[width=0.3\linewidth]{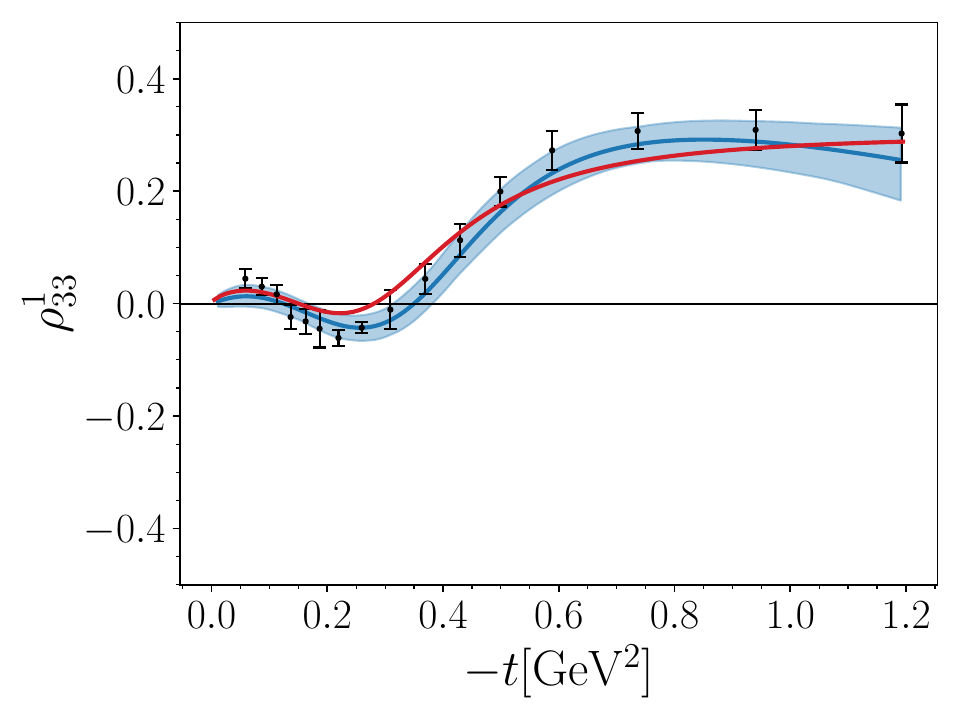}~\includegraphics[width=0.3\linewidth]{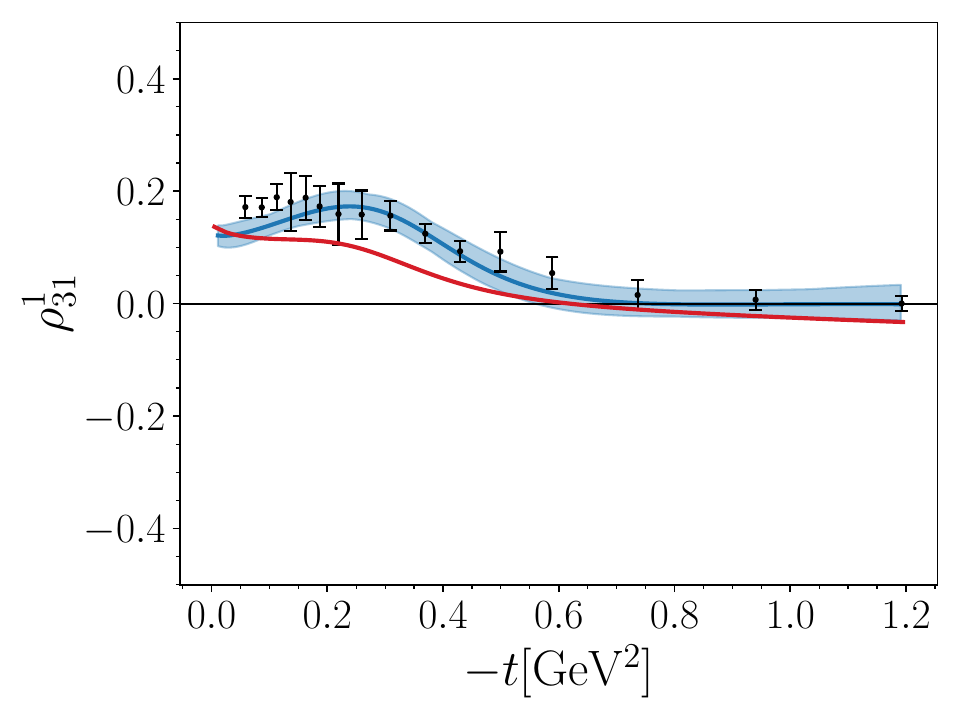}\\
    \includegraphics[width=0.3\linewidth]{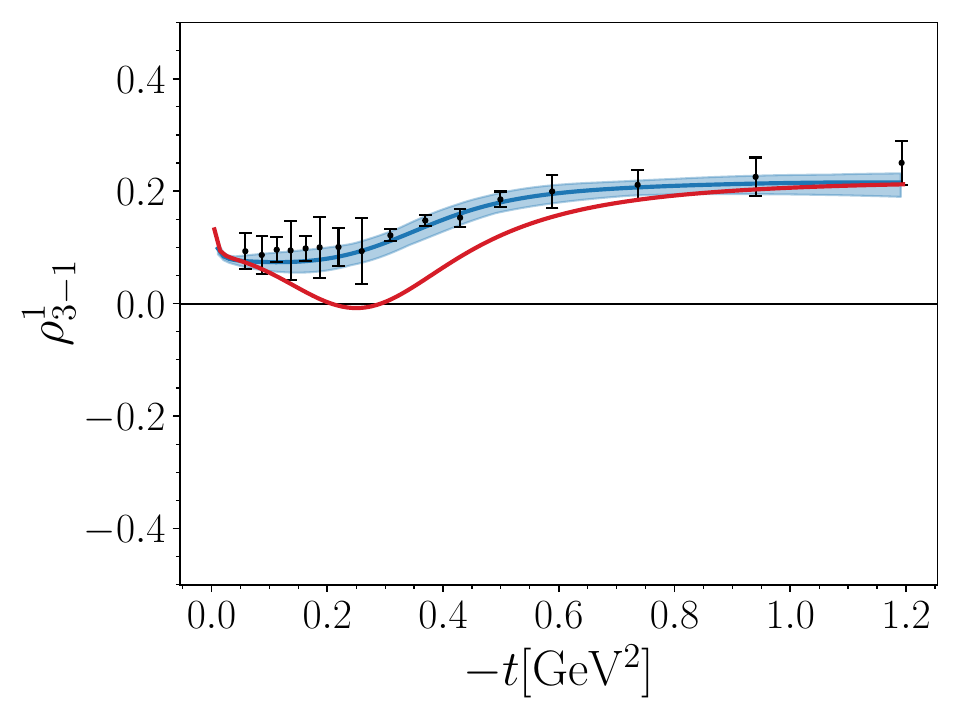}~\includegraphics[width=0.3\linewidth]{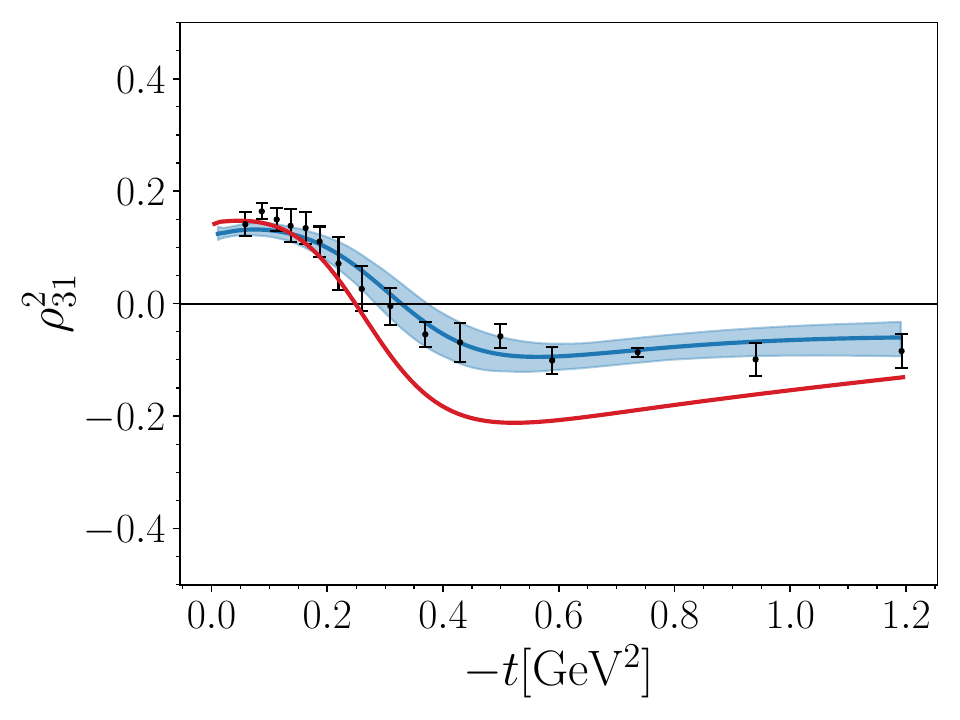}~\includegraphics[width=0.3\linewidth]{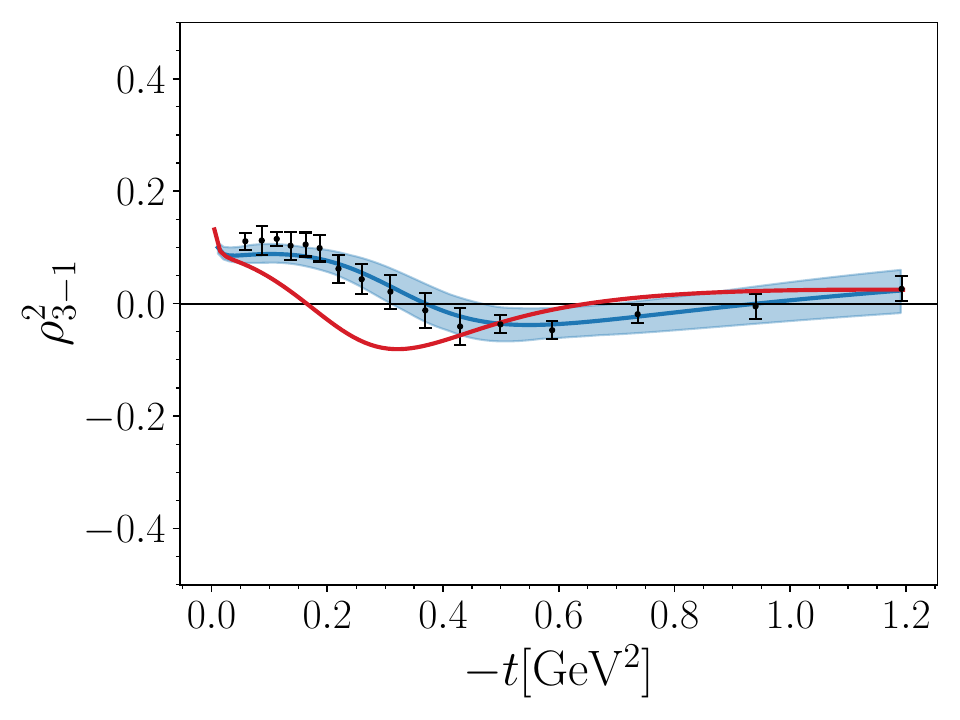}\\
    \caption{SDMEs of the $\Delta$ in the GJ frame from the current fit compared with the GlueX data \cite{GlueX:2024dbr} and the model from Ref.~\cite{JointPhysicsAnalysisCenter:2017del}. Note that we had to change the signs of amplitudes for $\helDel = 3/2$ and $\helDel = -1/2$ of Ref.~\cite{JointPhysicsAnalysisCenter:2017del} to better match the experimental data on the SDMEs.}
    \label{fig:sdmes}
\end{figure*}

It is pertinent to mention that this parameterization has the least number of parameters that are needed to explain the data. \Eg, having an additional parameter for $\beta^\pi_{\oneh\oneh} (= p_3^\pi + p_6^\pi\,t/s_0)$ leads to its parameter ($p_6^\pi$) being consistent with zero. Further, the $b_1$ contribution to unnatural exchange is so small, that keeping its polynomial coefficients independent from the ones of the pion makes the fit unstable. 

The values of the parameters and their uncertainties are listed in Table~\ref{tab:parvals}. From the Table we see that the parameters for unnatural parity exchanges ($b_\pi$ and $p^\pi_i$'s) are quite well constrained, driven by the dominance of pion exchange, while the natural parity exchanges are not as well determined.

The plot of the differential cross section is shown in Fig.~\ref{fig:xsec}. One of the important features that we see is that cross section is non-zero in the $t\to 0$ limit, as discussed in Sec.~\ref{sec:model}.

The plot of the SDMEs from the current fit in the GJ frame are shown in Fig.~\ref{fig:sdmes} in comparison with the GlueX data and the model from Ref.~\cite{JointPhysicsAnalysisCenter:2017del}.

To begin with, we note that the diagonal elements of the unpolarised SDMEs ($\rho^0_{\helDel\helDel}$) represent the fraction of the $\DelBar$ produced in the helicity state $\helDel$. Closer to the forward limit, the majority of the $\DelBar$ that are produced are in the $\helDel = \pm \oneh$ configuration and this fraction reduces slowly as $-t$ increases. The opposite is true for the $\helDel = \pm \threeh$ states. An immediate conclusion is that the amplitudes for $\helDel-\helproton = \pm2$ must be suppressed in the small-$t$ region since they contribute only to $\rho^0_{33}$. This feature of the SDMEs and the information present in there is encoded in the $\sqrt{-t}^{|\helproton - \helDel|}$ behavior of the half-angle factor in the amplitude.

Continuing to the polarised SDMEs, the diagonal elements of the $\rho^1$ add up to the BSA shown in Fig.~\ref{fig:xsec}. The interpretation of the BSA is straight forward in the reflectivity basis. Using the definition of the $\rho^1_{\helDel\helDel'}$ given in Eq.~\eqref{eq:1ref}, we see that the BSA, 
\begin{align}
   \Sigma &= \dfrac{\sum_\sigma (|N_\sigma|^2 - |U_\sigma|^2)}{\sum_\sigma (|N_\sigma|^2 + |U_\sigma|^2)}
\end{align}
Thus, the sign of the BSA represents the dominance of the respective reflectivity amplitudes which in turn implies the dominance of the respective naturality exchanges~\cite{GlueX:2024dbr}. Comparing with Fig.~\ref{fig:xsec} we see that the unnatural parity exchange dominate in the small-$t$ region and the natural parity exchanges dominate for larger values of $|t|$.
This behavior is consistent with the predictions from the Stichel theorem~\cite{Stichel:1964yvk,Ravndal:1970xe}.\par
\begin{table*}[t]
    \centering
    \begin{tabular}{|c|c|c|c|c|c|c|c|c|c|}
        \hline
        $\helprotont$ & $\helDelt$ & \multicolumn{2}{c|}{$\mathcal{R}^\pi_{1\helprotont\helDelt}$} & \multicolumn{2}{c|}{$\mathcal{R}^\rho_{1\helprotont\helDelt}$} & \multicolumn{2}{c|}{$\mathcal{R}^{b_1}_{1\helprotont\helDelt}$} & \multicolumn{2}{c|}{$\mathcal{R}^{a_2}_{1\helprotont\helDelt}$} \\
        \cline{3-10}
        & & This work & Ref.~\cite{JointPhysicsAnalysisCenter:2017del} & This work  & Ref.~\cite{JointPhysicsAnalysisCenter:2017del} & This work & Ref.~\cite{JointPhysicsAnalysisCenter:2017del} & This work & Ref.~\cite{JointPhysicsAnalysisCenter:2017del} \\\hline
        $\frac{1}{2}$ & $\frac{3}{2}$ & ---  & --- & $-8.0 \pm 2.4$ & $-12.6$ & $4 \pm 7$ & $140.1$ & $45 \pm 16$ & $31.7$ \\
        $\frac{1}{2}$ & $\frac{1}{2}$ & $-57.0 \pm 2.2$ & $-60.2$ & $-5.3 \pm 1.5$ & $-0.05$ & $14.7 \pm 14.5$ & $84.5$ & $38 \pm 15$ & $3.74$\\
        $-\frac{1}{2}$ & $\frac{3}{2}$ & --- & --- & --- & --- & --- & --- & $5.3 \pm 4.0$ & $0.0$ \\
        $-\frac{1}{2}$ & $\frac{1}{2}$ & --- & --- & $16.0 \pm 3.7$ & $-6.62$ & $57 \pm 30$ & $144.0$ & $-17 \pm 9$ & $11.7$ \\\hline
    \end{tabular}
    \caption{Residues of $\pi$, $\rho$, $b_1$, and $a_2$ exchanges extracted from the present fits compared to the values extracted from the models given in Refs.~\cite{JointPhysicsAnalysisCenter:2017del}. All numbers are expressed in the appropriate powers of GeV.}
    \label{tab:resTab}
\end{table*}
\begin{figure*}[t]
    \centering
    \includegraphics[width=0.25\linewidth]{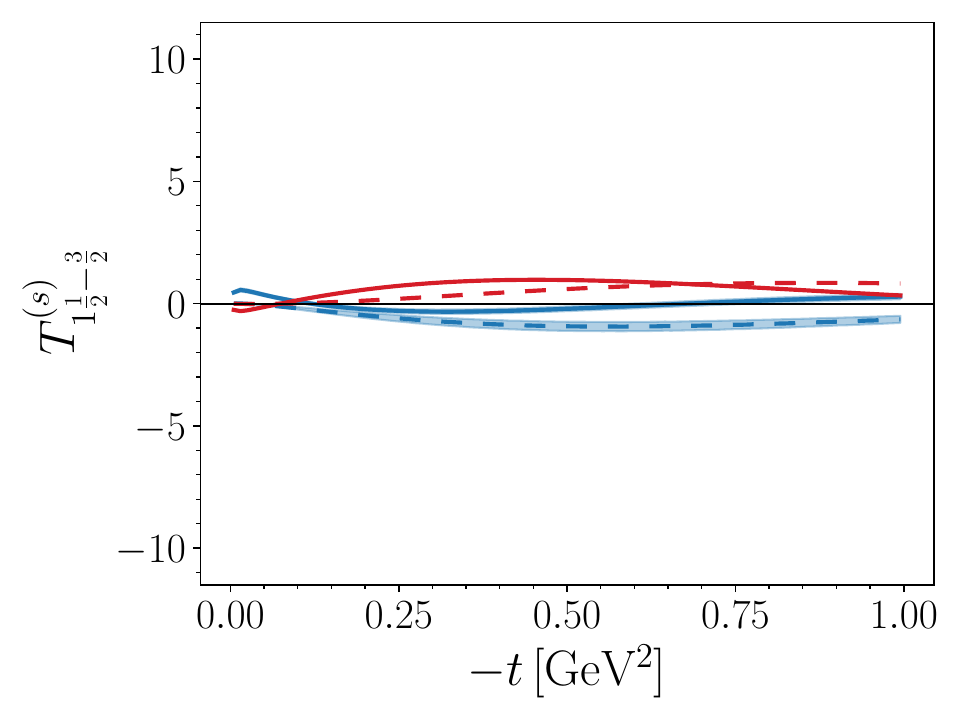}~
    \includegraphics[width=0.25\linewidth]{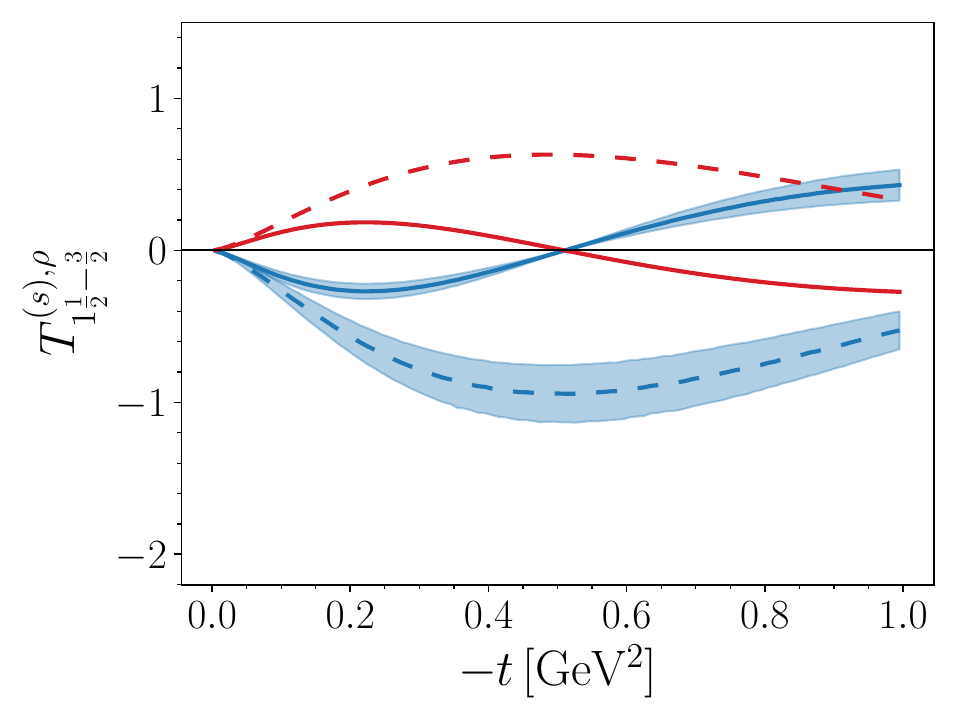}~\includegraphics[width=0.25\linewidth]{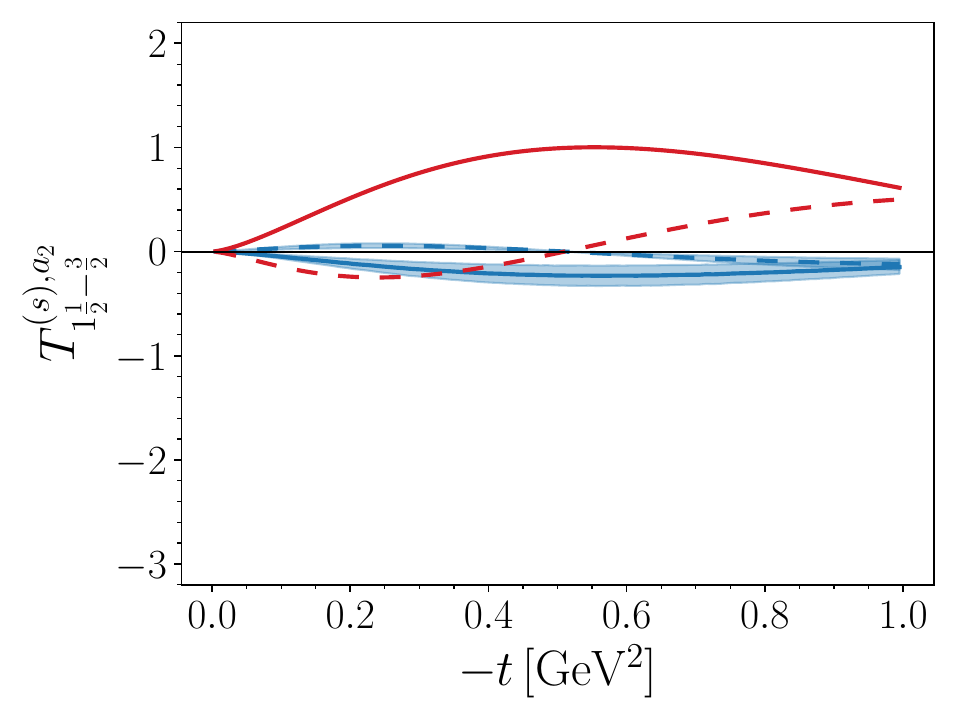}~\includegraphics[width=0.25\linewidth]{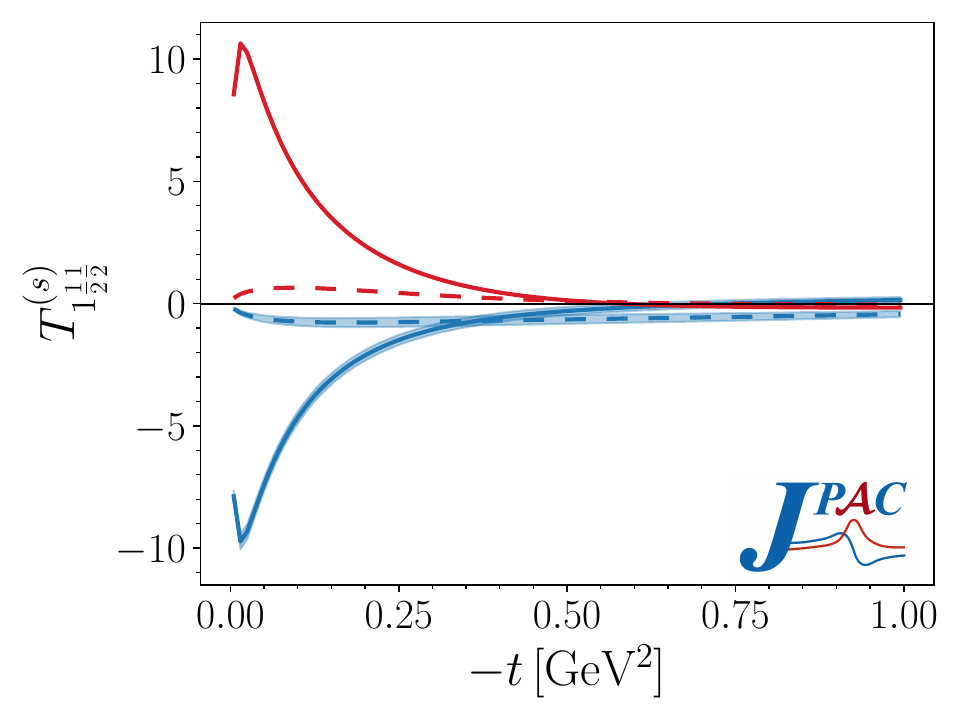}\\
    \includegraphics[width=0.25\linewidth]{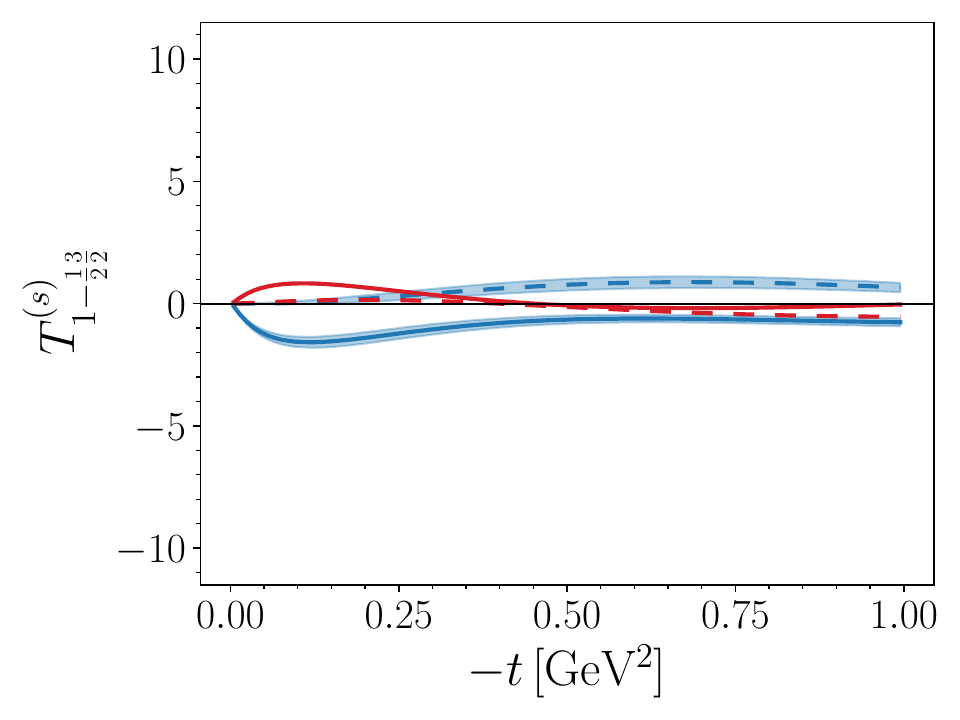}~\includegraphics[width=0.25\linewidth]{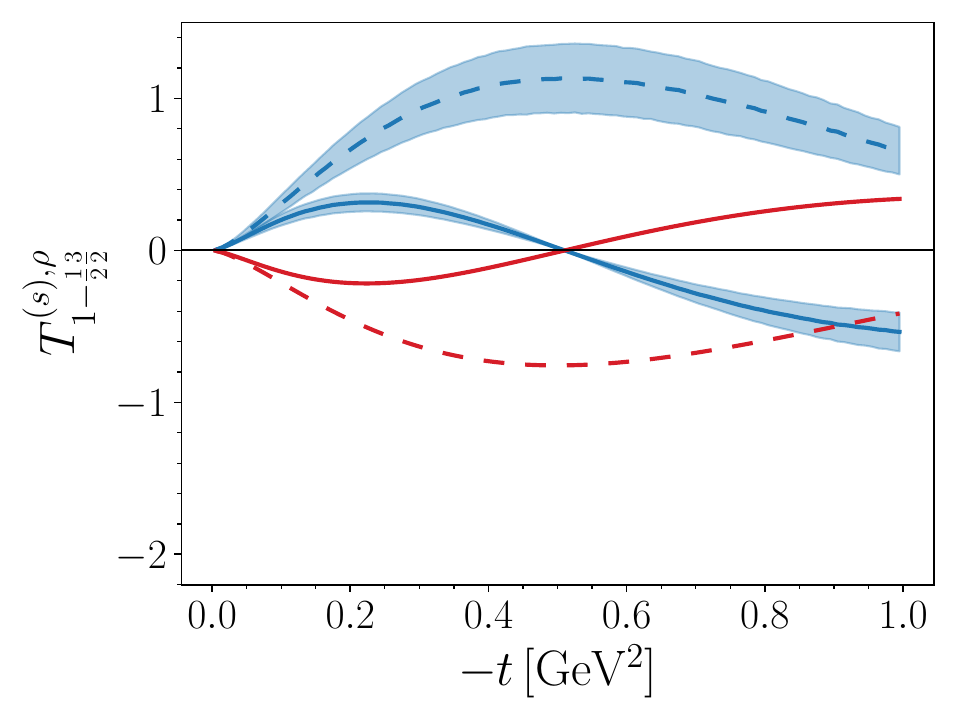}~
    \includegraphics[width=0.25\linewidth]{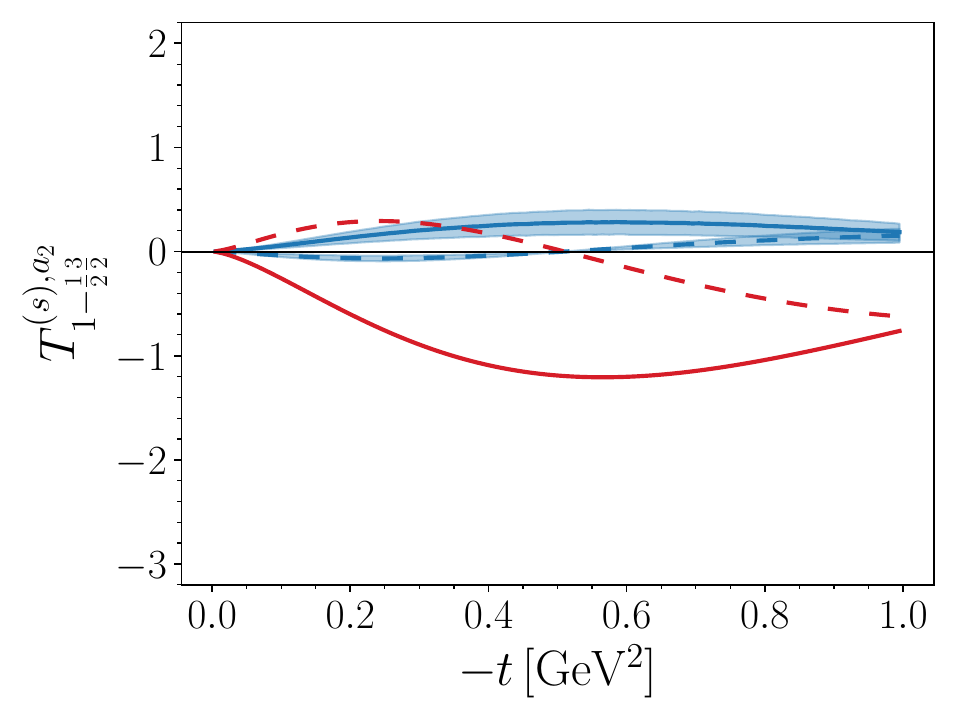}~\includegraphics[width=0.25\linewidth]{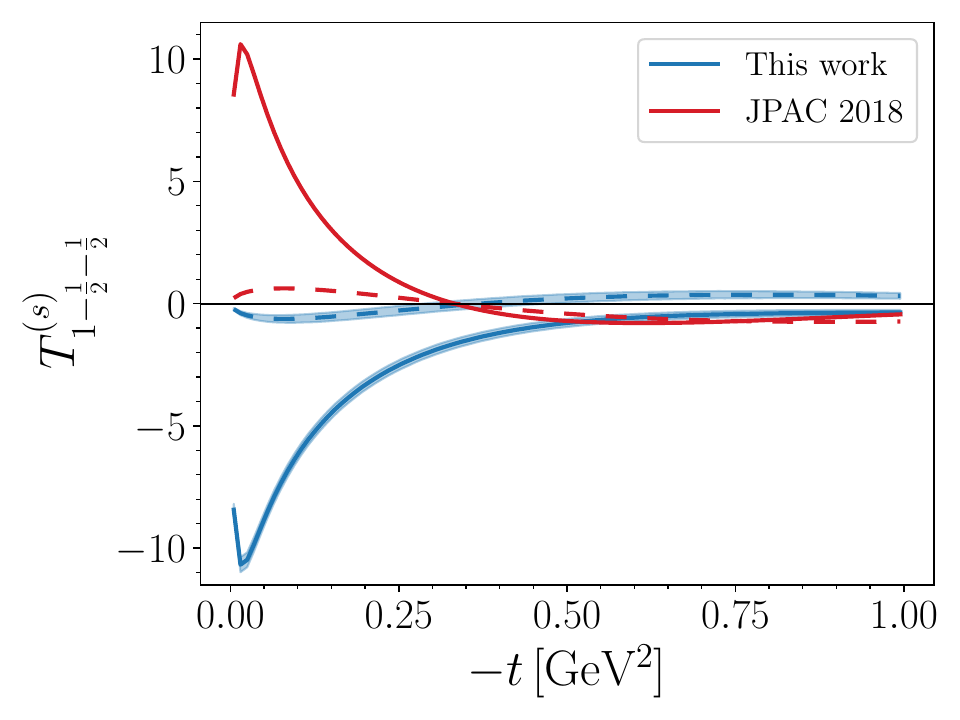}\\
    \caption{Some of the $s$-channel amplitudes from the present work compared to Ref.~\cite{JointPhysicsAnalysisCenter:2017del} (JPAC 2018). Solid lines are the real parts and dashed lines are the imaginary parts. The shaded bands represent $1\sigma$ variations.}
    \label{fig:amps}
\end{figure*}
Finally, we note that the off-diagonal elements of $\rho^{(0,1)}_{31}$ and $\rho^2_{\helDel\helDel'}$ are negligibly small in the large-$t$ region. As we can see from Eq.~\eqref{eq:intensity}, this behavior stems from the approximate $\cos(2\phi)$ dependence of the distribution of the decay products of the $\DelBar$ in the large-$t$ region. Consequently, since $\rho^2_{\helDel\helDel'}$ is given by the interference between the amplitudes of different reflectivities, this implies that one reflectivity dominates over the other.

\subsection{Amplitudes}
We now discuss the amplitudes. We show the plots of some of the amplitudes in Fig.~\ref{fig:amps}. A few observations on the helicity amplitudes are in order. We see from Fig.~\ref{fig:amps} that the amplitudes corresponding to $|\helproton - \helDel| = 2$ (double helicity flip) are suppressed by one order of magnitude in the small-$t$ region compared to the ones for $\helproton = \helDel$ (helicity nonflip) as expected from the half-angle factors.
 
In general, both $\rho$ and $a_2$ exchanges appear with similar strength in the natural exchange component, in the entire $t$ range. The exception to this are the $\helDel = \pm\frac{3}{2}$ amplitudes, where $\rho$ and $a_2$ dominate the double flip and the single flip amplitudes, respectively. Furthermore, the two double flip amplitudes show two peculiar features. The $T_{1-\oneh\threeh}$ has a dip around $t\simeq -0.5\gev^2$ whereas the $T_{1\oneh-\threeh}$ has a bump in the same vicinity.  In the former case, the $a_2$ exchange interferes with rest of the amplitudes to produce the dip where as the bump in the latter is due to the dominance of $\rho$ exchange over the $\pi$ and $a_2$ exchanges. Due to the interplay between the $\rho$ and $a_2$ exchanges mentioned earlier, the $\helDel = \pm\frac{1}{2}$ amplitudes do not show these features and a significant contribution from the pion exchange amplitude masks any such feature that might be present in the $T_{1\oneh\threeh}$ and $T_{1-\oneh-\threeh}$ amplitudes.
\begin{table*}[t]
    \centering
    \begin{tabular}{|c|c|c|c|c|c|c|}
         \hline
         \multirow{2}{*}{\backslashbox{$g^{(i)}_{R N \Delta}$}{$R\quad$}} & \multicolumn{2}{c|}{$\rho$} & \multicolumn{2}{c|}{$b_1$} & \multicolumn{2}{c|}{$a_2$}\\
         \cline{2-7}
         & Present & Ref.~\cite{JointPhysicsAnalysisCenter:2017del} & Present & Ref.~\cite{JointPhysicsAnalysisCenter:2017del} & Present & Ref.~\cite{JointPhysicsAnalysisCenter:2017del} \\\hline
         $g^{(1)}_{R N\Delta}$ & $33 \pm 10$ & $53$ & $-1.1 \pm 1.9$ & $-45$ & $-69 \pm 27$ & $53$\\
         $g^{(2)}_{R N\Delta}$ & $-522 \pm 82$ & $-132$ & $-199 \pm 98$ & $-518$ & $627 \pm 151$ & $205$\\
         $g^{(3)}_{R N\Delta}$ & $53 \pm 19$ & $-66$ & $66 \pm 37$ & $68$ & $53 \pm 40$ & $77$\\
         $g^{(4)}_{R N\Delta}$ & --- & & --- & & $6.6 \pm 5.0$ & $0.0$\\\hline
    \end{tabular}
    \caption{Couplings of the various meson to $\Delta\bar{p}$ extracted from the residues using effective Lagrangians. Also listed are the corresponding values from Ref.~\cite{JointPhysicsAnalysisCenter:2017del}. The coupling constants for the upper vertices are: $g_{\rho\pi\gamma} = 0.170$, $g_{b_1\pi\gamma} = 0.130$, and $g_{a_2\pi\gamma} = 1.007$.}
    \label{tab:couplings}
\end{table*}
On the other hand, the pion dominates among unnatural exchanges, as expected from the proximity of its pole to the physical region.
The corresponding amplitude does not present any significant features, and comes very close to the one derived in Ref.~\cite{JointPhysicsAnalysisCenter:2017del}, up to an overall sign. The amplitude for $b_1$ exchange is suppressed by at least an order of magnitude. An interesting factor to note is that the slope of the forma factor for pion exchange is positive but those for $b_1$ exchange and natural exchange are negative. The complete set of plots can be found in Appendix~\ref{app:figs}.

\subsection{Residues}
The values of the residues extracted using Eq.~\eqref{eq:residueFinal} are listed in Table~\ref{tab:resTab}. The uncertainties are estimated out of the resampled bootstrapped data sets discussed above. For comparison, we list the residues evaluated using the model presented in Ref.~\cite{JointPhysicsAnalysisCenter:2017del}.

We need to note here the peculiar pattern of uncertainties in the residues. The residues of $\pi$ and $\rho$ exchanges can be determined to very good accuracies. This is partly because the $\pi$-exchange is the dominant mechanism in the photoproduction process. In addition, the number of data points available in the GlueX measurements in the region where pion exchange dominates is also large (\cf Fig.~\ref{fig:sdmes}). Hence the $\pi$-exchange residues can be determined to a very high degree of accuracy (\cf Table~\ref{tab:parvals}).

The residues derived above can be expressed also in the form of coupling constants of the respective mesons to the $\pi\gamma$ and $\Delta\bar{p}$ systems. For the purpose of demonstration, let us consider the pion exchange. The amplitude can be constructed using the prescription of Ref.~\cite{JointPhysicsAnalysisCenter:2024kck}, and is given by,
\begin{align}
    T^\pi_{\helgammat\helprotont\helDelt} (t) &= -2i e \frac{f_{\pi N\Delta}}{m_\pi}~ p^t~\helgammat \sqrt{t - t_{th}}~\sqrt{\frac{2}{3}}\frac{t}{m_\Delta} \frac{\delta_{\helprotont,\helDelt}}{t-m_\pi^2}
\end{align}
where $f_{\pi N\Delta}$ has been normalized using the pion mass. The residue for pion exchange follows as,
\begin{align}
    \mathcal{R}^\pi_{\helgammat\helprotont\helDelt} &= -16\pi \lim_{t \to m_\pi^2}  \frac{e f_{\pi N\Delta}}{m_\pi m_\Delta}\sqrt{\frac{2}{3}} t\helgammat (p^t)^2~(t-t_{th})\delta_{\helprotont,\helDelt}
\end{align}
From this expression and the value listed in Table~\ref{tab:resTab}, we get $f_{\pi N\Delta} = -2.18 \pm 0.08$, which is consistent with the value obtained from the decay width of the $\DelBar$~\cite{ParticleDataGroup:2024cfk}, quark model estimates~\cite{Elster:1988zu,Janssen:1996kx}, and the values from chiral perturbations theory~\cite{Hemmert:1997tj,Fettes:2000bb}. To extract the coupling constants for other exchanges, we need to first model the upper vertex. These can be extracted from the radiative decays of the respective resonances. The values are listed in the caption of Table ~\ref{tab:couplings}.
The effective Lagrangians for the various meson exchange amplitudes and the expressions for the corresponding residues are described in Appendix~\ref{app:EffLag}. The values of those coupling constants are listed in Table~\ref{tab:couplings}.

There have been a few attempts to determine the $\rho N\Delta$ coupling constants, \eg, from quark model calculations~\cite{Arenhovel:1975vf,Janssen:1996kx,Nam:2011np}. 
These calculations lead to $g^{(2)}_{\rho N\Delta} \simeq 23$, an order of magnitude smaller than our extraction, and neglect the other two couplings. Similar value was found by a low-energy global fit \cite{Ronchen:2012eg}. However, the $\rho$-exchange appearing in the potential is crucially dependent on the choice of the form factor, making the corresponding coupling prone to large uncertainties. To the best of our knowledge, no attempts have been made to extract either the $\rho N\Delta$ coupling constants using the complete Lagrangian or the $b_1 N\Delta$ and $a_2 N\Delta$ couplings. Hence, our work serves as their first estimation.

\section{Conclusions\label{sec:Conclusions}}

In this work, we have performed an amplitude analysis of high-energy polarized photoproduction of $\pi^- \Delta^{++}$ within a Regge framework, with the goal of constraining the underlying production mechanisms and extracting the residues of various meson exchanges. The amplitude was constructed in the $s$-channel frame and includes contributions from both natural and unnatural parity Reggeons. The $s$-channel Regge couplings are parametrized as polynomials in $t$ and determined through a combined fit to the SDMEs and cross section. The inclusion of SDME data allows us to constrain not only the magnitudes but also the relative phases of the helicity amplitudes, leading to a quantitatively good description of the data across the accessible kinematic range. The results confirm the dominance of pion exchange in the small-$t$ region, while natural parity exchanges become important in the large-$t$ region.\par

The $s$-channel amplitude is crossed to the $t$-channel frame, where the poles of definite parity partial waves can be identified with the physical states. To extract the residues, we isolate the kinematical singularities from the $t$-channel PCHAs and evaluate the residual function at the meson pole. The value of the $\pi N\Delta$ coupling constant extracted from the pion exchange residue is in agreement with value from the decay 
width of the $\Delta(1232)$. In the case of $\rho,$ $b_1,$ and $a_2$ exchanges, we provide the first extraction of coupling constants for the respective vertices. This study demonstrates that high-energy photoproduction, combined with analyticity and crossing symmetry, offers a powerful tool for determining hadronic couplings in regimes where direct measurements are not feasible.

\begin{acknowledgments}
The authors thank M.~Shepherd, F.~Afzal, A.~Schertz, and J.~Stevens from the GlueX collaboration for useful discussions.

This work was supported by the U.S. Department of Energy contract \mbox{DE-AC05-06OR23177}, under which Jefferson Science Associates, LLC operates Jefferson Lab, by U.S. Department of Energy Grant Nos.~\mbox{DE-FG02-87ER40365}, and \mbox{DE-SC0011090}, and it contributes to the aims of the U.S. Department of Energy \mbox{ExoHad} Topical Collaboration, contract \mbox{DE-SC0023598}. RJP acknowledges support by the Simons Foundation award Simons Collaboration on Confinement and \mbox{QCD StringsMPS-QCD-00994314} and by MIT. GM and VM have been supported by projects \mbox{CEX2024-001451-M} (Unidad de Excelencia ``María de Maeztu''), \mbox{PID2023-147112NB-C21}, all financed by \mbox{MICIU/AEI/10.13039/501100011033/} and FEDER, UE. GM is a Beatriu de Pin\'os Fellow. VM is a Professor Serra H\'unter. VM acknowledges support from \mbox{CNS2022-136085}. {\L}B was partially financed by the Faculty of Physics
and Applied Computer Science AGH UST statutory tasks within subsidy of
Ministry of Science and Higher Education.
\end{acknowledgments}

\appendix

\section{Derivation of the Parity-conserving helicity amplitudes}
\label{app:pcha}
In the center-of-mass frame of the $t$-channel process, the amplitude (henceforth, $t$-channel amplitude) can be decomposed into partial waves as,
\begin{align}
    T^{(t)}_{\helgammat\helprotont\helDelt} (t,z_t) &= \sum_{J\geq M} \frac{2J+1}{4\pi}  a^J_{\helgammat\helprotont\helDelt} (t) d^J_{\helgammat,\helprotont-\helDelt} (z_t)\label{eq:pwa}
\end{align}
While such a decomposition separates the various spin-$J$ exchanges, it does not distinguish between the different naturalities since the Wigner-$d$ functions are not parity definite.
To identify the individual poles with physical states, we need to construct the partial wave amplitudes that are eigenstates of parity. This automatically guaranties the separation of the naturalities~\cite{JPAC:2018dfc,Collins:1971ff}. The parity definite combination of the Wigner-$d$ function is defined as,
\begin{align}
    \hat{d}^{J\pm}_{\helgammat,\helprotont-\helDelt}(z_t) &= \hat{d}^{J}_{\helgammat,\helprotont-\helDelt} \mp (-1)^{\helprotont - \helDelt + M} \hat{d}^{J}_{-\helgammat,\helprotont-\helDelt}\label{eq:PCWd}
\end{align}
where
\begin{align}
    \hat{d}^{J}_{\helgammat,\helprotont-\helDelt}(z_t) &= \frac{d^{J}_{\helgammat,\helprotont-\helDelt}}{\xi^{(t)}_{\helgammat\helprotont\helDelt}}\label{eq:redWd}
\end{align}
with $\xi^{(t)}_{\helgammat\helprotont\helDelt}$ being the $t$-channel half-angle factor. This construction leads to definite-parity polynomials in $z_t$. In the large-$s$ limit they reduce to,
\begin{subequations}
\begin{align}
    \hat{d}^{J+}_{\helgammat,\helprotont-\helDelt} (z_t) &\simeq F^J_{\helgammat\helprotont\helDelt} \frac{s^{J-M}}{(2q^t p^t)^{J-M}} + \mathcal{O}\left(s^{J-M-2}\right)\label{eq:dhat}\\
    \hat{d}^{J-}_{\helgammat,\helprotont-\helDelt} (z_t) &\simeq \mathcal{O}\left(s^{J-M-1}\right)
\end{align}
\end{subequations}
and 
\begin{align}
    F^J_{\helgammat\helprotont\helDelt} &= \frac{1}{2^{J-1}} \frac{(2J)!\, (-1)^{\oneh(|\helgammat-\helprotont+\helDelt|+\helgammat-\helprotont+\helDelt)}}{\sqrt{(J-M)!\,(J+M)!\,(J-N)!\,(J+N)!}}\text{ .} \label{eq:TheF}
\end{align}
where, $N=\text{Min}(|\helgammat|,|\helprotont-\helDelt|)$. The $\hat{d}^{J-}(z_t)$ can therefore always be neglected.

The $t$-channel half angle factor contains all possible kinematical singularities in $s$. Therefore, we define PCHAs~\cite{JPAC:2018dfc},
\begin{align}
    \hat{T}^{\eta}_{\helgammat\helprotont\helDelt} (t,s) &= \hat{T}^{(t)}_{\helgammat\helprotont\helDelt} \notag\\
    &  - \eta\,\eta_\pi\,\eta_\gamma\,\eta_\Delta\,\eta_{\bar{N}} (-1)^{\helprotont - \helDelt + M}~\hat{T}^{(t)}_{-\helgammat\helprotont\helDelt}\text{ ,}\label{eq:ksfpchadef}
\end{align}
and in the large-$s$ limit,

\begin{align}
& \hat{T}^{\eta}_{\helgammat\helprotont\helDelt} (t,s) =\frac{1}{4\pi} \sum_J (2J+1) \Big[a^{J\eta}_{\helgammat\helprotont\helDelt}(t)~\hat{d}^{J+}_{\helgammat,\helprotont-\helDelt} (z_t) \nonumber\\
    &\qquad\qquad + \mathcal{O}\left( s^{J-M-1}\right) \Big] \notag\\
     &\quad= \sum_J \Bigg[\frac{(2J+1)}{4\pi}s^{J-M}  \frac{\,F^J_{\helgammat\helprotont\helDelt}}{\,(2q^t p^t)^{J-M}} a^{J\eta}_{\helgammat\helprotont\helDelt}(t)\nonumber\\
    &\qquad\qquad + \mathcal{O}\left( s^{J-M-1}\right)\Bigg] \label{eq:Lagksfpcha}
\end{align}
where $a^{J\eta}_{\helgammat\helprotont\helDelt}(t) = \frac{1}{2}[a^{J}_{\helgammat\helprotont\helDelt}(t) + \eta\,a^{J}_{-\helgammat\helprotont\helDelt}(t)]$. 
The coefficient of $\dfrac{2J+1}{4\pi}s^{J-M}$ is the amplitude for the partial wave with spin-naturality $J\eta$.

These partial wave amplitudes possess kinematical singularities in $t$ 
at the $t$-channel threshold, pseudothreshold, and around $t=0$. 
Removing these singularities makes the resulting quantities analytic everywhere except for the dynamical singularities.
The kinematical singularities are determined entirely by the barrier factors of the lowest $L$, and hence can be constructed using angular momentum counting~\cite{Cohen-Tannoudji:1968lnm,Jackson:1968rfn,Collins:1971ff}. Taking into account an extra power of $E_\gamma^t = q_t$ arising from gauge invariance for unnatural parity~\cite{JointPhysicsAnalysisCenter:2024kck}, one gets
\begin{align}
    K^\eta_{\helgammat\helprotont\helDelt}(t) &= \left(t-m_\pi^2\right)^{M-J_\gamma-J_\pi+1}\notag \\
    &\times\sqrt{t-t_{th}}^{M-J_\Delta-J_N + \frac{1+\eta}{2}} \notag\\
    &\times \sqrt{t-t_{pth}}^{M-J_\Delta-J_N + \frac{1-\eta}{2}} t^{-\frac{M-N}{2}}\notag \\ 
 &= (2q^t\sqrt{t})^{M} (2p^t\sqrt{t})^{M-\threeh} \sqrt{\frac{t - t_{th}}{t - t_{pth}}}^\frac{\eta}{2} t^{-\frac{M-N}{2}} \label{eq:tBF}
\end{align}
Note that $\frac{M-N}{2}$ is either $0$ or $\frac{1}{2}$.

Dividing the PCHA in Eq.~\eqref{eq:Lagksfpcha} by the barrier factors gives us the PCHA free of kinematical singularities in $t$. The corresponding partial wave amplitude is given by,
\begin{align}
    \hat{a}^{J\eta}_{\helgammat\helprotont\helDelt}(t) &= \frac{1}{K^\eta_{\helgammat\helprotont\helDelt}(t)}\frac{F^J_{\helgammat\helprotont\helDelt}}{(2q^t p^t)^{J-M}}\,a^{J\eta}_{\helgammat\helprotont\helDelt}(t)\text{ .}\label{eq:ahat}
\end{align}
$\hat{a}^{J\eta}$ can in principle possess additional simple poles at $t=0$. For a discussion on their relation to daughter trajectories, see~\cite{Collins:1971ff}.

\section{Pion exchange and gauge invariance\label{app:pionEx}}

To extract the pion exchange residue correctly, the partial wave decomposition and hence the residue function must be analytically continued to the $J\to 0$ limit. The formalism to achieve this was described in detail in Ref.~\cite{JointPhysicsAnalysisCenter:2024kck} for the case of single pion photoproduction process. We recall from Eq.~\eqref{eq:dhat} that $\hat{d}^{J+}_{\helgammat,\helprotont-\helDelt} \propto F^J_{\helgammat\helprotont\helDelt}$. For a spin-0 exchange, $\hat{d}^{J=0,+}_{\helgammat,\helprotont-\helDelt}\equiv 0$, which makes the corresponding partial wave undefined. However, in the $J\to 0$ limit, 

\begin{align}
    \lim_{J\to 0}~F^J_{\helgammat\helprotont\helDelt} &\to -2\sqrt{J}~\delta_{\helprotont,\helDelt}\text{ .}
\end{align}
This gives a singularity in the Eq.~\eqref{eq:dnpwa}. To demonstrate how this cancels, we first sketch the derivation given in Ref.~\cite{JointPhysicsAnalysisCenter:2024kck} by making use of an effective Lagrangian given explicitly in Appendix~\ref{app:EffLag}. The amplitude is given by,
\begin{align}
    T^{J}_{\helgammat\helprotont\helDelt} &= -2 g^J_{\pi\gamma} g^J_{\bar{N}\Delta}\sqrt{t - t_{th}}~E_\gamma^{J} p^{J+1} \sqrt{t}~ c_J c_{J-1} \nonumber\\
    &\qquad C^{J\helgammat}_{J-10;1\helgammat}\sqrt{\frac{2}{3}}\frac{\sqrt{t}}{m_\Delta} \frac{1}{t-m_J^2} d^J_{\helgammat,\helprotont - \helDelt}(\theta_t)
\end{align}
where we have omitted the label for the $t$-channel momenta. Using the relations,
\begin{align}
    &c_{J-1} = \sqrt{\frac{2J-1}{J}} c_J,  \quad C^{J\helgammat}_{J-10;1\helgammat} = \sqrt{\frac{J+1}{2(2J-1)}}\,,\nonumber\\
     & d^J_{\helgammat,0}(\theta_t) = \sqrt{\frac{J+1}{2J}} d^1_{\helgammat,0}(\theta_t) P^{11}_{J-1} (z_t)
\end{align}
we get for $\helDelt = \helprotont$,
\begin{align}
    T^{J}_{\helgammat\helprotont\helDelt}\Big|_{\helprotont=\helDelt} &= -2 g^J_{\pi\gamma} g^J_{\bar{N}\Delta}\sqrt{t - t_{th}}~E_\gamma^{J} p^{J+1}~\sqrt{\frac{2}{3}}\frac{t}{m_\Delta} c_J^2\nonumber\\
    &\qquad  \left(\frac{J+1}{2J}\right) d^1_{\helgammat,0}(\theta_t) P^{11}_{J-1} (z_t) \frac{1}{t-m_J^2}
\end{align}
To impose definite signature, the symmetrization of $P_{J-1}^{11}(z_t) \to \frac{1}{2} \left[P_{J-1}^{11}(z_t) - P_{J-1}^{11}(-z_t)\right]$ is understood. Note that the symmetrized Jacobi polynomial is proportional to $J$, thus canceling the divergence seen above. Implementing this philosophy to a generic amplitude of the form given in Eq.~\eqref{eq:pwa} we get,
\begin{align}
    T^{J}_{\helgammat\helprotont\helDelt} &= \frac{2J+1}{4\pi}  a^J_{\helgammat\helprotont\helDelt} (t) d^J_{\helgammat,\helprotont-\helDelt} (z_t)\\
    &= \frac{2J+1}{4\pi}  a^J_{\helgammat\helprotont\helDelt} (t) \sqrt{\frac{J+1}{2J}} d^1_{\helgammat,0}(\theta_t) P^{11}_{J-1} (z_t)\\
    &= \frac{2J+1}{4\pi}  \tilde{a}^J_{\helgammat\helprotont\helDelt} (t) \left(\frac{J+1}{2J}\right) d^1_{\helgammat,0}(\theta_t) P^{11}_{J-1} (z_t)
\end{align}
where, $\tilde{a}^J_{\helgammat\helprotont\helDelt} (t) = \sqrt{2J/(J+1)}\,a^J_{\helgammat\helprotont\helDelt} (t)$. Since the factor $F^J_{\helgammat,\helprotont\helDelt}$ is already a part of $d^J_{\helgammat,\helprotont-\helDelt} (z_t)$, it is sufficient to redefine
\begin{align}
    \lim_{J\to 0} \hat{F}^J_{\helgammat,\helprotont\helDelt} &= \lim_{J\to 0} F^J_{\helgammat\helprotont\helDelt}\sqrt{\frac{J+1}{2J}}\\
    &= -\sqrt{2}~\delta_{\helprotont,\helDelt}\text{ .}
\end{align}

\section{Cancellation of kinematical singularities in crossing relation\label{app:xingapp}}
From the general form of the $s$-channel helicity amplitude (\cf Eq.~\eqref{eq:amp}) we see that the half-angle factor contains singularities in $t$. 
From Eq.~\eqref{eq:halfangleslarges} one can see that such singularities appear as branch points at \mbox{$t=0$}. 
Combining Eqs.~\eqref{eq:xingrel}, \eqref{eq:ksftchannel}, and~\eqref{eq:amp}, we cross to the $t$-channel and make the kinematical singularity structure explicit,
\begin{align}
    \hat{T}^{(t)}_{\helgammat\helprotont\helDelt}\!\! &=\!\! \sum_{\helgamma,\helproton,\helDel}\delta_{\helgammat,\helgamma} d^\oneh_{\helprotont,\helproton}(\omega_N)\, d^\threeh_{\helDelt,\helDel}(\omega_\Delta)\nonumber\\
    &\qquad\qquad\times\frac{\xi^{(s)}_{\helgamma\helproton\helDel}}{\xi^{(t)}_{\helgammat\helprotont\helDelt}} \hat{T}^{(s)}_{\helgamma\helproton\helDel}\label{eq:ksfamp}
\end{align}
In the large-$s$ limit, the $t$-channel half-angle factor takes the form,
\begin{align}
    \xi^{(t)}_{\helgammat\helprotont\helDelt}  &= i^{|\helgammat + \helprotont - \helDelt|} \frac{s^M}{(2q^t p^t)^M}~ + \mathcal{O}\left(s^{M-1} \right)\label{eq:HAt}
\end{align}
The $(2q^t p^t)^M$ has branch singularities in $t$ at threshold and pseudothreshold of the final state ($\Delta\bar{p}$ system), and a zero at $t=m_\pi^2$, coming from the collapse of initial-state threshold and pseudotreshold due to the massless photon.
We now proceed with defining the analogs of Eq.~\eqref{eq:Lagksfpcha} from the crossed amplitudes of Eq.~\eqref{eq:ksfamp}. 
Substituting \cref{eq:ksfamp,eq:Bsumpoles} in Eq.~\eqref{eq:ksfpchadef} we get after some algebra

\begin{align}
    \hat{T}^{J\eta}_{\helgammat\helprotont\helDelt} &= (2q^t p^t)^M \sum_{\helgamma,\helproton,\helDel} \frac{s^{J-M}}{t-m_J^2}\delta_{\helgammat,\helgamma}\, \nonumber\\
    &\qquad\mathcal{C}^\eta_{\helprotont\helDelt;\helproton\helDel} \frac{\sqrt{-t}^{|\helgamma - \helproton + \helDel|}}{i^{|\helgammat + \helprotont - \helDelt|}} \mathcal{B}^{J\eta}_{\helgamma\helproton\helDel}\label{eq:ksfpchaX}
\end{align}
where,
\begin{align}
    \mathcal{C}^\eta_{\helprotont\helDelt;\helproton\helDel} &= d^\oneh_{\helprotont,\helproton}(\omega_N)\, d^\threeh_{\helDelt,\helDel}(\omega_\Delta) \nonumber\\
    &- \eta\,(-1)^{\helproton - \helDel} d^\oneh_{\helprotont,-\helproton}(\omega_N)\, d^\threeh_{\helDelt,-\helDel}(\omega_\Delta)\text{ .}
\end{align}
In arriving at the expression above we have made use of the symmetry properties of Wigner-$d$ matrices as well as the $\hat{T}^{(s)}_{\helgamma\helproton\helDel}$, and the large-$s$ expressions for the half-angle factors. 
Comparing Eq.~\eqref{eq:ksfpchaX} with Eq.~\eqref{eq:Lagksfpcha}, we get the partial wave amplitude as,
\begin{align}
     a^{J\eta}_{\helgammat\helprotont\helDelt}(t) &= \frac{4\pi}{2J+1} \frac{(2q^t p^t)^J}{F^J_{\helgammat\helprotont\helDelt}} \frac{1}{t-m_J^2} \nonumber\\
     &\sum_{\helgamma,\helproton,\helDel} \delta_{\helgammat,\helgamma}\mathcal{C}^\eta_{\helprotont\helDelt;\helproton\helDel} \nonumber\\
     &\qquad \frac{\sqrt{-t}^{|\helgamma - \helproton + \helDel|}}{i^{|\helgammat + \helprotont - \helDelt|}} \mathcal{B}^{J\eta}_{\helgamma\helproton\helDel} \label{eq:dnpwa}
\end{align}
These partial wave amplitudes contain kinematical singularities in $t$ which they inherit from Eq.~\eqref{eq:ksfpchaX}. 
The threshold and pseudothreshold behaviors arise entirely out of the crossing matrices, where as the singularity at $t=0$ comes from the both crossing matrices and $s$-channel half-angle factor. 
The crossing matrix can be put in the form,
\begin{align}
    \mathcal{C}^\eta_{\helprotont\helDelt;\helproton\helDel} &\to \frac{1}{(p^t)^2}\left(\sqrt{t-t_{th}}~(1+\eta)~\mathscr{P}(t)\right.\nonumber\\
    &\qquad\left. + \sqrt{t-t_{pth}}~(1-\eta)~\mathscr{Q}(t)\right)
\end{align}

where $\mathscr{P}(t)$ and $\mathscr{Q}(t)$ are some (helicity dependent) functions of $t$ regular 
near the threshold ($t_{th}$) and pseudothreshold ($t_{pth}$) but not necessarily near $t=0$. The function $\mathcal{C}^\eta_{\helprotont\helDelt;\helproton\helDel}$ has branch singularities at either threshold or pseudothreshold, depending on the naturality only.
Indeed, the fact that these singularities have kinematical origin requires that they cannot involve an interplay between different $\mathcal{B}_{\helgamma\helproton\helDel}$ (which are driven by dynamics), and therefore must factor out for any helicity.
The $t=0$ singularity is controlled by the requirement of factorization of the $s$-channel amplitude (originally called `evasion'~\cite{Cohen-Tannoudji:1968eoa}). Due to this, the $\sqrt{-t}$ factor in Eq.~\eqref{eq:dnpwa} arising from the $s$-channel half-angle factor cancels the corresponding branch point in $\mathcal{C}^\eta_{\helprotont\helDelt;\helproton\helDel}$ for each value of $\{\helproton,\helDel\}$.\par

The singularity structure just discussed perfectly matches the one obtained above from the barrier factors in Eqs.~\eqref{eq:tBF} and~\eqref{eq:ahat}. The following expression is thus free of kinematical singularities. 
The simple poles at threshold and pseudothreshold follow opposite association \ie natural parity exchange is associated with pseudothreshold pole and unnatural exchange with the threshold pole. This behavior is consistent with the expected structure of the kinematic factors shown in Eq.~\eqref{eq:tBF}. 
The factor $1\big/(p^t)^2$ arises only for external states with spin high enough (when $J_\Delta + J_N \ge 2$)~\cite{Cohen-Tannoudji:1968lnm}. 
Following the prescription described in Sec.~\ref{sec:formalism}, we remove these kinematical singularities and obtain the expression shown in Eq.~\eqref{eq:residueLag}.

\section{Spin-density matrix elements\label{app:sdmeref}}
\begin{figure}[b]
    \centering
    \includegraphics[width=\linewidth]{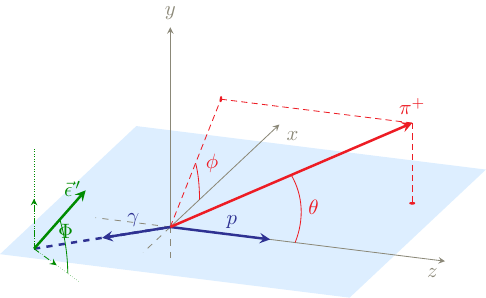}
    \caption{The GJ frame of reference. Here $p$ stands for the target proton, and $\vec{\epsilon}'$ the polarization vector of the photon. The direction of the $3$-momentum of the $\pi^-$ is not shown and can be obtained by adding the corresponding vectors of $\gamma$ and the proton.}
    \label{fig:GJ}
\end{figure}

In this appendix we review some of the properties of the SDMEs that are relevant for the present work. The angular distribution of the pion coming from the decay of the $\DelBar$ in the photoproduction process is given by~\cite{Yu:2017kng,GlueX:2021pcl},
\begin{widetext}
\begin{align}
 W(\Omega_{\pi^+},\Phi) & = 2N\bigg\{  \rho^0_{33} \sin^2\theta + \rho^0_{11}\left(\frac{1}{3} + \cos^2\theta \right) -\frac{2}{\sqrt{3}} \re\rho^0_{31} \sin2\theta\cos\phi-\frac{2}{\sqrt{3}} \re\rho^0_{3-1} \sin^2\theta\cos2\phi \nonumber
 \\
 & -P_\gamma \cos2 \Phi \left[  \rho^1_{33} \sin^2\theta + \rho^1_{11}\left(\frac{1}{3} + \cos^2\theta \right) -\frac{2}{\sqrt{3}} \re\rho^1_{31} \sin2\theta\cos\phi-\frac{2}{\sqrt{3}} \re\rho^1_{3-1} \sin^2\theta\cos2\phi  \right]
 \nonumber
 \\
 &-P_\gamma \sin2 \Phi \left[ \frac{2}{\sqrt{3}} 
 \im \rho^2_{31} \sin2\theta\sin\phi +  \frac{2}{\sqrt{3}} \im \rho^2_{3-1} \sin^2\theta\sin2\phi
 \right]
 \bigg\}\label{eq:intensity}
\end{align}
\end{widetext}
where, $P_\gamma$ is the degree of polarization of the photon, $\Phi$ is the angle made by the photon polarization vector with respect to the production plane of $\Delta^{++}$, $(\theta,\phi)$ are the polar and azimuthal angles of the $\pi^+$ decaying from the $\DelBar$ respectively~\cite{GlueX:2024dbr},  
and the $\Delta$ SDMEs are defined as,
\bsub\label{eq:rho_all}\begin{align}
    \rho^0_{\helDel\helDel^\prime} & = \frac{1}{2 N} \sum_{\helgamma,\helproton} T_{\helgamma \helproton \helDel}(s,t) T_{\helgamma\helproton \helDel^\prime}^*(s,t) \label{eq:rho0}\\
\rho^1_{\helDel\helDel^\prime} & = \frac{1}{2 N} \sum_{\helgamma,\helproton} T_{\helgamma \helproton \helDel}(s,t) T_{-\helgamma\helproton \helDel^\prime}^*(s,t)\label{eq:rho1}\\
 \rho^2_{\helDel\helDel^\prime} & = \frac{-i}{2 N} \sum_{\helgamma,\helproton} \helgamma T_{\helgamma \helproton \helDel}(s,t) T_{-\helgamma\helproton \helDel^\prime}^*(s,t)\label{eq:rho2}\\
 N &= \frac{1}{2}\sum_{\helgamma,\helproton, \helDel} |T_{\helgamma \helproton \helDel}(s,t)|^2 \label{eq:norm}
\end{align}\esub
where, $T_{\helgamma\helproton\helDel}$ are the production amplitudes\footnote{Throughout the paper we use only the sign and numerator of helicity in the notation for SDMEs. Thus, \eg $\rho^{0}_{3-1}$ should be read as $\rho^{0}_{\threeh-\oneh}$ and so on. This is {\it not} the case for the amplitudes.}. The SDMEs are frame dependent quantities since the helicity eigen states are dependent on the orientation of the spin with respect to the $z$-axis. The Gottfried-Jackson (GJ) frame of reference is shown in Fig.~\ref{fig:GJ}. These SDMEs govern the angular distribution of the decay products of the $\DelBar$ and hence contain information about the corresponding production mechanism. Specifically, the azimuthal distribution ($\phi$-dependence) is governed entirely by the off-diagonal SDMEs ($\helDel\! \neq \!\helDel^\prime$) where as the diagonal elements influence only the polar distribution.
The SDMEs follow the symmetry relations $\sum_{\helDel}\rho^0_{\helDel\helDel}=1$, $\rho^{0,1,2}_{-\helDel-\helDel^\prime}=(-1)^{\helDel-\helDel^\prime}\rho^{0,1,2}_{\helDel\helDel^\prime}$.

\subsection{Reflectivity and SDMEs}
The helicity states possess a symmetry under reflection about the plane containing their momenta~\cite{Chung:1974fq}. At large values of $s$, this symmetry is realized by the production amplitude. Making use of this symmetry we can define the amplitude in the photon reflectivity basis as~\cite{Chung:1974fq},
\begin{align}
    \,T^{(\epsilon)}_{\helproton,\helDel}(s,t) &= \oneh\left( T_{1,\helproton,\helDel}(s,t)+\epsilon~T_{-1,\helproton,\helDel}(s,t)\right) \label{eq:refAmp}
\end{align}
where, $\epsilon=\pm$ is the reflectivity quantum number. If one further restricts to the leading term in the expansion over $1/s$, we see that the reflectivity quantum number corresponds to the naturality of the state exchanged~\cite{Chung:1974fq,Mathieu:2019fts}. Since the realization of the approximate reflection symmetry by the production amplitude is dependent on the factorization, any effects that break the factorization like absorption, rescattering, and/or cross-channel interferences also break the reflection symmetry.  The reflectivity amplitudes also follow the parity relation,
\begin{align}
    T^{(\epsilon)}_{-\helproton-\helDel} &= -\epsilon(-1)^{\helproton-\helDel}~T^{(\epsilon)}_{\helproton\helDel}.
\end{align}
In the reflectivity basis, the SDMEs $\rho^{0,1}_{\helDel\helDel^\prime}$ separate into the respective $\epsilon$ components as,
\begin{align}
    \rho^0_{\helDel\helDel^\prime} & = \frac{1}{N} \sum_{\helproton} \left(T^{(+)}_{\helproton \helDel} T^{(+)*}_{\helproton \helDel^\prime}  + T^{(-)}_{\helproton \helDel} T^{(-)*}_{\helproton \helDel^\prime}\right)\label{eq:0ref} \\
    \rho^1_{\helDel\helDel^\prime} & =  \frac{1}{N} \sum_{\helproton} \left(T^{(+)}_{\helproton \helDel} T^{(+)*}_{\helproton \helDel^\prime}  - T^{(-)}_{\helproton \helDel} T^{(-)*}_{\helproton \helDel^\prime}\right)\label{eq:1ref} 
\end{align}

The sum and difference of these two SDMEs are reflectivity definite. From Eq.~\eqref{eq:1ref} we see that the sign of the diagonal elements of $\rho^1$ is given by the reflectivity component that dominates the production process at the given value of $t$. Since in the large-$s$ limit the reflectivity amplitudes are dominated by the exchanges of specific naturality,  this is by the far the clearest demonstration of the separation of the naturalities. 
Continuing further, the differential cross section for production by {\it in-plane} and {\it out-of-plane} polarized photon is given by $d\sigma_\parallel/dt$ and $d\sigma_\perp/dt$ respectively. In the reflectivity basis their definitions become,
\begin{align}
    \frac{d\sigma_{\parallel/\perp}}{dt} &= \kappa~\sum_{\helproton,\,\helDel} |T^{-/+}_{\helproton\helDel}|^2
\end{align}
where $\kappa$ is the phase space factor and includes any symmetry factors.
From this definition it is clear that the production from {\it in-plane} ({\it out-of-plane}) polarized photon is associated with negative (positive) reflectivity amplitudes and in turn, the dominance of the unnatural (natural) parity exchanges. Therefore, the diagonal elements of the reflectivity definite combinations of SDMEs ($\rho^0_{\helDel\helDel^\prime} + \epsilon \rho^1_{\helDel\helDel^\prime}$) provide the production probability of $\Delta$ in a given helicity state from either {\it in-plane} or {\it out-of-plane} polarized photon. It should be noted that this association of the {\it in-plane} ({\it out-of-plane}) polarization with $-(+)$ reflectivity is specific to the reaction where the net naturality of the external states is positive. In the opposite case (\eg, $\gamma +p \to a_0(980) \Delta$) {\it in-plane} ({\it out-of-plane}) polarization will be associated with $+(-)$ reflectivity. In addition, if an exchange of a specific naturality dominates in any region of $t$, then $\rho^{0,1}_{\helDel\helDel^\prime}$ acquire the approximate symmetry, $\rho^{0}_{\helDel\helDel^\prime} \approx \epsilon~\rho^{1}_{\helDel\helDel^\prime}$.\par

Since in the $s\to\infty$ limit the phase gained by the amplitude under reflection depends on the naturality of the exchanged state as well, these SDMEs in the reflectivity basis provide a direct measure of the contributions of the exchanges of the respective naturalities. Under this symmetry, the reflectivity definite SDME combinations mentioned above are be dominated by natural (for $\epsilon=+$) or unnatural Reggeon exchanges ($\epsilon=-$). Whether a given naturality exchange is suppressed or dominant depends also on the naturality of the external states involved. In general, one should look at the product of the naturalities of all the states (external as well as exchange).\par 

The other remaining SDME does not factorize into reflectivity components. In fact, $\rho^2_{\helDel\helDel^\prime}$ gives the interference between the amplitudes of two reflectivity as,
\begin{align}
    \rho^2_{\helDel\helDel^\prime} &=  \frac{i}{N} \sum_{\helproton} \left(T^{(-)}_{\helproton \helDel} T^{(+)*}_{\helproton \helDel^\prime} - T^{(+)}_{\helproton \helDel} T^{(-)*}_{\helproton \helDel^\prime}\right)\text{ .} \label{eq:2ref}
\end{align}
Again, in regions where exchange of one particular naturality dominates all the reflectivity amplitudes, $\rho^{2}_{\helDel\helDel^\prime}$ will be vanishingly small. In the following sections, we use both the helicity and reflectivity bases to discuss the physics extracted from the SDMEs. The notations $N_\sigma$ and $U_\sigma$ will be used to represent $T^{(+)}_{\helproton\helDel}$ and $T^{(-)}_{\helproton\helDel}$ respectively, with $\sigma = \helDel - \helproton$. While this gives us $8$ possible values for $\sigma$ and hence $16$ reflectivity amplitudes, the parity relations (Eq. (4)) reduce the number of independent reflectivity amplitudes to $8$ - four for each reflectivity. We consider the case where $\helproton = \frac{1}{2}$ thereby restricting the values of $\sigma = \{-2,-1,0,1\}$.

\section{Reflectivity amplitudes\label{app:refAmp}}
In this appendix, we give the explicit expressions for the reflectivity amplitudes and discuss some of their properties. 
Substituting Eq.~\eqref{eq:sAmpwAbs} in Eq.~\eqref{eq:refAmp} we get,
\begin{widetext}
\begin{align}
    U_\sigma &= \oneh \sum_{R \in \{\pi, b_1, \ldots\}} \beta^R_1 (t) \beta^R_{\helproton\helDel} (t) \mathcal{P}_R(s,t) \mathcal{S}_R(t)\sqrt{-m_\pi^2}^{1+|\sigma|} \left(\sqrt{\frac{t}{m_\pi^2}}^{|1+\sigma|} + \sqrt{\frac{t}{m_\pi^2}}^{|1-\sigma|} \right) \label{eq:refAmpUnnat}\\
    N_\sigma &= \sum_{R \in \{\rho, a_2, \ldots\}} \beta^R_1 (t) \beta^R_{\helproton\helDel} (t) \mathcal{P}_R(s,t) \mathcal{S}_R(t) \sqrt{-t}^{1+|\sigma|} \nonumber\\
    &\qquad + \oneh \sum_{R \in \{\pi, b_1, \ldots\}} \beta^R_1 (t) \beta^R_{\helproton\helDel} (t) \mathcal{P}_R(s,t) \mathcal{S}_R(t) \sqrt{-m_\pi^2}^{1+|\sigma|} \left(\sqrt{\frac{t}{m_\pi^2}}^{|1+\sigma|} - \sqrt{\frac{t}{m_\pi^2}}^{|1-\sigma|} \right) \text{ .} \label{eq:refAmpNat}
\end{align}
\end{widetext}
The negative reflectivity amplitudes given in Eq.~\eqref{eq:refAmpUnnat} contain only the unnatural parity meson exchanges. The peculiarity in the $\sqrt{-t}$ dependence is due to the PMA, which also generates the polynomial corrections to the positive reflectivity amplitude as seen in the second line of Eq.~\eqref{eq:refAmpNat}. Upon removing the PMA corrections (\ie, replacing $m_\pi^2 \to t$), the second line of the Eq.~\eqref{eq:refAmpNat} vanishes, and the form of Eq.~\eqref{eq:refAmpUnnat} reduces to that of the first line of Eq.~\eqref{eq:refAmpNat}. Further, absorption effects are present only for $\sigma \neq 0$.

\section{Plots of $s$-channel amplitudes\label{app:figs}}
\begin{figure*}
    \centering
    \includegraphics[width=0.24\linewidth]{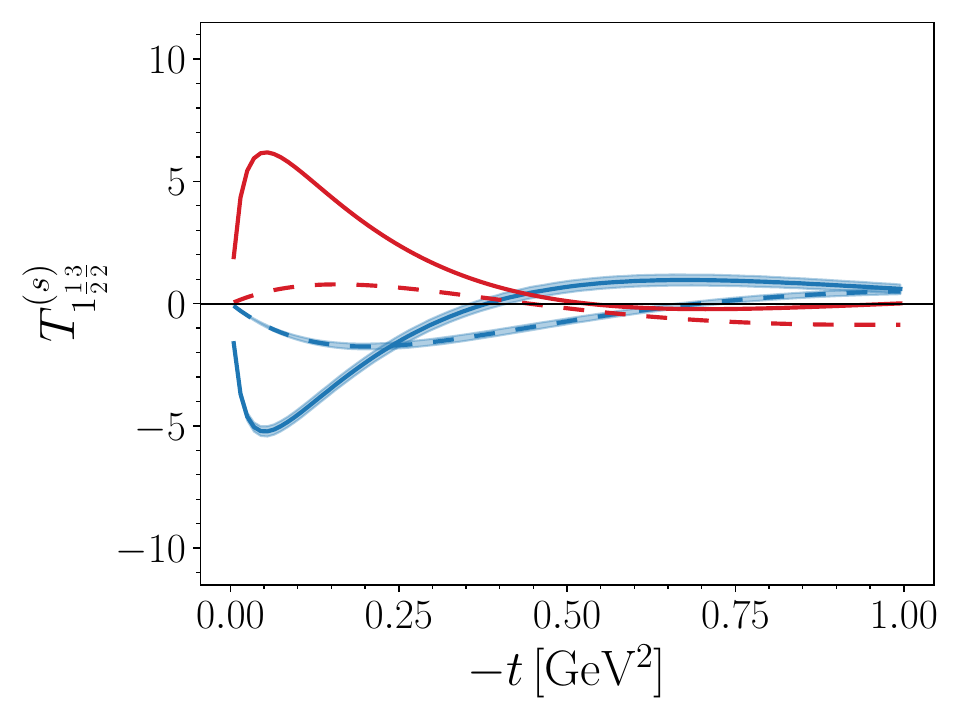}~
    \includegraphics[width=0.24\linewidth]{Amps_1m3_Full.pdf}~
    \includegraphics[width=0.24\linewidth]{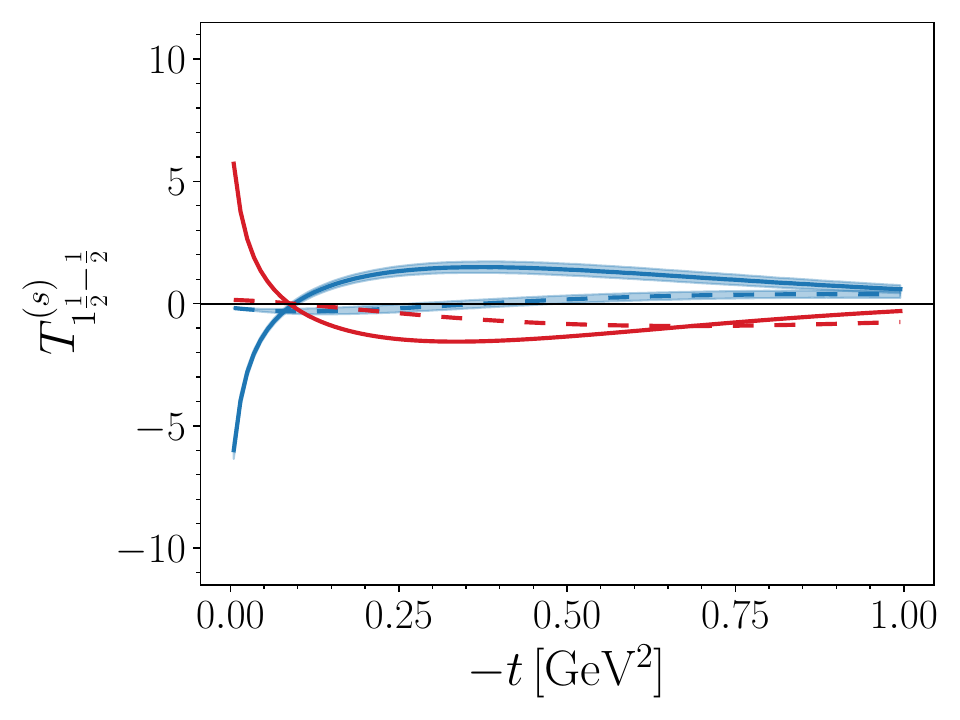}~
    \includegraphics[width=0.24\linewidth]{Amps_11_Full.pdf}\\
    \includegraphics[width=0.24\linewidth]{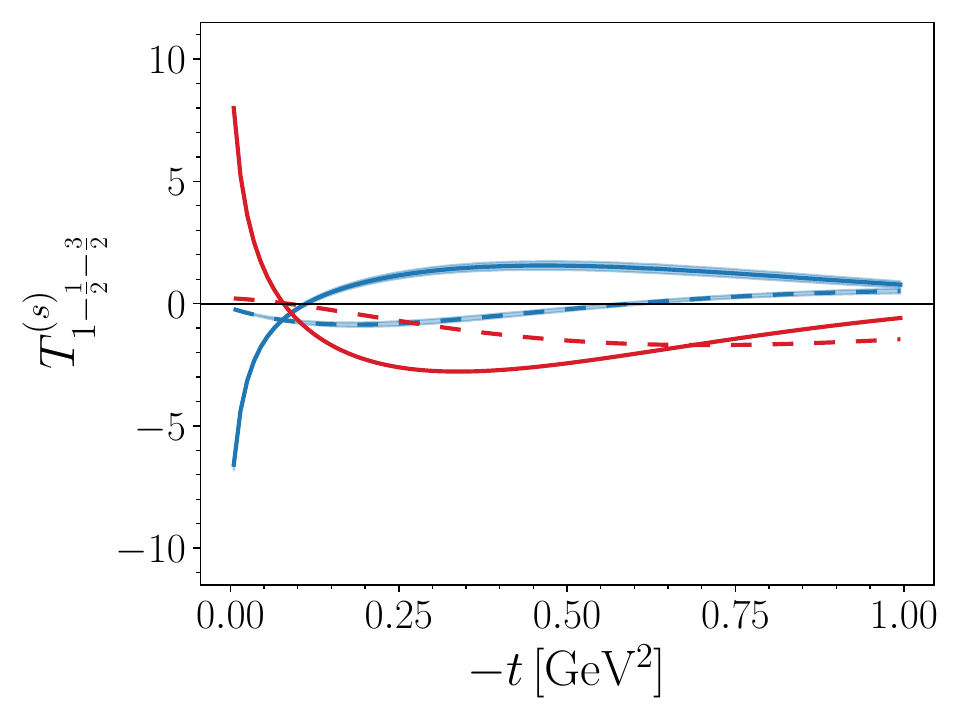}~
    \includegraphics[width=0.24\linewidth]{Amps_m13_Full.pdf}~
    \includegraphics[width=0.24\linewidth]{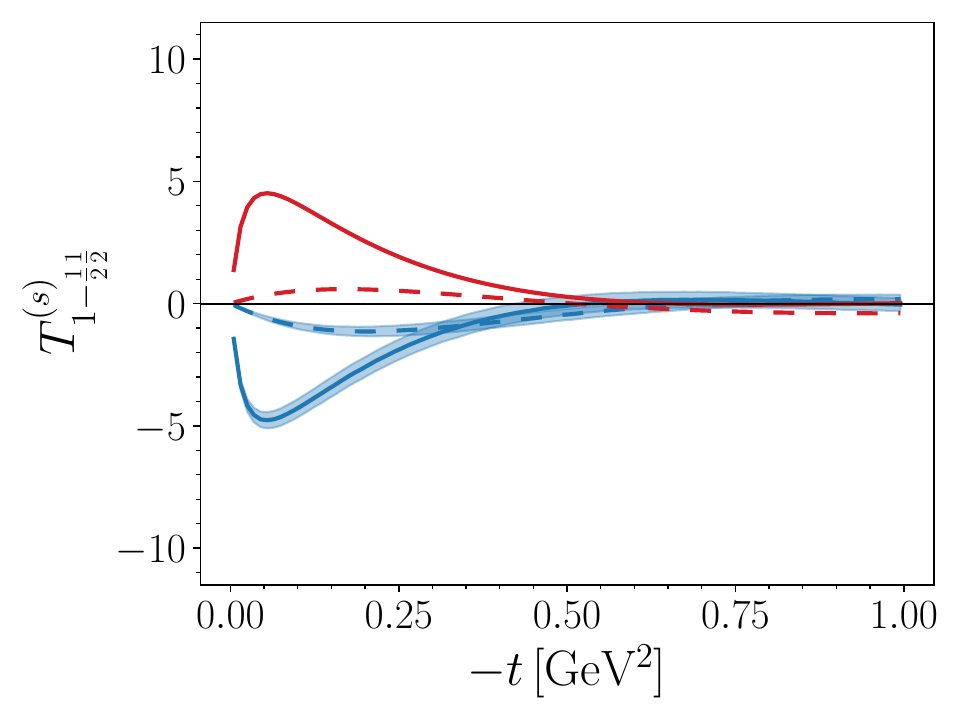}~
    \includegraphics[width=0.24\linewidth]{Amps_m1m1_Full.pdf}
    \caption{The $s$-channel helicity amplitudes. Solid lines are the real parts and dashed lines are the imaginary parts. The shaded bands represent $1\sigma$ variations.}
    \label{fig:ampind}
\end{figure*}

\begin{figure*}
    \centering
    \includegraphics[width=0.24\linewidth]{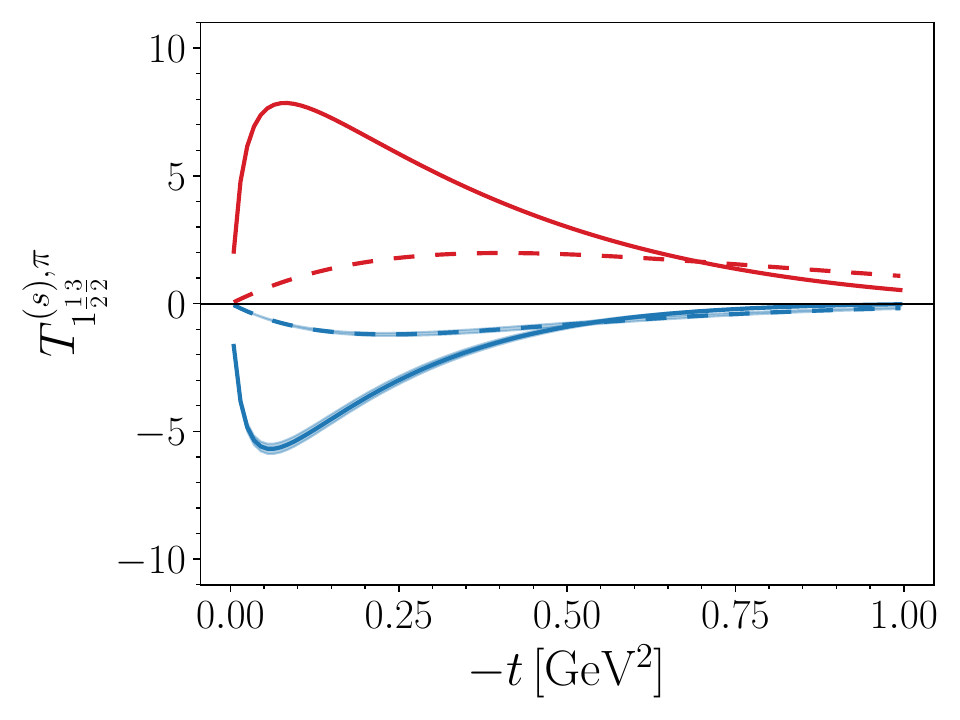}~
    \includegraphics[width=0.24\linewidth]{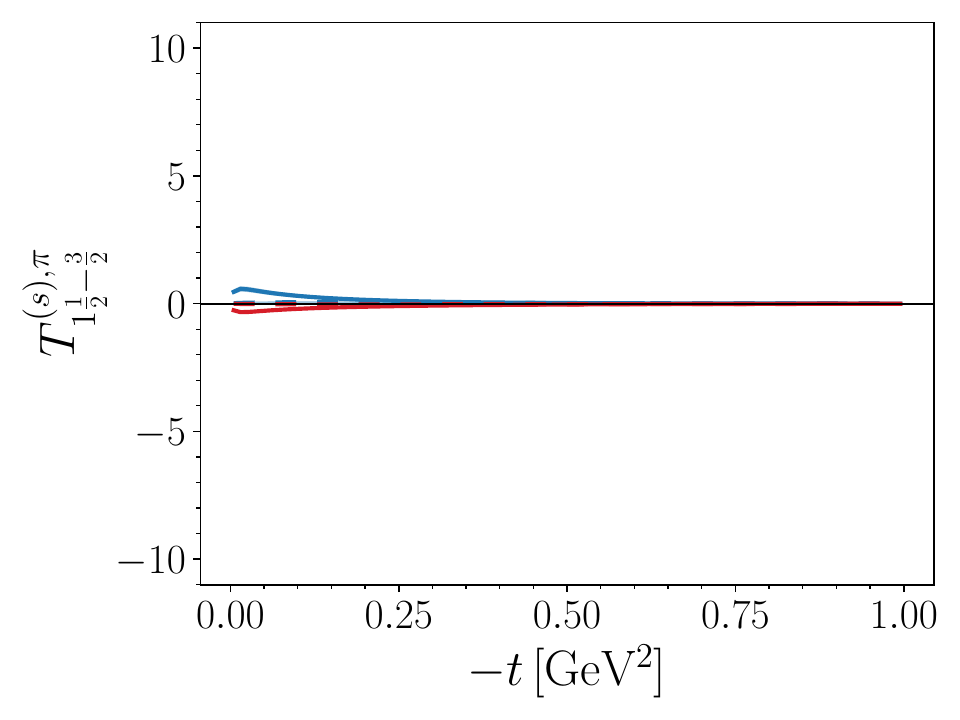}~
    \includegraphics[width=0.24\linewidth]{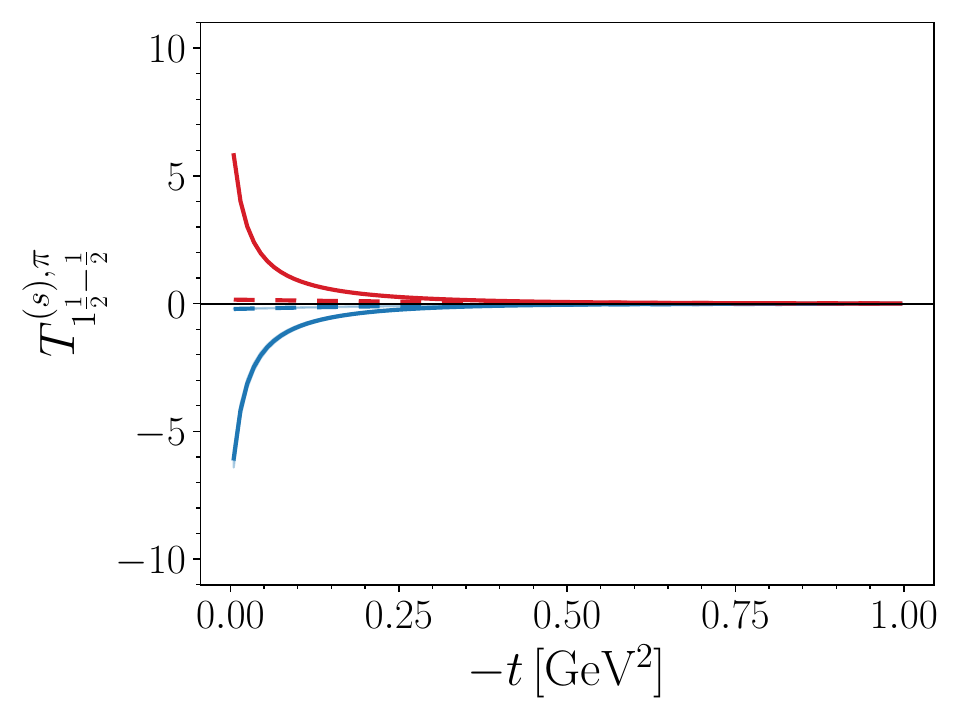}~
    \includegraphics[width=0.24\linewidth]{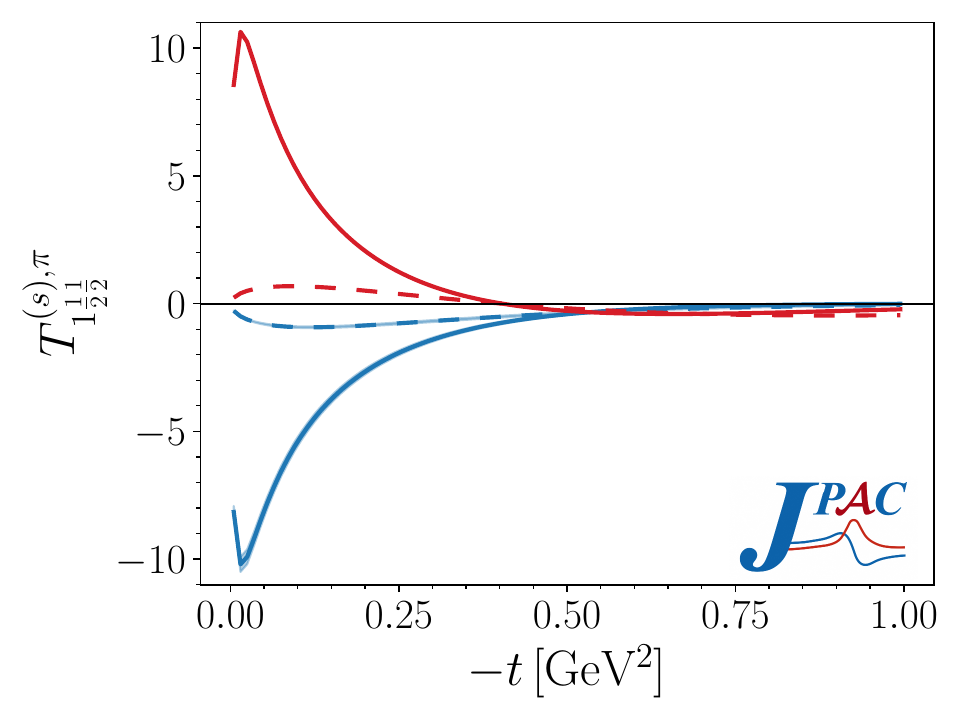}\\
    \includegraphics[width=0.24\linewidth]{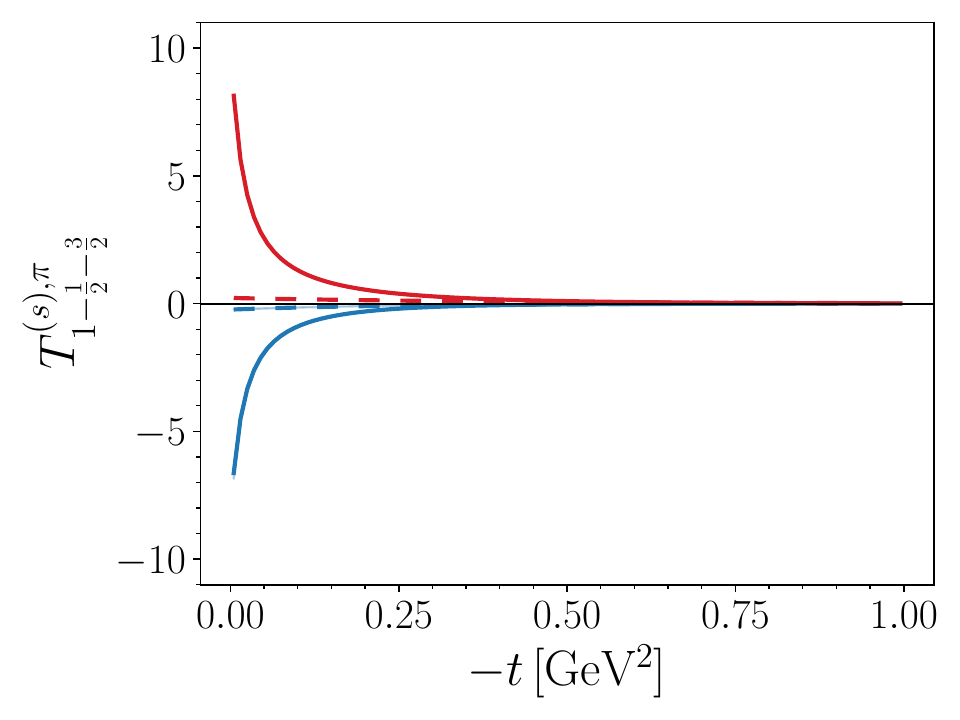}~
    \includegraphics[width=0.24\linewidth]{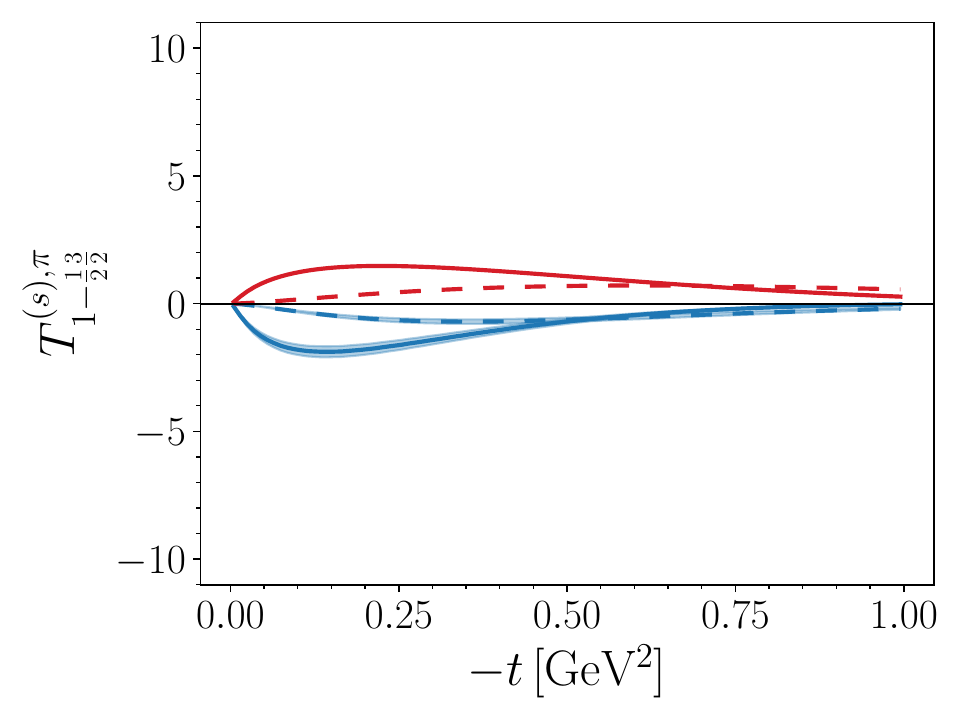}~
    \includegraphics[width=0.24\linewidth]{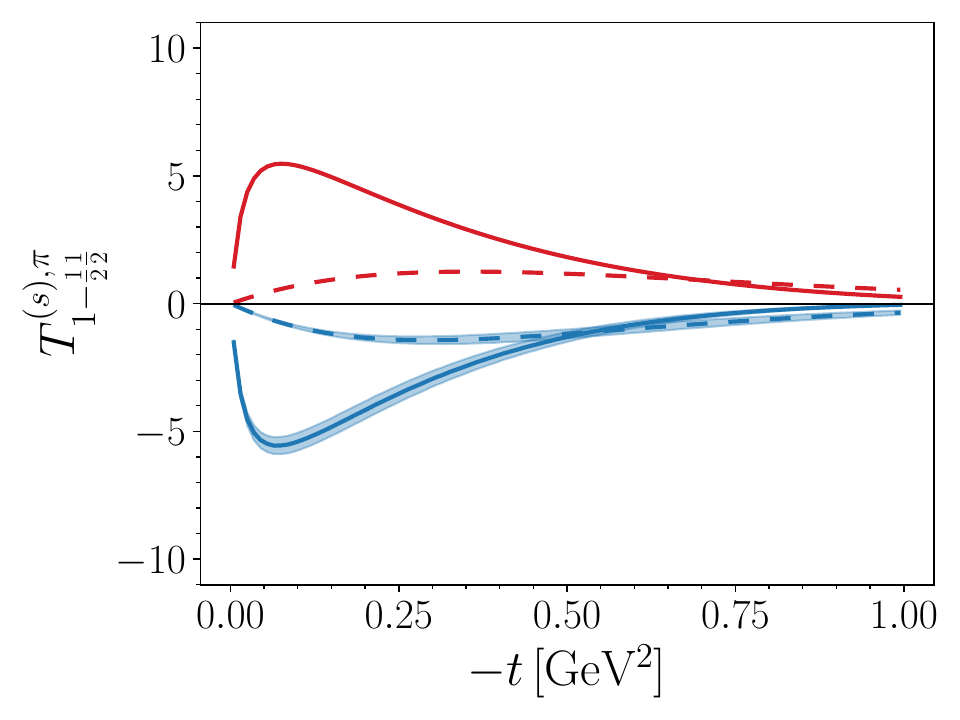}~
    \includegraphics[width=0.24\linewidth]{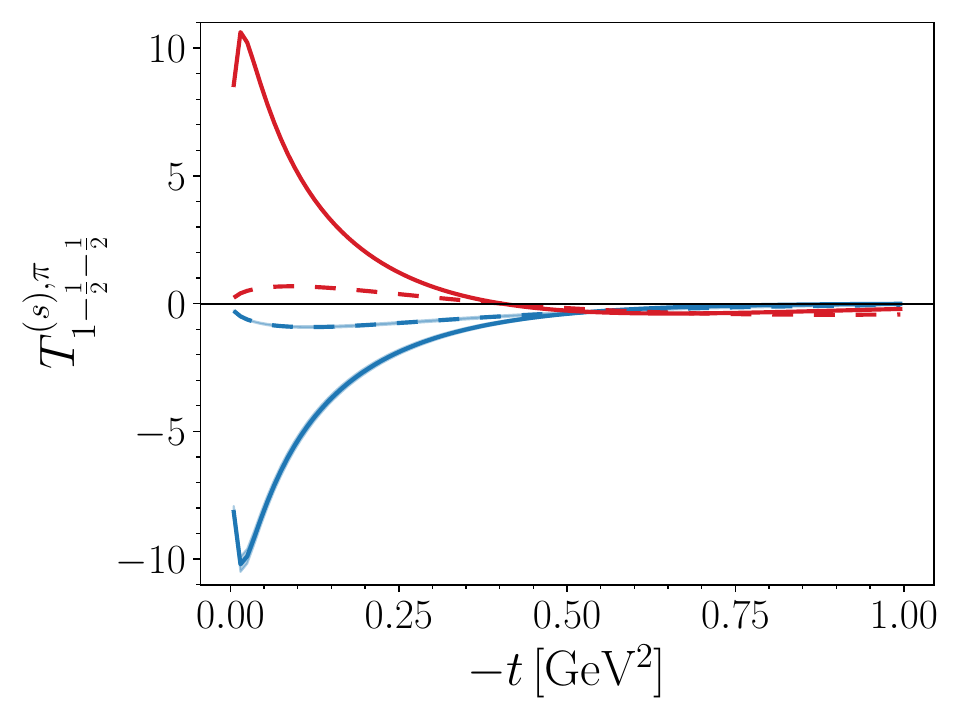}
    \caption{Pion exchange contributions to the $s$-channel amplitude. Solid lines are the real parts and dashed lines are the imaginary parts. The shaded bands represent $1\sigma$ variations.}
    \label{fig:ampindpi}
\end{figure*}

\begin{figure*}
    \centering
    \includegraphics[width=0.24\linewidth]{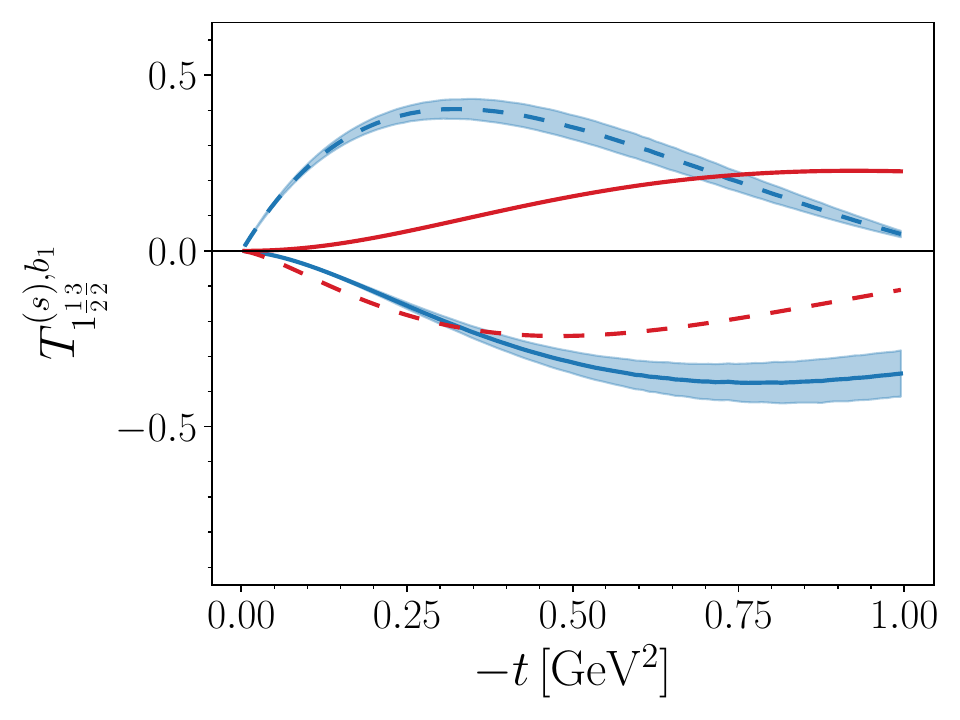}~
    \includegraphics[width=0.24\linewidth]{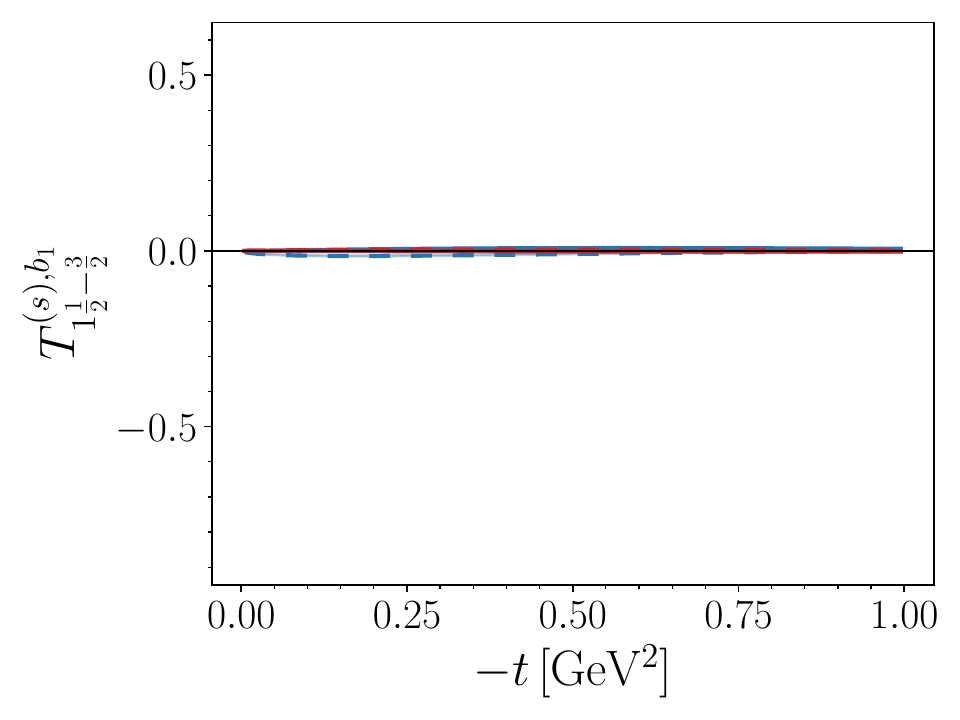}~
    \includegraphics[width=0.24\linewidth]{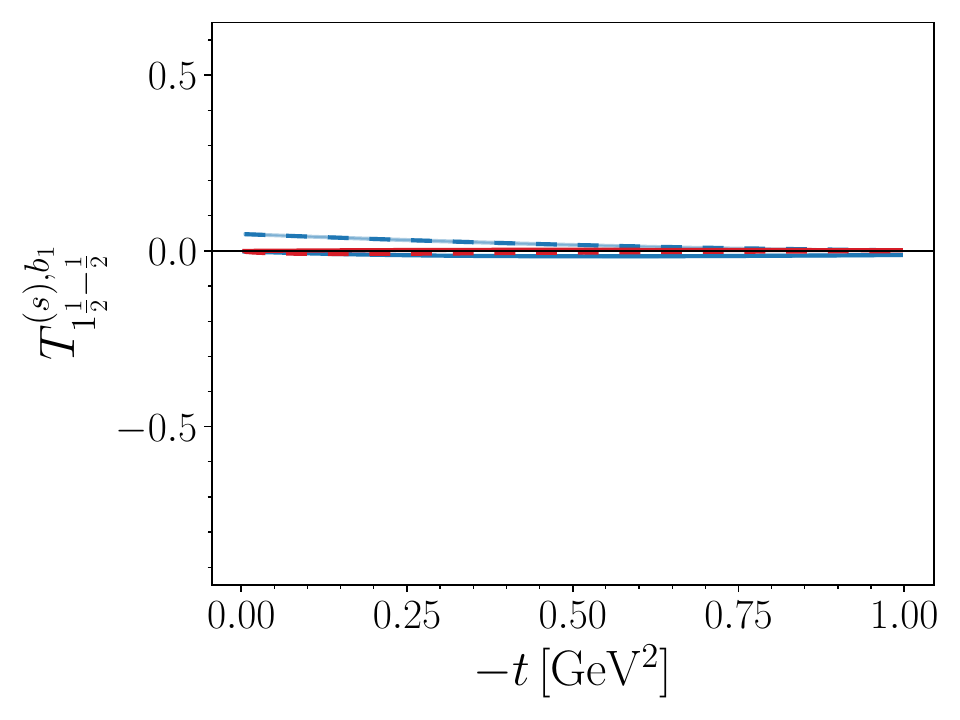}~
    \includegraphics[width=0.24\linewidth]{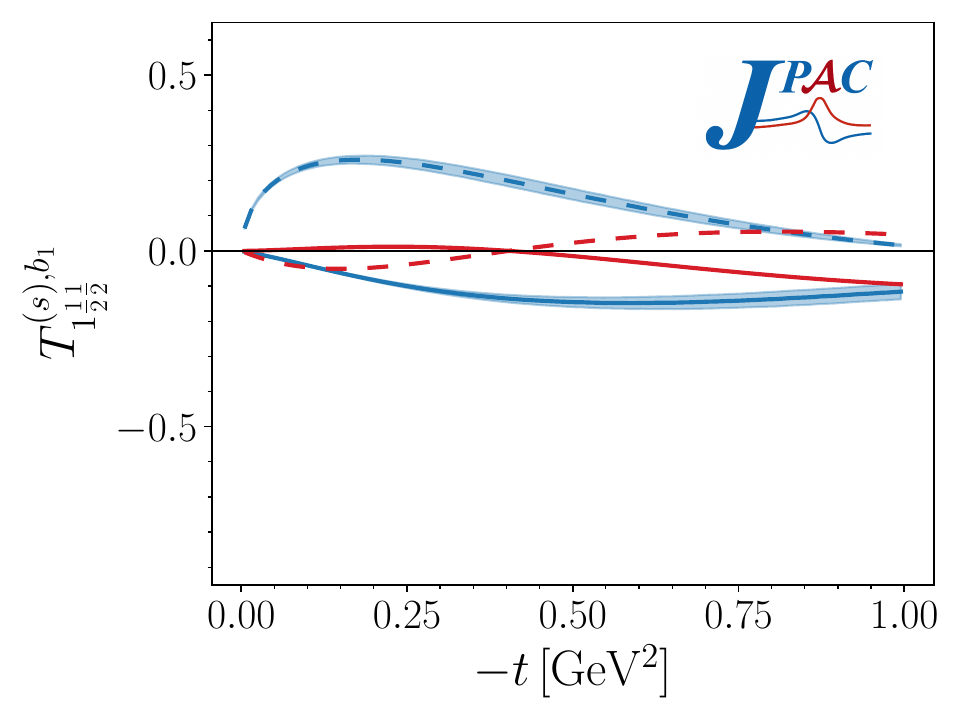}\\
    \includegraphics[width=0.24\linewidth]{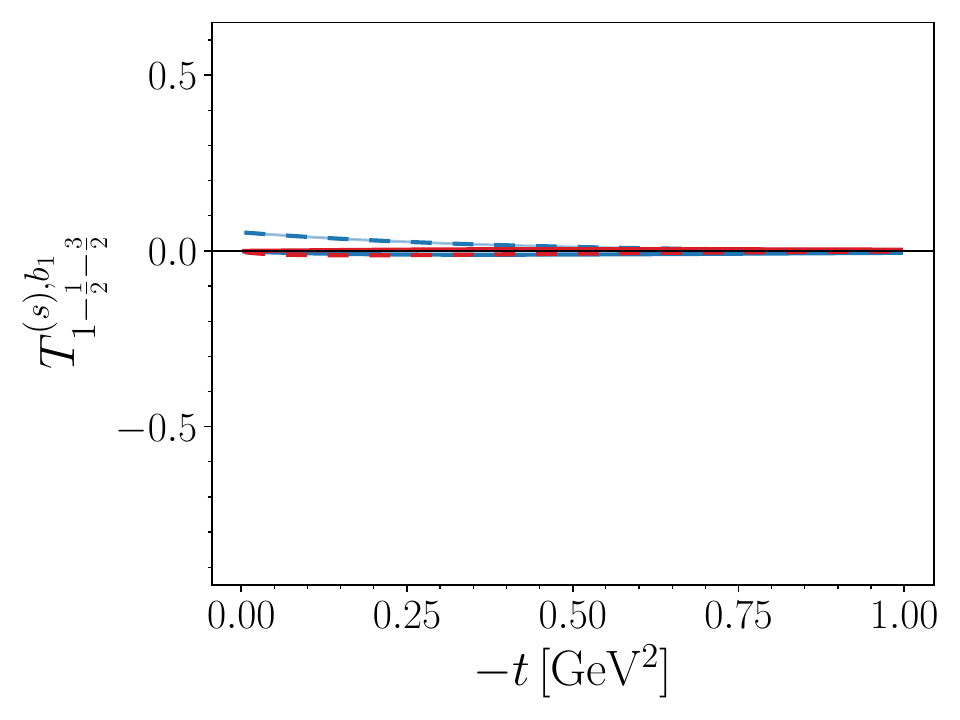}~
    \includegraphics[width=0.24\linewidth]{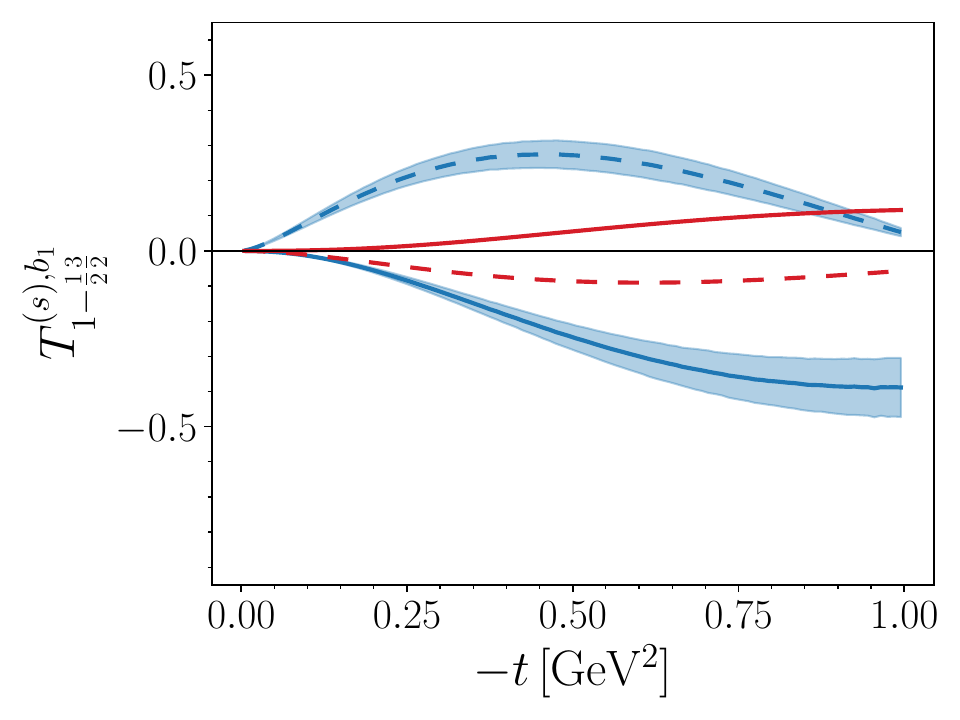}~
    \includegraphics[width=0.24\linewidth]{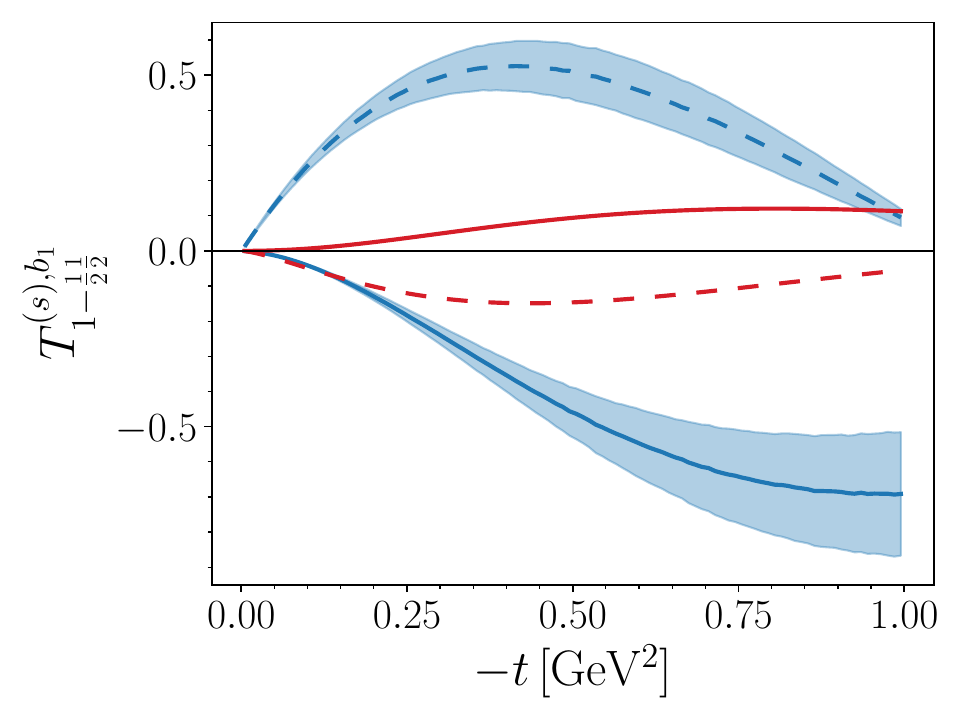}~
    \includegraphics[width=0.24\linewidth]{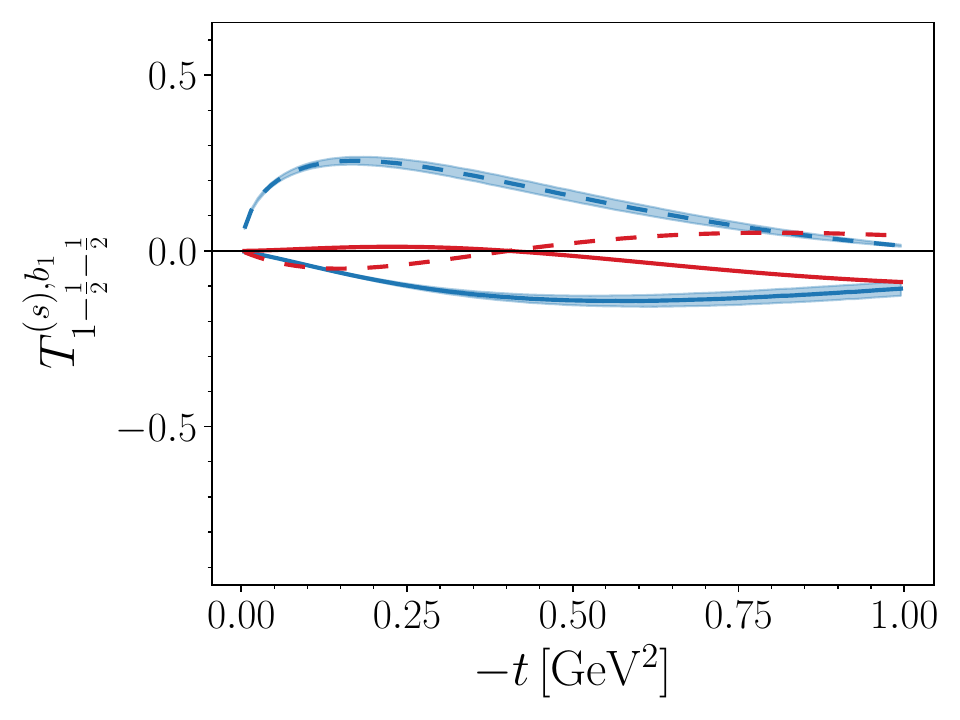}
    \caption{Same as Fig.~\ref{fig:ampindpi}, but for $b_1$ exchange.}
    \label{fig:ampindb1}
\end{figure*}

\begin{figure*}
    \centering
    \includegraphics[width=0.24\linewidth]{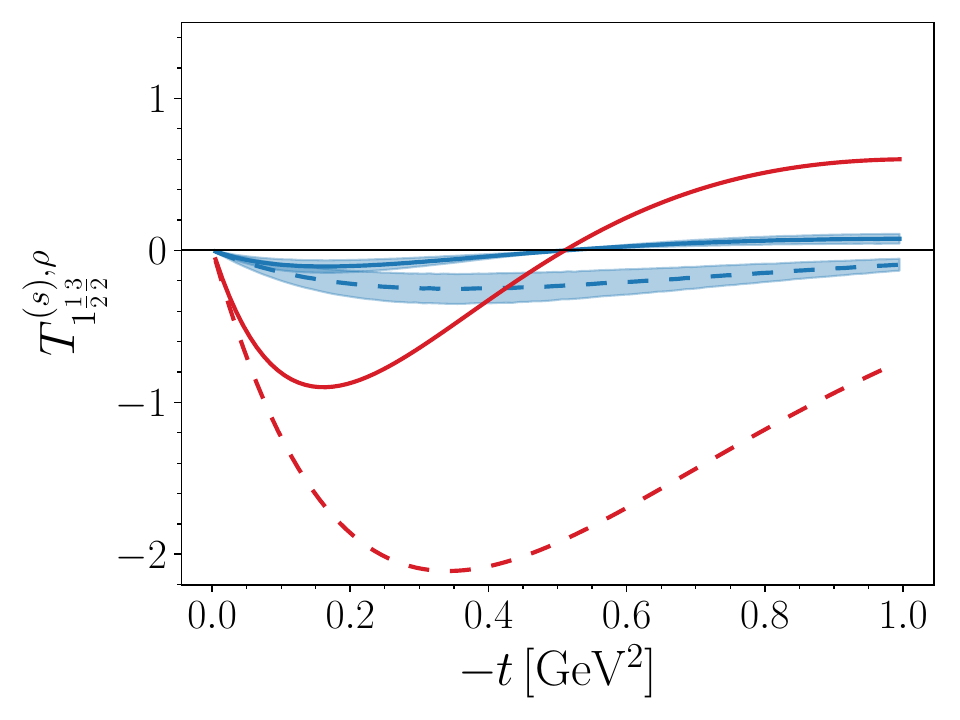}~
    \includegraphics[width=0.24\linewidth]{Amps_1m3_rho.pdf}~
    \includegraphics[width=0.24\linewidth]{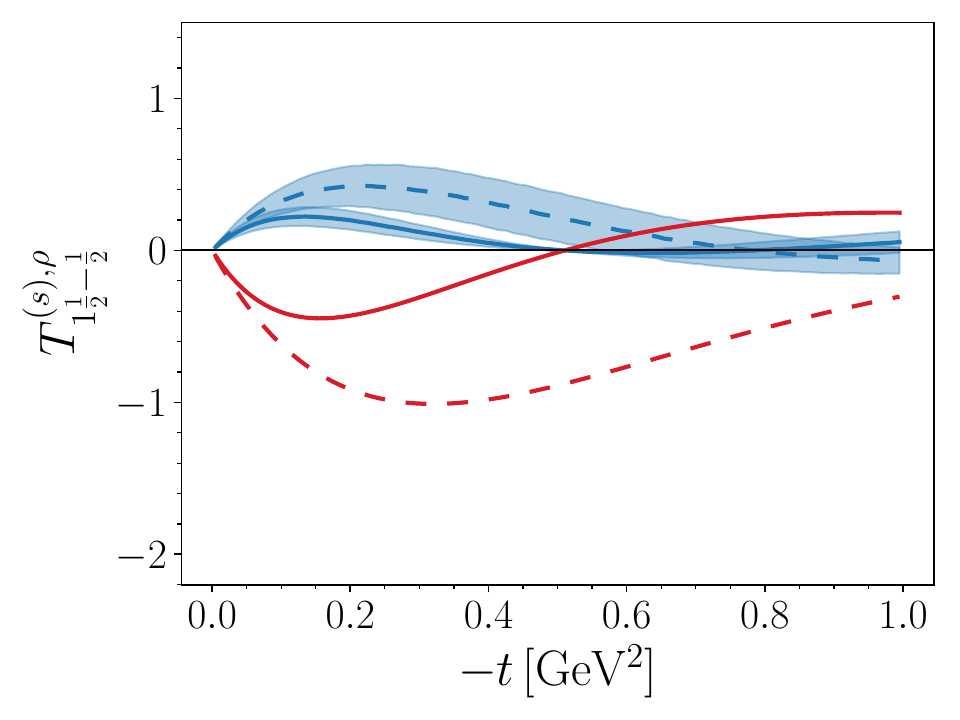}~
    \includegraphics[width=0.24\linewidth]{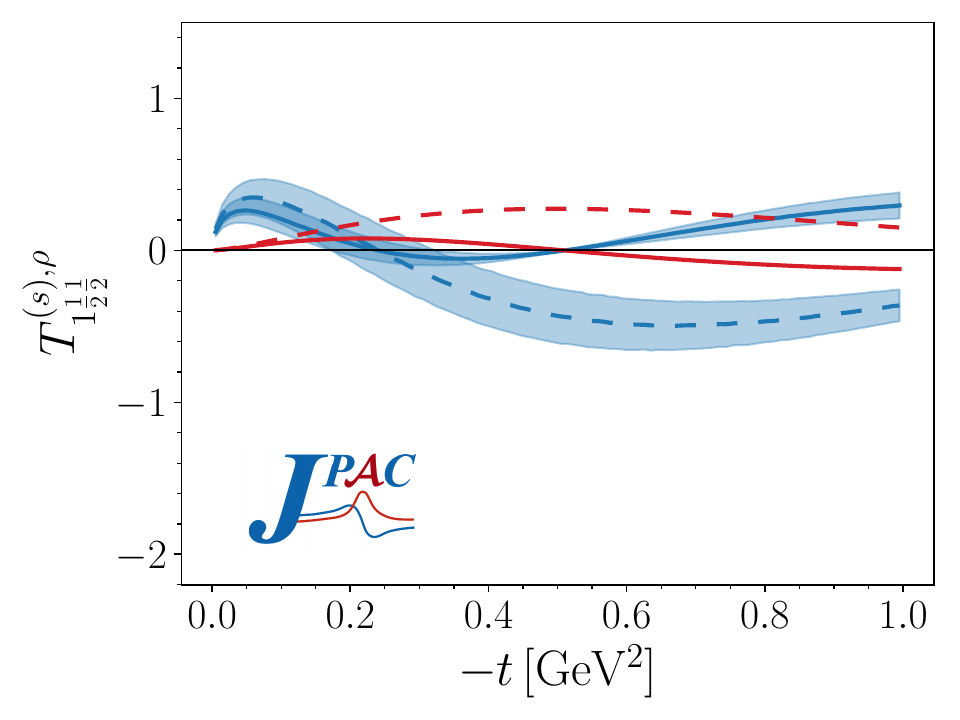}\\
    \includegraphics[width=0.24\linewidth]{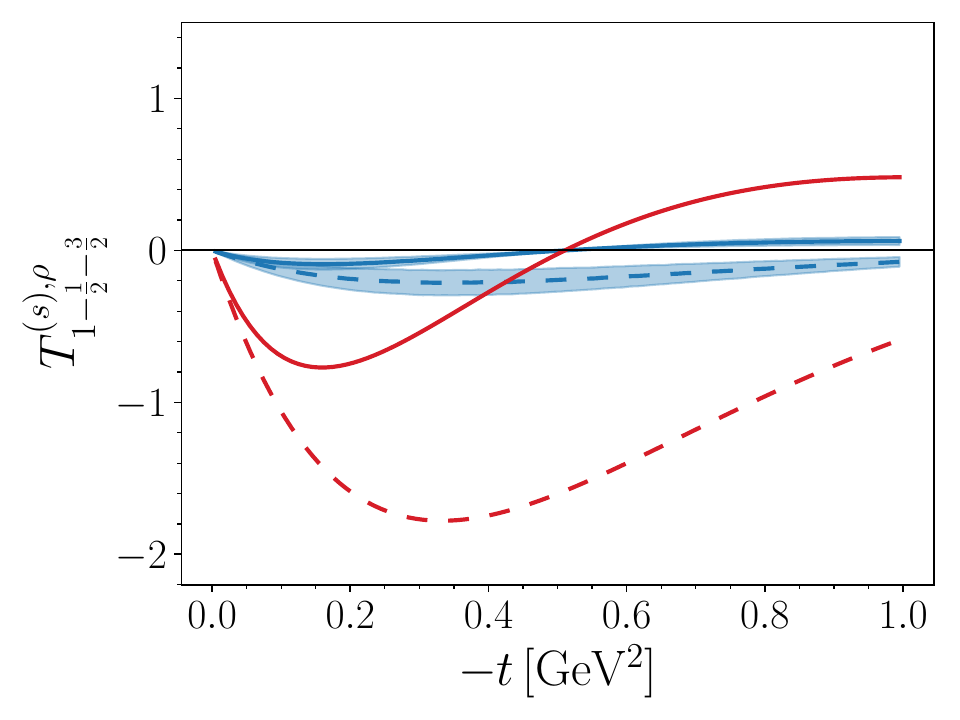}~
    \includegraphics[width=0.24\linewidth]{Amps_m13_rho.pdf}~
    \includegraphics[width=0.24\linewidth]{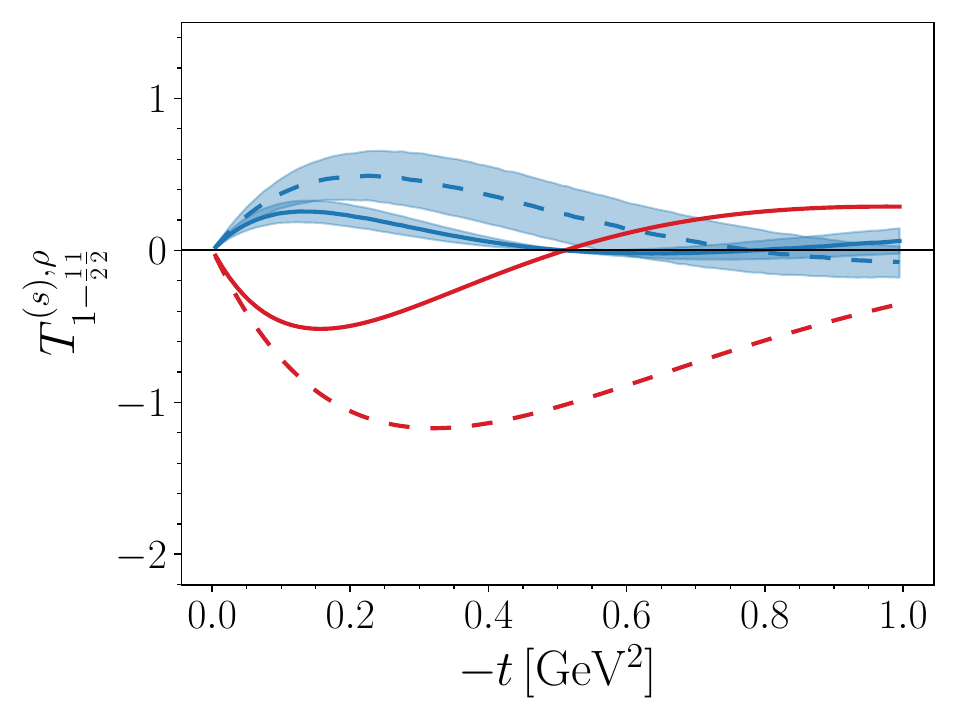}~
    \includegraphics[width=0.24\linewidth]{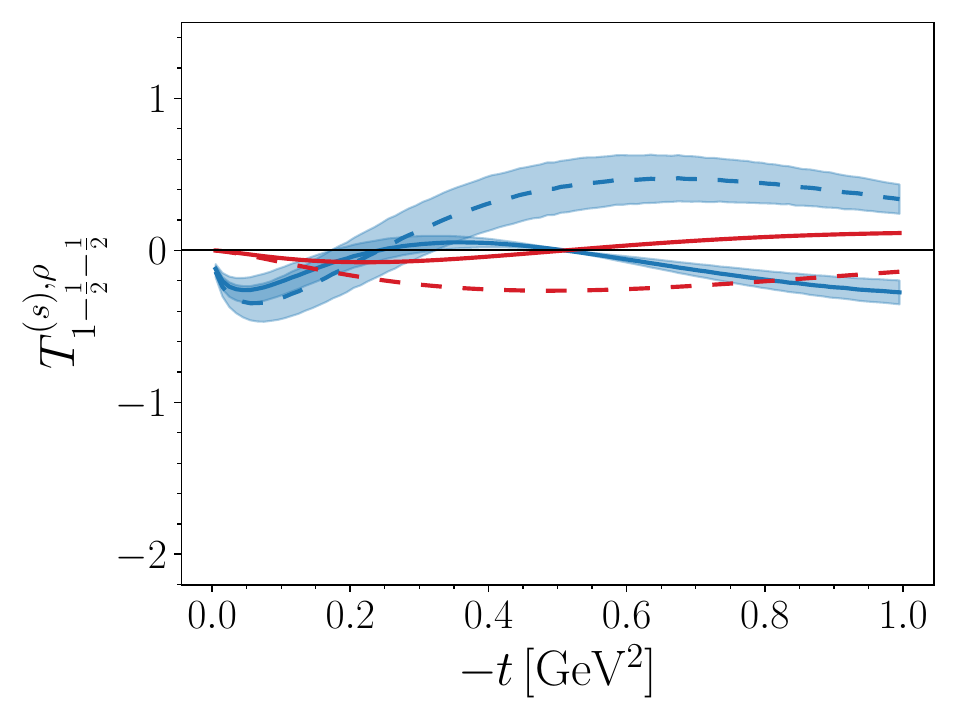}
    \caption{Same as Fig.~\ref{fig:ampindpi}, but for $\rho$ exchange.}
    \label{fig:ampindrho}
\end{figure*}

\begin{figure*}
    \centering
    \includegraphics[width=0.24\linewidth]{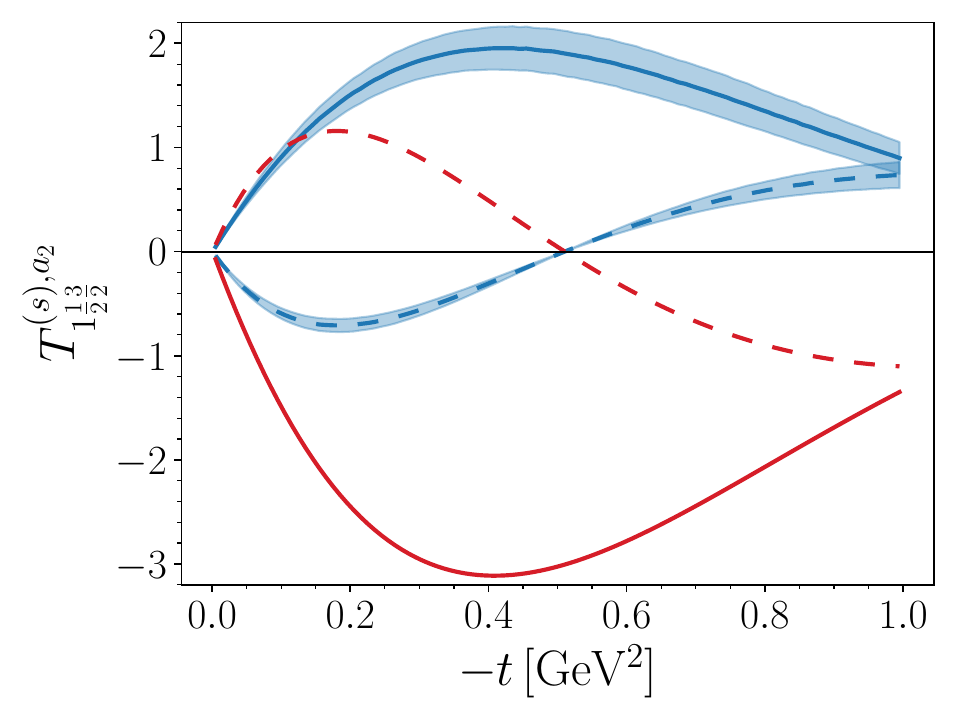}~
    \includegraphics[width=0.24\linewidth]{Amps_1m3_a2.pdf}~
    \includegraphics[width=0.24\linewidth]{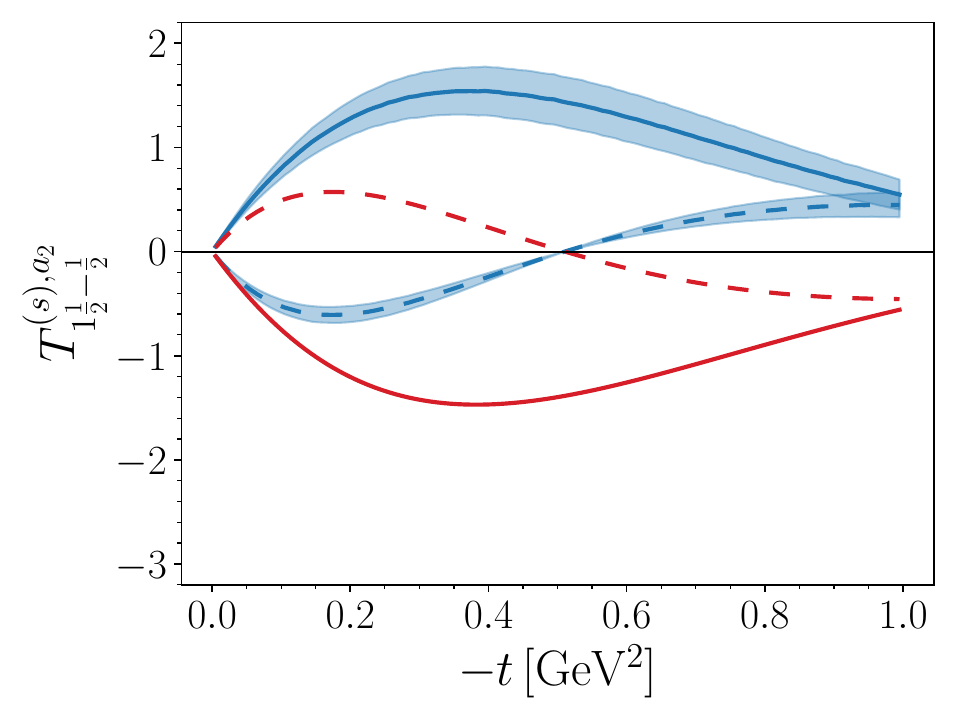}~
    \includegraphics[width=0.24\linewidth]{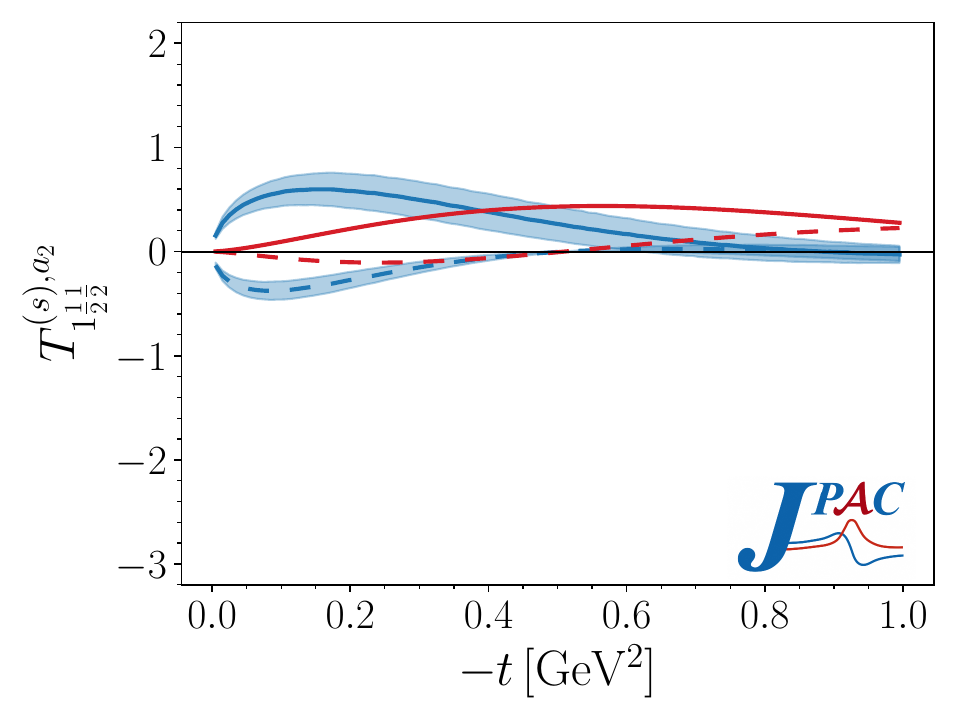}\\
    \includegraphics[width=0.24\linewidth]{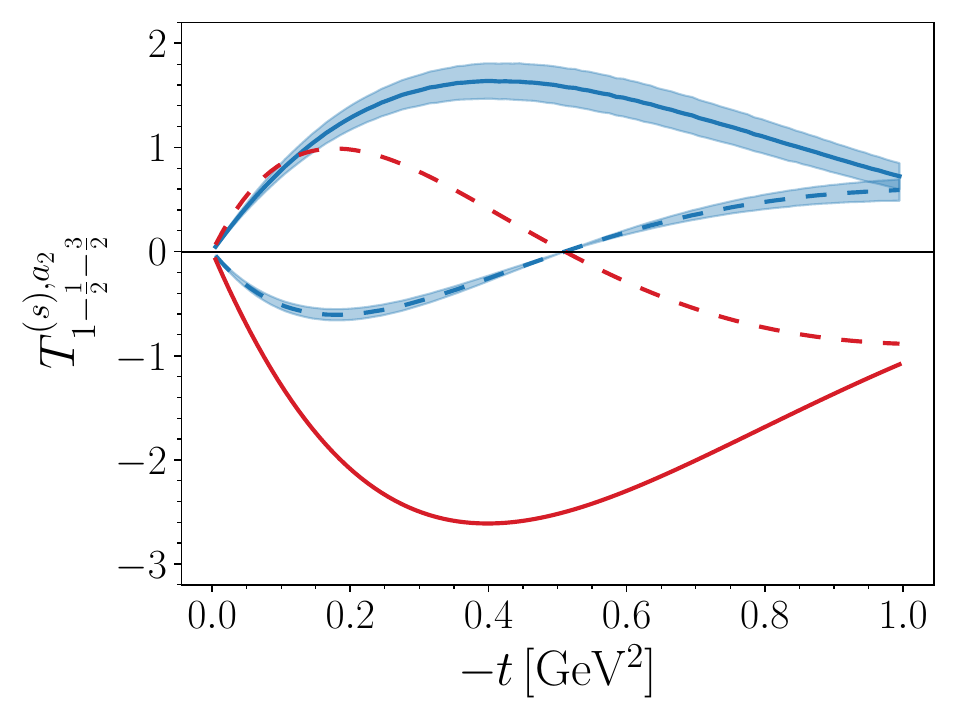}~
    \includegraphics[width=0.24\linewidth]{Amps_m13_a2.pdf}~
    \includegraphics[width=0.24\linewidth]{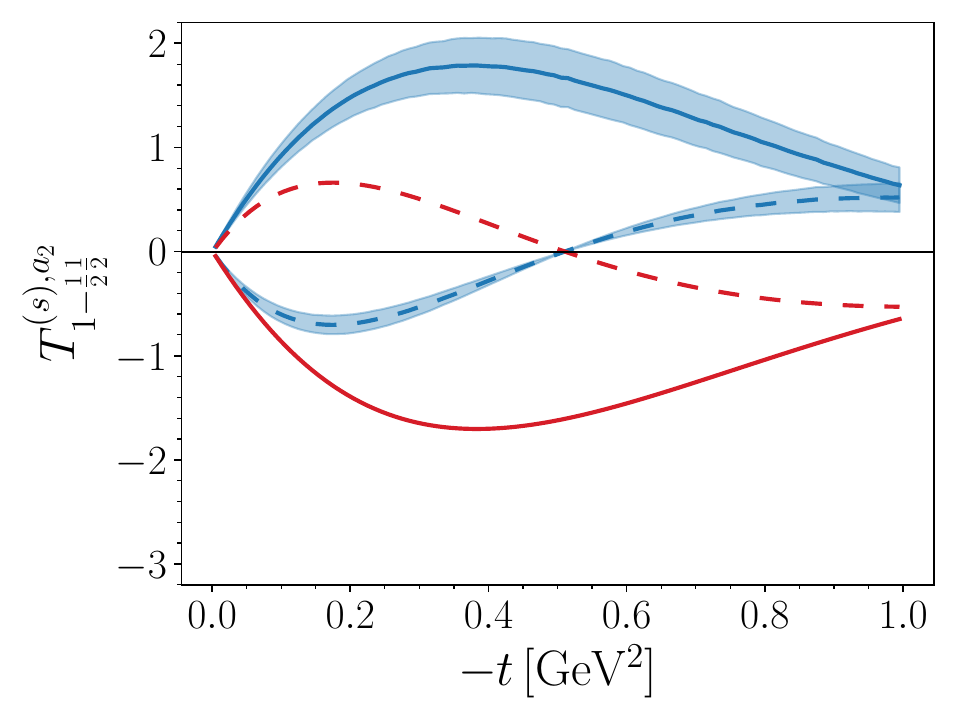}~
    \includegraphics[width=0.24\linewidth]{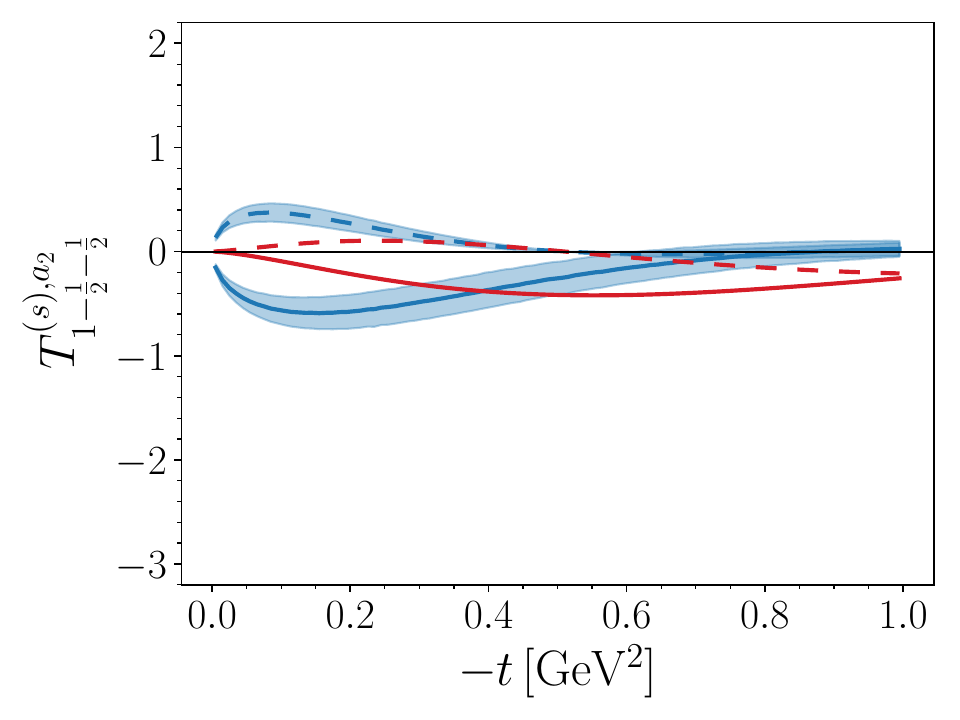}
    \caption{Same as Fig.~\ref{fig:ampindpi}, but for $a_2$ exchange.}
    \label{fig:ampinda2}
\end{figure*}

In this Appendix we give the plots of the $s$-channel helicity amplitudes as well as the contributions of different Reggeon exchanges. Fig.~\ref{fig:ampind} shows the $s$-channel helicity amplitudes.  Figs.~\ref{fig:ampindpi},~\ref{fig:ampindb1},~\ref{fig:ampindrho}, and ~\ref{fig:ampinda2} show the contributions of $\pi$, $b_1$, $\rho$, and $a_2$ exchanges respectively.
Each plot is for a given helicity with the solid lines representing the real parts of the amplitudes and the dashed lines giving the imaginary parts.

\section{Effective Lagrangians\label{app:EffLag}}
If an individual meson exchange amplitude in the $t$-channel frame is written in the form,
\begin{align}
    T^{J\eta}_{\helgammat\helprotont\helDelt} (t,s) &= \frac{2J+1}{4\pi} a^{J\eta}_{\helgammat\helprotont\helDelt} (t) d^J_{\helgammat,\helprotont-\helDelt} (z_t) \label{eq:AmpJLagt}
\end{align}
then we get the PCHAs free of kinematical singularities as (\cf Sec.~\ref{sec:formalism}),
\begin{align}
    \hat{a}^{J\eta}_{\helgammat,\helprotont,\helDelt} &= \frac{(2q^t p^t)^{M-J}}{K^\eta_{\helgammat\helprotont\helDelt}(t)} \frac{4\pi}{2J+1}~F^J_{\helgammat\helprotont\helDelt}~a^{J\eta}_{\helgammat,\helprotont,\helDelt}\text{ .}\label{eq:BSFKSFPCHALagt}
\end{align}
The residue can be evaluated using the formalism presented in Sec.~\ref{sec:formalism}. The residue extracted thuswise must be equal to the one from Eq.~\eqref{eq:residueFinal} since both represent the same process. This allows us to extract values of the coupling constants that represent the $RN\Delta$ vertex. To perform this extraction, we need to model the corresponding upper vertex, the coupling constants of which can be extracted from the respective radiative decay widths.\par

The amplitude for the decay of a natural parity spin-$J$ state to $\pi\gamma$ is given by~\cite{JointPhysicsAnalysisCenter:2024kck},
\begin{align}
    V^{RN}_{\helgammat \mu_R} &= g_{R\pi\gamma}(-q^t)^{J}\sqrt{t}~\helgammat~c_{J-1}C^{J\mu_R}_{J-1 0,1\helgammat}\delta_{\mu_R\helgammat} \label{eq:NradV}
\end{align}
and that for an unnatural parity state is~\cite{JointPhysicsAnalysisCenter:2024kck},
\begin{align}
    V^{RU}_{\helgammat \mu_R} &= -2g_{R\pi\gamma}(-q^t)^{J}\sqrt{t}~c_{J-1}C^{J\mu_R}_{J-1 0,1\helgammat}\delta_{\mu_R\helgammat} \label{eq:UradV}
\end{align}
where $\mu_R$ is the helicity of the parent state and $c_{k} = k!\,\sqrt{2^k/(2k)!}$.
At the lower vertex, a spin-$J$ state couples to the $\Delta\bar{p}$ system via the angular momentum channels,
\begin{align}
    |J-1|, J+1 &\to \text{Natural parity},\\
    |J-2|, J, J+2 &\to \text{Unnatural parity.}
\end{align}
The $L=J$, $L=|J-1|$, and $L=J+1$ channels are doubly degenerate giving a maximum of 4 channels (and hence, couplings) for each spin-$J$. We take these Lorentz structures to be,
\begin{align}
    \Gamma^{\alpha \mu_1\ldots\mu_J} &=
    \left\{\begin{matrix}
    g^{\mu_1\alpha} p^{\mu_2}_N \ldots p^{\mu_J}_N\\
    p_J^\alpha p^{\mu_1}_N \ldots p^{\mu_J}_N\\
    p_J^\alpha \gamma^{\mu_1} p^{\mu_2}_N \ldots p^{\mu_J}_N\\
    g^{\mu_1\alpha}\gamma^{\mu_2} p^{\mu_3}_N \ldots p^{\mu_J}_N
    \end{matrix} \right\}\otimes\begin{pmatrix}
        \gamma_5\\1
    \end{pmatrix}
\end{align}
where, the $\gamma_5$ is present when spin-$J$ state is of natural parity. Note that for $J<2$, only a subset of these vertices contribute. 
The amplitude for the $R\bar{N}\Delta$ vertex is given by,
\begin{align}
    V^{J}_{\mu_J\helprotont\helDelt} &= \bar{U}_\alpha (p_\Delta, \helDelt) \Gamma^{\alpha \nu_1\ldots\nu_J} v(p_{\bar{N}},\helprotont) \epsilon_{\nu_1\ldots\nu_J} (p_J,\mu_J)
\end{align}
The individual expressions can be read off of the final amplitude that we present in a moment. The full amplitude for a spin-$J$ exchange is given by,
\begin{align}
    T^J_{\helgammat\helprotont\helDelt} (t,z_t) &= \sum_{\mu_J} V^{J}_{\helgammat \mu_J}V^{J}_{\mu_J\helprotont\helDelt}\frac{\alpha^\prime}{J-\alpha(t)}\\
    &= a^J_{\helgammat\helprotont\helDelt} (t) d^J_{\helgammat,\helprotont-\helDelt}(z_t)
\end{align}
where the partial wave $a^J_{\helgammat\helprotont\helDelt} (t)$ is given by, 
\onecolumngrid
\begin{widetext}
\begin{align}
    a^{J}_{\helgammat\helprotont\helDelt} (t) &= 2\helprotont\helgammat(-E_\gamma)^{J}g^R_{\pi\gamma}\sqrt{t}C^{J\helgammat}_{J-1 0,1\helgammat}(-p)^{J-1} c_{J-1}^2 \sqrt{t-t_{pth}}\left[g^{(1)}_{R\bar{N}\Delta} C^{J\mu'}_{J-10;1\mu'} C^{\frac{3}{2}\helDelt}_{\frac{1}{2}\helprotont;1(-\mu')}\left(\delta_{\mu',\pm1}+\delta_{\mu',0}\frac{E_\Delta}{m_\Delta}\right)\right.\nonumber\\
    &\left. - g^{(2)}_{R\bar{N}\Delta}\frac{\sqrt{t}}{m_\Delta}p^2 \sqrt{\frac{J}{2J-1}} \sqrt{\frac{2}{3}} \delta_{\mu',0} - g^{(3)}_{R\bar{N}\Delta} \frac{(t-t_{th})}{2 m_\Delta} \sqrt{\frac23} \delta_{\helDelt,\pm\frac{1}{2}}C^{J\mu'}_{J-10;1\mu'} \left( \sqrt{\frac{t_{pth}}{t}} \delta_{\mu'0}+\sqrt{2} \delta_{\mu'\pm 1}\right)\right.\nonumber\\
    &\left. - g^{(4)}_{R\bar{N}\Delta}\Bigg\{ C^{J\mu'}_{J-1\mu';10}C^{J-1\mu'}_{J-20;1\mu'}C^{\frac32\helDelt}_{\frac12\mu_{\bar N};1(-\mu')} \sqrt{\frac{t_{pth}}{t}} \Big(\delta_{\mu_{\bar N},(\helDelt\pm 1)}+\delta_{\mu_{\bar N},\helDelt}\frac{E_\Delta}{m_\Delta}\Big) \right.\nonumber\\
    &\left. + C^{J\mu'}_{J-1(-\mu_{\bar N}-\helDelt);1(2\mu_{\bar N})}C^{J-1(-\mu_{\bar N}-\helDelt)}_{J-20;1(-\mu_{\bar N}-\helDelt)}C^{\frac32\helDelt}_{\frac12(-\mu_{\bar N)};1(\mu_{\bar N}+\helDelt)} \sqrt{2}\Big(\delta_{-\mu_{\bar N},(\helDelt\pm 1)}+\delta_{-\mu_{\bar N},\helDelt}\frac{E_\Delta}{m_\Delta}\Big)\Bigg\}\right.\nonumber\\
    &\left. \times\frac{2\sqrt{t}c_{J-2}}{(t-t_{pth})~c_{J-1}} \right] \frac{\alpha^\prime}{J-\alpha(t)}
\end{align}
for natural exchange and
\begin{align}
    a^{J}_{\helgammat\helprotont\helDelt} (t) &= -2g^J_{\pi\gamma}~(-E_\gamma)^J(-p)^{J-1}\sqrt{t}~c_{J-1}^2 C^{J\helgammat}_{J-10;1\helgammat}\sqrt{t-t_{th}}~\left[-g^{(1)}_{R\bar{N}\Delta}C^{J\mu'}_{J-10;1\mu'} C^{\frac{3}{2}\helDelt}_{\frac{1}{2}\helprotont;1(-\mu')} \left(\delta_{\mu',\pm1} + \delta_{\mu',0}\frac{E_\Delta}{m_\Delta} \right)\right.\nonumber\\
    &\left. - g^{(2)}_{R\bar{N}\Delta} p^2 \sqrt{\frac{J}{2J-1}}\sqrt{\frac{2}{3}}\frac{\sqrt{t}}{m_\Delta} \delta_{\mu',0} + g^{(3)}_{R\bar{N}\Delta} \sqrt{\frac{2}{3}} \frac{t-t_{pth}}{2m_\Delta} \delta_{\helDelt,\pm\frac{1}{2}} C^{J,\mu'}_{J-10;1\mu'}\left(\sqrt{\frac{t_{th}}{t}}\delta_{\mu',0} + \sqrt{2} \delta_{\mu',\pm1}\right)\right.\nonumber\\
    &\left. + g^{(4)}_{R\bar{N}\Delta} \left\{C^{J\mu'}_{J-1\mu';10} C^{J-1\mu'}_{J-20;1\mu'} C^{\frac{3}{2}\helDelt}_{\frac{1}{2}\helprotont;1(-\mu')} \sqrt{\frac{t_{th}}{t}} \left(\delta_{\helprotont,\helDelt\pm1} + \delta_{\mu',0} \frac{E_\Delta}{m_\Delta} \right) \right.\right.\nonumber\\
    &\left.\left. +\sqrt{2} C^{J\mu'}_{J-1(-\helprotont-\helDelt);1(2\helprotont)} C^{J-1(-\helprotont-\helDelt)}_{J-20;1(-\helprotont-\helDelt)} C^{\frac{3}{2}\helDelt}_{\frac{1}{2}-\helprotont;1(\helprotont+\helDelt)} \left(\delta_{-\helprotont,\helDelt\pm1} + \delta_{-\helprotont,\helDelt}\frac{E_\Delta}{m_\Delta} \right) \right\}\right.\nonumber\\
    &\left. \times \frac{2\sqrt{t} c_{J-2}}{c_{J-1} (t-t_{th)}} \right] \frac{\alpha^\prime}{J-\alpha(t)}
\end{align}
for unnatural exchange, and $\mu' = \helprotont-\helDelt$. The residues can be evaluated by substituting Eq.~\eqref{eq:BSFKSFPCHALagt} in Eq.~\eqref{eq:residueLag} and taking the limit $t \to m_J^2$. Below  we list the residues for $b_1\,,\,\rho\,,\&\,a_2$ exchanges.
\begin{align}
    \mathcal{R}^{b_1}_{\helgammat,\helprotont,\helDelt} &= \Bigg[\frac{g_{b_1\pi\gamma}}{m_{b_1}}\frac{16\pi}{2^{M}3\sqrt{t}^{M+N-2}} (t-t_{th})~\left[-g^{(1)}_{b_1\bar{N}\Delta} C^{\frac{3}{2}\helDelt}_{\frac{1}{2}\helprotont;1(-\mu')} \left(\delta_{\mu',\pm1} + \delta_{\mu',0}\frac{E_\Delta}{m_\Delta} \right) - \frac{g^{(2)}_{b_1\bar{N}\Delta}}{t_{th}} p^2 \sqrt{\frac{2}{3}}\frac{\sqrt{t}}{m_\Delta} \delta_{\mu',0}\right.\nonumber\\
    &\left. + \frac{g^{(3)}_{b_1\bar{N}\Delta}}{W_{th}} \sqrt{\frac{2}{3}} \frac{t-t_{pth}}{2m_\Delta} \delta_{\helDelt,\pm\frac{1}{2}}\left(\sqrt{\frac{t_{th}}{t}}\delta_{\mu',0} + \sqrt{2} \delta_{\mu',\pm1}\right) \right] \Bigg]_{t\to m_{b_1}^2} F^1_{\helgammat,\helprotont,\helDelt}\label{eq:lagb1Res}\\
    \mathcal{R}^{\rho}_{\helgammat,\helprotont,\helDelt} &= \Bigg[-\frac{g_{\rho\pi\gamma}}{m_\rho}\frac{t~4\pi(2\helprotont)\helgammat}{2^{M}3\sqrt{t}^{N+1}}(t-t_{pth})\left[g^{(1)}_{\rho\bar{N}\Delta} C^{\frac{3}{2}\helDelt}_{\frac{1}{2}\helprotont;1(-\mu')} \left(\delta_{\mu',\pm1}+\delta_{\mu',0}\frac{E_\Delta}{m_\Delta}\right)  - \frac{g^{(2)}_{\rho\bar{N}\Delta}}{t_{th}}\frac{\sqrt{t}}{m_\Delta}p^2 \sqrt{\frac{2}{3}} \delta_{\mu',0} \right.\nonumber\\
    &\left. - \frac{g^{(3)}_{\rho\bar{N}\Delta}}{W_{th}} \frac{(t-t_{th})}{2 m_\Delta} \sqrt{\frac23} \delta_{\helDelt,\pm\frac{1}{2}} \left( \sqrt{\frac{t_{pth}}{t}} \delta_{\mu',0}+\sqrt{2} \delta_{\mu',\pm 1}\right) \right]\Bigg]_{t\to m_\rho^2} F^1_{\helgammat,\helprotont,\helDelt}\label{eq:LagRhoRes}\\
    \mathcal{R}^{a_2}_{\helgammat,\helprotont,\helDelt} &= \Bigg[-\frac{ g_{a_2\pi\gamma}}{m_{a_2}^2}\frac{t~2\pi(2\helprotont)\helgammat}{2^{M}5\sqrt{t}^{M+N}}C^{2\helgammat}_{1 0,1\helgammat} (t-t_{pth})\left[g^{(1)}_{a_2\bar{N}\Delta} C^{2\mu'}_{10;1\mu'} C^{\frac{3}{2}\helDelt}_{\frac{1}{2}\helprotont;1(-\mu')}\left(\delta_{\mu',\pm1}+\delta_{\mu',0}\frac{E_\Delta}{m_\Delta}\right) \right.\nonumber\\
    &\left.  - \frac{g^{(2)}_{a_2\bar{N}\Delta}}{t_{th}}\frac{\sqrt{t}}{m_\Delta}p^2 \frac{2}{3} \delta_{\mu',0}  - \frac{g^{(3)}_{a_2\bar{N}\Delta}}{W_{th}} \frac{(t-t_{th})}{2 m_\Delta} \sqrt{\frac23} \delta_{\helDelt,\pm\frac{1}{2}}C^{2\mu'}_{10;1\mu'} \left( \sqrt{\frac{t_{pth}}{t}} \delta_{\mu'0}+\sqrt{2} \delta_{\mu'\pm 1}\right)\right.\nonumber\\
    &\left. - g^{(4)}_{a_2\bar{N}\Delta} W_{th}\Bigg\{ C^{2\mu'}_{1\mu';10} C^{\frac32\helDelt}_{\frac12\mu_{\bar N};1(-\mu')} \sqrt{\frac{t_{pth}}{t}} \Big(\delta_{\mu_{\bar N},(\helDelt\pm 1)}+\delta_{\mu_{\bar N},\helDelt}\frac{E_\Delta}{m_\Delta}\Big) \right.\nonumber\\
    &\left. + C^{2\mu'}_{1(-\mu_{\bar N}-\helDelt);1(2\mu_{\bar N})} C^{\frac32\helDelt}_{\frac12(-\mu_{\bar N)};1(\mu_{\bar N}+\helDelt)} \sqrt{2}\Big(\delta_{-\mu_{\bar N},(\helDelt\pm 1)}+\delta_{-\mu_{\bar N},\helDelt}\frac{E_\Delta}{m_\Delta}\Big)\Bigg\}\frac{2\sqrt{t}}{(t-t_{pth})} \right]\Bigg]_{t\to m_{a_2}^2} F^2_{\helgammat,\helprotont,\helDelt}\text{ .}\label{eq:laga2Res}
\end{align}
Note that the coupling constants have been scaled by the appropriate powers of the masses or thresholds so that they are all dimensionless.\par
\end{widetext}
\twocolumngrid

\bibliographystyle{apsrev4-2}
\bibliography{Ref}

\end{document}